\RequirePackage{fix-cm}

\documentclass[twocolumn]{svjour3}          %
\smartqed  %
\journalname{IJCV}%

\usepackage{booktabs}
\usepackage{float}
\usepackage{multicol}
\usepackage{multirow}

\usepackage[dvipsnames]{xcolor}
\usepackage{graphicx}
\usepackage{float}

\usepackage[pagebackref=true,breaklinks=true,colorlinks,bookmarks=false]{hyperref}

\usepackage{algorithm}
\usepackage{algpseudocode}

\newcommand{\fontsf}[1]{\textsf{\small #1}}
\newcommand{\selfdeblur}{\textsf{SelfDeblur}}
\newcommand{\explore}{\textsf{Explore}}

\usepackage[leqno,fleqn,intlimits]{empheq}

\usepackage{graphicx,amsmath,amsfonts,amscd,amssymb,bm,url,latexsym}
\usepackage{physics}
\allowdisplaybreaks
\usepackage[capitalize,nameinlink]{cleveref}

\renewcommand{\mathbf}{\boldsymbol}

\newcommand{\mb}{\mathbf}
\newcommand{\mc}{\mathcal}

\newcommand{\bb}{\mathbb}

\newcommand{\paren}{\pqty}

\newcommand{\wh}{\widehat}

\begin{document}

\title{Blind Image Deblurring with Unknown Kernel Size and Substantial Noise}

\author{
Zhong Zhuang \and
Taihui Li \and
Hengkang Wang \and
Ju Sun
}

\institute{
Z. Zhuang \at
Electrical and Computer Engineering \\
University of Minnesota \\
\email{zhuan143@umn.edu}
\and
T. Li \at
Computer Science and Engineering \\
University of Minnesota \\
\email{lixx5027@umn.edu}
\and
H. Wang \at
Computer Science and Engineering \\
University of Minnesota \\
\email{wang9881@umn.edu}
\and
J. Sun \at
Computer Science and Engineering \\
University of Minnesota \\
\email{jusun@umn.edu}
}

\maketitle
\hbadness=10000
\hfuzz=\maxdimen
\newcount\hbadness
\newdimen\hfuzz

\abstract{
Blind image deblurring (BID) has been extensively studied in computer vision and adjacent fields. Modern methods for BID can be grouped into two categories: single-instance methods that deal with individual instances using statistical inference and numerical optimization, and data-driven methods that train deep-learning models to deblur future instances directly. Data-driven methods can be free from the difficulty in deriving accurate blur models, but are fundamentally limited by the diversity and quality of the training data---collecting sufficiently expressive and realistic training data is a standing challenge. \emph{In this paper, we focus on single-instance methods that remain competitive and indispensable}. However, most such methods do not prescribe how to deal with unknown kernel size and substantial noise, precluding practical deployment. Indeed, we show that several state-of-the-art (SOTA) single-instance methods are unstable when the kernel size is overspecified, and/or the noise level is high. On the positive side, we propose a practical BID method that is stable against both, \emph{the first of its kind}. Our method builds on the recent ideas of solving inverse problems by integrating physical models and structured deep neural networks, \emph{without extra training data}. We introduce several crucial modifications to achieve the desired stability. Extensive empirical tests on standard synthetic datasets, as well as real-world \texttt{NTIRE2020} and \texttt{RealBlur} datasets, show the superior effectiveness and practicality of our BID method compared to SOTA single-instance as well as data-driven methods. The code of our method is available at \url{https://github.com/sun-umn/Blind-Image-Deblurring}. 
}


\section{Introduction}\label{sec:introduction}
\begin{figure*}[!htbp]
    \centering  
    \includegraphics[width=0.95\linewidth]{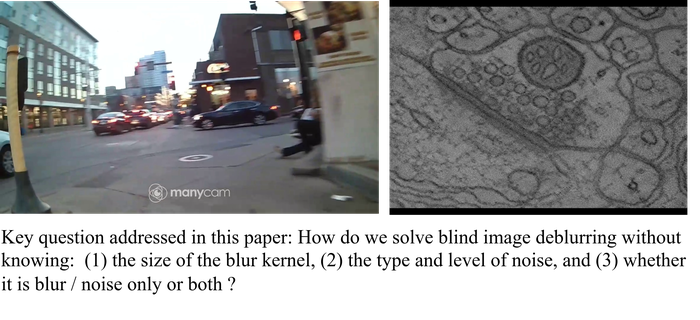}
    \vspace{-1em}
    \caption{Given a blurry and potentially also noisy image, how to perform reliable blind image deblurring? The kernel size and the noise type/level are typically unknown, and the image may contain blur or noise only, or both. \textbf{Left}: A street scene captured by a camera mounted on a rapidly moving e-scooter (image captured by Le Peng and Wenjie Zhang of the authors' group; permission granted); \textbf{Right}: A biological specimen captured by a realistic microscopy system (Image CCDB:3684 from the Cell Image Library; source url: \url{http://cellimagelibrary.org/images/CCDB_3684}; created by Mark Ellisman, Gina Sosinsky, Ying Jones licensed under CC BY 3.0). }
    \label{fig:question}
\end{figure*}
\begin{figure*}[!htbp]
  \centering

  \includegraphics[width=0.95\linewidth]{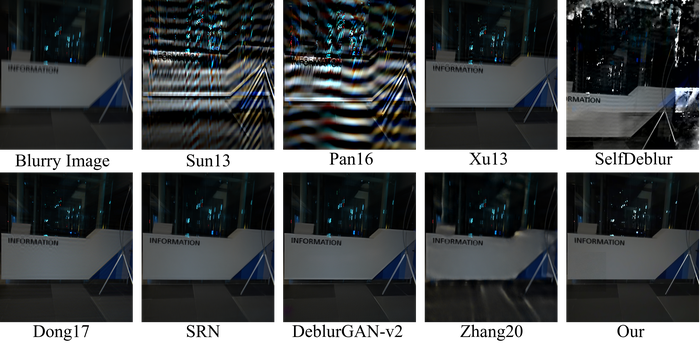}
  \vspace{-1em}
  \caption{Deblurring results of several SOTA single-instance and data-driven BID methods on a real-world blurry image taken from \cite{RimEtAl2020Real}. The $6$ single-instance methods are: Sun13~\cite{SunEtAl2013Edge}, Pan16~\cite{PanEtAl2016Blind}, Xu13~\cite{XuEtAl2013Unnatural}, Dong17~\cite{DongEtAl2017Blind}, \selfdeblur~\cite{RenEtAl2020Neural}, and our method proposed in this paper (see \cref{sec:method_pipeline}); $3$ data-driven methods are: SRN~\cite{tao2018scale}, DeblurGAN-v2~\cite{kupyn2019deblurgan}, Zhang20~\cite{zhang2020deblurring}, for which we directly take their pretrained models. }
  \label{fig:bd_results_intro_overview}
\end{figure*}
Image blur is mostly caused by the optical nonideality of the camera (e.g., defocus, lens distortion), i.e., \emph{optical blur}, and relative motions between the scene and the camera, i.e., \emph{motion blur}~\cite{Szeliski2021Computer,KundurHatzinakos1996Blind,JoshiEtAl2008PSF,LevinEtAl2011Understanding,KoehlerEtAl2012Recording,LaiEtAl2016Comparative,KohEtAl2021Single,SunDonoho2021Convex}. It is often coupled with noticeable sensory noise, e.g. when one images fast-moving objects in low-light environments. Thus, in the simplest form, image blur is often modeled as 
\begin{equation} \label{eq:bd_model}
    \mb y = \mb k \ast \mb x + \mb n,  
\end{equation}
where $\mb y$ is the observed blurry and noisy image, and $\mb k$, $\mb x$, $\mb n$ are the blur kernel, clean image, and additive sensory noise, respectively. The notation $\ast$ here is linear convolution, which encodes the assumption that the blur effect is uniform over the spatial domain. When there are complicated 3D motions (e.g., multiple independently moving objects, and 3D in-plane rotations), or substantial depth variations, this model can be upgraded to account for the non-uniform blur effect~\cite{LevinEtAl2011Understanding,KoehlerEtAl2012Recording,LaiEtAl2016Comparative,KohEtAl2021Single}. \emph{In this paper, we focus on the uniform setting and leave the non-uniform setting as future work}. 

Assume the model in \cref{eq:bd_model}. Given $\mb y$ and $\mb k$, estimating $\mb x$ is called (non-blind) \emph{deconvolution}, a linear inverse problem that is relatively easy to solve. However, in practice, $\mb k$---including its size and numerical value---is often unavailable. For example, neither defocus nor motions can be reliably estimated in wild environments~\cite{KundurHatzinakos1996Blind} (see, e.g., \cref{fig:question}). This leads to \emph{blind deconvolution} (BD), where $\mb k$ and $\mb x$ are estimated together from $\mb y$. 

Over the past decades, a rich set of ideas have been developed to tackle BID and BD, evolving from \emph{single-instance methods} that rely on analytical processing or statistical inference and numerical optimization to solve one instance each time, to modern \emph{data-driven methods} that aim to train deep learning (DL) models to solve all future instances. The sequence of landmark review articles~\cite{KundurHatzinakos1996Blind,LevinEtAl2011Understanding,KoehlerEtAl2012Recording,LaiEtAl2016Comparative,KohEtAl2021Single,ZhangEtAl2022Deep} chronicle these developments; see also \cref{sec:bd_background} below. Evaluation has also moved from synthetic to real-world data, best exemplified by the recent NTIRE 2020/2021 challenges on real-world image deblurring ~\cite{NahEtAl2020NTIRE,NahEtAl2021NTIRE}. 

\emph{In this paper, we focus on single-instance methods for BID}. Although recent data-driven methods have shown great promise, as statistical learning methods, they are intrinsically limited by the training data: if trained with sufficiently diverse and realistic data, these methods are likely to generalize well. However, the collection of high-quality training sets that meet the demand has been identified as a continuing challenge~\cite{KohEtAl2021Single,ZhangEtAl2022Deep}. Therefore, single-instance methods will likely be a mainstay alongside data-driven methods for practical BID, especially for scenarios where relevant data are rare or expensive to collect. 

Prior single-instance methods for BID seem vague on three critical issues toward practicality: (1) \emph{unknown kernel ($\mb k$) size}: Except for methods that directly estimate $\mb x$ only (e.g., the inverse filtering approach to BD~\cite{Wiggins1978Minimum,Donoho1981MINIMUM,Cabrelli1985Minimum,SunDonoho2021Convex}), a nearly-optimal estimate of the kernel size is needed~\cite{SiYaoEtAl2019Understanding}. But it is practically unclear how such an accurate estimate can be reliably obtained, and how sensitive the existing methods are to kernel-size misspecification;  (2) \emph{substantial noise ($\mb n$)}: Sensory noise after convolution may still be substantial, while most previous methods assume noise-free or low-noise settings in their evaluations~\cite{TaiLin2012Motion,ZhongEtAl2013Handling,PanEtAl2016Robust,DongEtAl2017Blind,GongEtAl2017Self,ChenEtAl2020OID}; and (3) \emph{model stability}: The image may be blurry only, noisy only, or both. Whatever the case, in practice, an ideal BID method should work seamlessly across the different regimes. This has rarely been tested for prior methods. These three issues are summarized in \cref{fig:question}. 

To quickly confirm these practicality issues, we pick $6$ state-of-the-art (SOTA) single-instance BID methods (plus $3$ representative data-driven methods by taking their pretrained models), and test them on a real-world image taken in a low-light setting, \emph{with unknown kernel size and unknown noise type/level}. We specify a kernel size that is half of the image size in each dimension to provide a loose upper bound. \cref{fig:bd_results_intro_overview} shows how miserably these single-instance methods can fail; more failures can be checked in \cref{sec:expriments}. 

This paper aims to address these practicality issues. We follow the major modeling ideas in the statistical inference and optimization approach to BID, but parametrize both the kernel and the image using trainable structured deep neural networks (DNNs). This idea has recently been independently introduced to BID by \cite{WangEtAl2019Image}, \cite{RenEtAl2020Neural} (\fontsf{SelfDeblur}), and \cite{TranEtAl2021Explore}, inspired by the remarkable success of deep image prior (DIP)~\cite{UlyanovEtAl2020Deep} and its variants~\cite{HeckelHand2019Deep,SitzmannEtAl2020Implicit} in solving a variety of inverse problems in computer vision and imaging~\cite{DarestaniHeckel2021Accelerated,GandelsmanEtAl2019Double,SitzmannEtAl2020Implicit,TancikEtAl2020Fourier,QayyumEtAl2021Untrained} and beyond~\cite{RavulaDimakis2019One,MichelashviliWolf2019Speech}. Our key contributions include 
\begin{itemize}
    \item \textbf{identifying three practicality issues of SOTA single-instance BID methods, including \selfdeblur}. As far as we are aware, this is the first time these three practicality issues have been discussed and addressed together in the BID literature.  \textbf{BID with these three issues is a more difficult but practical version than what SelfDeblur and most classical BID methods target}. This is also the first time both classical and SOTA data-driven BID methods are systematically evaluated in the simultaneous presence of the three issues; see \cref{sec:method_ingredients} and \cref{sec:exp_synthetic_data}; 
    \item \textbf{revamping \selfdeblur\,with six crucial modifications to address the three issues}. In \cref{sec:method_ingredients}, we clearly describe our modifications, as well as the rationale and intuitions behind them. Figuring out these modifications and their right combination is a highly nontrivial task, making our algorithm pipeline sufficiently different from \selfdeblur.  
    \item \textbf{systematic evaluation of our method against SOTA single-instance BID methods on synthetic SOTA datasets, and against SOTA data-driven BID methods on real world datasets}, confirming the superior effectiveness and practicality of our method (\cref{sec:expriments}; \cref{fig:bd_results_intro_overview} gives a quick preview). We also pinpoint the failure modes and limitations of our method in \cref{sec:failure}.  
\end{itemize}
\section{Background}\label{sec:related_work}

\subsection{Blind deconvolution (BD)}  \label{sec:bd_background}
BD refers to the nonlinear inverse problem of estimating $(\mb k, \mb x)$ from $\mb y$ according to the model in \cref{eq:bd_model}, and finds applications in numerous fields such as seismology~\cite{Wiggins1978Minimum,Donoho1981MINIMUM}, digital communications~\cite{VembuEtAl1994Convex,DingLuo2000fast}, neuroscience~\cite{Lewicki1998review,EkanadhamEtAl2011blind}, microscopy~\cite{CheungEtAl2020Dictionary}, and computer vision.

Due to the bilinear mapping $(\mb k, \mb x) \mapsto \mb k \ast \mb x$, $(\mb \delta, \mb k\ast \mb x)$ is always a trivial solution, where $\mb \delta$ is the Dirac delta function. Therefore, without further restrictions to $\mb k$ and $\mb x$, recovery is hopeless. To ensure identifiability, different domain-specific priors have been proposed over time. A popularly used prior across these domains is that $\mb x$ is (approximately) ``sparse'' in an appropriate sense~\cite{Wiggins1978Minimum,Donoho1981MINIMUM,Cabrelli1985Minimum,SunDonoho2021Convex,VembuEtAl1994Convex,DingLuo2000fast,Lewicki1998review,EkanadhamEtAl2011blind,CheungEtAl2020Dictionary}. 
For BID, $\mb x$ as the natural image to be recovered is often assumed to be sparse in the gradient domain. Furthermore, $\mb k$ is often ``short'' or ``small'', as characteristic patterns are often narrowly confined in their temporal or spatial extents~\cite{Lewicki1998review,EkanadhamEtAl2011blind,CheungEtAl2020Dictionary}. For BID, the blur kernel, either optical or motion, tends to be smaller in support, if not much, than the size of the blurry image itself. Therefore, the goal of many BD applications is to solve this \emph{short-and-sparse BD} (SSBD). 

Another notable feature of BD caused by the bilinear mapping $(\mb k, \mb x) \mapsto \mb k \ast \mb x$ is trivial symmetries. If we assume $\mb k$ and $\mb x$ are 1-dimensional infinite sequences---they can still have finite supports, then $\mb k \ast \mb x = (\frac{1}{\alpha} \mb k_{-\tau}) \ast (\alpha \mb x_{\tau})$ for any $\alpha \ne 0$ and $\tau \in \bb Z$, where $\mb v_\tau$ for any $\mb v$ means shifting $\mb v$ by $\tau$ time step. In other words, we have scale and shift symmetries. So, recovery is up to these symmetries, which often suffices for practical purposes. When we take a finite-window observation of $\mb k \ast \mb x$, a more faithful model is 
\begin{equation} \label{eq:bd_true_model}
    \mb y = \mc T\paren{\mb k \ast \mb x} + \mb n,  
\end{equation}
where $\mc T$ models the truncation effect of the window. The shift symmetry and the truncation effect together, if not handled appropriately, can easily lead to algorithmic failures, as discussed in~\cref{sec:method_over_k,sec:method_over_x}. 

On the theoretical front, \cite{Donoho1981MINIMUM,SunDonoho2021Convex,ChoudharyMitra2014Sparse,LiEtAl2015Unified,LiEtAl2017Identifiability,KechKrahmer2017Optimal} discuss the identifiability of BD under different priors. For guaranteed recovery, \cite{AhmedEtAl2014Blind,Chi2016Guaranteed,LiEtAl2019Rapid} assume $\mb k$ and/or $\mb x$ lying on random subspaces, and \cite{ZhangEtAl2017Global,ZhangEtAl2020Structured,KuoEtAl2020Geometry} work on SSBD under certain probabilistic generative models on $\mb x$. In addition, \cite{WipfZhang2014Revisiting} derives insights on different priors and formulations for BD from a Bayesian perspective. 

\subsection{BD specialized to blind image deblurring (BID)}
\label{sec:bd_bid}
For BID, SSBD is often solved with additional kernel- and/or image-specific priors. A subset of early BID methods write $\mb k$ in parametrized analytical forms, e.g., Gaussian shaped, and solve BID with simple analytical or computational steps~\cite{KundurHatzinakos1996Blind}. This has been largely superseded by the statistical inference and numerical optimization approach over the past decade, which formulates SSBD as regularized optimization problems, often interpreted as Maximum A Posterior (MAP) estimation: 
\begin{equation} \label{eq:deblur_regfit_generic}
    \min_{\mb k, \mb x}\; \underbrace{\ell\paren{\mb y, \mb k \ast \mb x}}_{\text{data fitting}} + \underbrace{\lambda_{\mb k} R_{\mb k}\paren{\mb k}}_{\text{regularizing $\mb k$}} + \underbrace{\lambda_{\mb x} R_{\mb x}\paren{\mb x}}_{\text{regularizing $\mb x$}}, 
\end{equation} 
where $\lambda_{\mb k}, \lambda_{\mb x}$ are regularization parameters. A canonical choice is $\ell\paren{\mb y, \mb k \ast \mb x} = \norm{\mb y - \mb k \ast \mb x}_2^2$, and $R_{\mb x}\paren{\mb x} = \norm{\nabla \mb x}_1$ (i.e., total-variation, or TV, norm on $\mb x$) to encode sparsity in the gradient. But since $\mb k \ast \mb x = \paren{\frac{1}{\alpha} \mb k} \ast (\alpha \mb x)$ and $\norm{\nabla (\alpha \mb x)}_1 = \abs{\alpha} \norm{\nabla \mb x}_1$ for any $\alpha \ne 0$, without any further constraint the global solution is when $\mb x = \mb 0$. So a considerable chunk of recent research is about dealing with the scaling issue together with better sparsity encoding: 
\begin{itemize}
    \item $\mb k \ge \mb 0, \sum_i k_i = 1, R_{\mb x}\paren{\mb x} = \norm{\nabla \mb x}_1$: This is a classical remedy~\cite{ChanWong1998Total}, but is shown to prefer the trivial solution with $\mb k = \mb \delta$ in certain regimes~\cite{LevinEtAl2011Understanding}. In fact, the trivial solution can occur even if one takes $R_{\mb x}\paren{\mb x} = \norm{\nabla \mb x}_q$ ($q \in (0, 1]$), considerably tighter sparsity proxies. Nonetheless, perhaps surprisingly, carefully chosen algorithms can find nontrivial local solutions that lead to good recovery~\cite{PerroneFavaro2014Total}. 
    
    \item $\mb k \ge \mb 0, \sum_i k_i = 1$, $R_{\mb x}\paren{\mb x} = \frac{\norm{\nabla \mb x}_1}{\norm{\nabla \mb x}_2}$ or $\norm{\nabla \mb x}_0$: The high-level intuition why the above may prefer the trivial solution $\paren{\mb \delta, \mb k \ast \nabla \mb x}$ (assuming $\mb n= \mb 0$) is: when $\mb k$ is non-sparse and satisfies the simplex constraint (i.e., $\mb k \ge \mb 0, \sum_i k_i = 1$), $\nabla (\mb k \ast \mb x) = \mb k \ast \nabla \mb x$ tends to have higher sparsity level that of $\nabla \mb x$ due to the potential smoothing effect of $\mb k$, but $\mb k \ast \nabla\mb x$ has a lower numerical scaling than that of $\nabla \mb x$\footnote{Indeed, by Young's convolution inequality and the fact $\norm{\mb k}_1 = 1$, $\norm{\mb k \ast \paren{\nabla\mb x}}_1 \le \norm{\mb k}_1 \norm{\nabla \mb x}_1 \le \norm{\nabla \mb x}_1$.}. The latter tends to outweigh the former as $\mb k$ becomes sufficiently dense~\cite{LevinEtAl2011Understanding,BenichouxEtAl2013fundamental}. So, a possible fix is to use scale-invariant sparsity measures such as $\ell_1/\ell_2$~\cite{KrishnanEtAl2011Blind,HurleyRickard2009Comparing}\footnote{See also similar ideas for the inverse filtering approach in~\cite{Cabrelli1985Minimum,SunDonoho2021Convex}.} or (near) $\ell_0$~\cite{XuEtAl2013Unnatural,PanEtAl2014Deblurring,WipfZhang2014Revisiting}.  
    
    \item $\norm{\mb k}_2 = 1$, $R_{\mb x}\paren{\mb x} = \norm{\nabla \mb x}_1$ or (near) $ \norm{\nabla \mb x}_0$: Recently, it has been shown under different settings~\cite{WipfZhang2014Revisiting,ZhangEtAl2020Structured,ZhangEtAl2017Global,KuoEtAl2020Geometry,JinEtAl2018Normalized} that $\ell_2$ normalization on $\mb k$ can change the optimization landscape and render true $(\mb k, \mb x)$ as a global solution, even with the scale-sensitive $\norm{\nabla \mb x}_1$. This is also related to the popularly used $\ell_2$ regularization in $\mb k$, which can be understood as the penalty form of such a constraint~\cite{XuEtAl2013Unnatural,PanEtAl2014Deblurring,PanEtAl2016Blind,YanEtAl2017Image,ChenEtAl2019Blind,TranEtAl2021Explore}.  

    \item Other priors: Other image-specific priors, such as color prior~\cite{JoshiEtAl2009Image}, Markov-random-field prior~\cite{KomodakisParagios2013MRF}, patch recurrence prior~\cite{MichaeliIrani2014Blind}, dark channel prior~\cite{PanEtAl2016Blind}, extreme channel prior~\cite{YanEtAl2017Image}, local maximum gradient prior~\cite{ChenEtAl2019Blind}, also help encode extra image structures and break the issue with the trivial solution. 
\end{itemize}
Another line of ideas works with the data-fitting loss $\norm{\nabla \mb y - \mb k \ast \nabla\mb x}_2^2$, combined with the different priors and regularizers discussed above~\cite{JoshiEtAl2008PSF,ChoLee2009Fast,XuJia2010Two,SunEtAl2013Edge,ZhongEtAl2013Handling,FangEtAl2014Separable,GongEtAl2016Blind,ZhangEtAl2017Global,ChoLee2017Convergence,LiuEtAl2018Deblurring,YangJi2019Variational}. Most of them employ explicit edge detection and filtering to improve kernel estimation at initialization and during iteration, but edge processing can be sensitive to noise~\cite{ZhongEtAl2013Handling,GongEtAl2016Blind}.  

\begin{figure}[!htbp]
    \centering
    \includegraphics[width=0.95\linewidth]{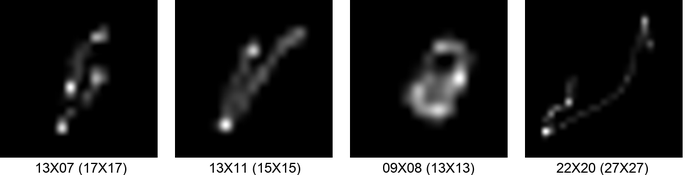}
    \includegraphics[width=0.95\linewidth]{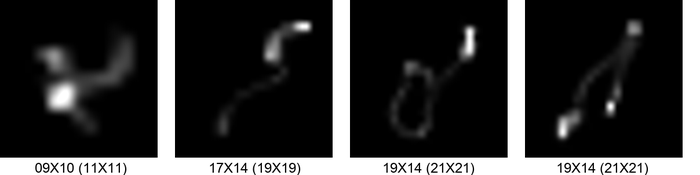}
    \includegraphics[width=0.95\linewidth]{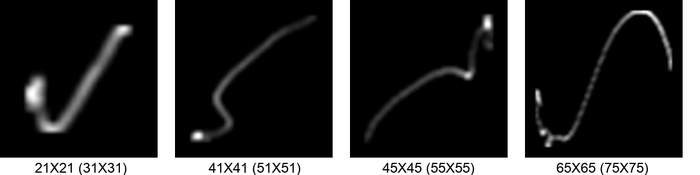}
    \caption{Kernels from the synthetic datasets in~\cite{LevinEtAl2011Understanding} and \cite{LaiEtAl2016Comparative}. Note that the true supports of the kernels are all slightly smaller than the specified kernel sizes, due to the presence of the black (zero) boundaries. Convention in the subcaptions: true size (specified size). }
    \label{fig:kernel_size_examples}
\end{figure}

Almost all the existing single-instance methods accept a user-specified kernel size, hopefully a tight upper bound of the true size, as a problem hyperparameter. For synthetic datasets such as those released by \cite{LevinEtAl2011Understanding,LaiEtAl2016Comparative}, the ``true'' kernel sizes---which are in fact slightly over-specified kernel sizes, as shown in~\cref{fig:kernel_size_examples}---are available. For real-world datasets, such as the real-world part of~\cite{LaiEtAl2016Comparative} and \cite{NahEtAl2020NTIRE}, kernel sizes are unknown, and most prior work is vague about how they choose appropriate kernel sizes. We suspect that their selections are probably based on trial-and-error combined with visual inspection of the recovery quality.  
As far as we are aware, \cite{SiYaoEtAl2019Understanding} is the first work explicitly addressing the kernel-size overspecification issue. They propose adding a low-rankness prior on the kernel: indeed, with increasing overspecification, the kernel becomes relatively sparse and low-rank, as is evident from \cref{fig:kernel_size_examples}.

\begin{figure}[!htbp]
    \centering
    \includegraphics[width=0.95\linewidth]{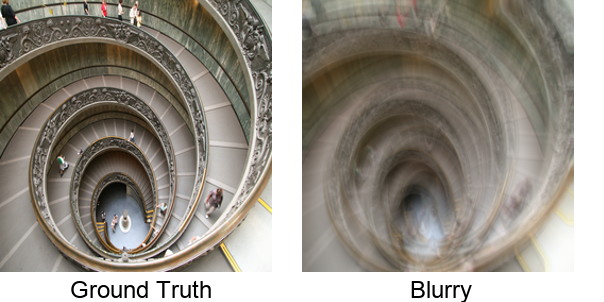}
    \includegraphics[width=0.95\linewidth]{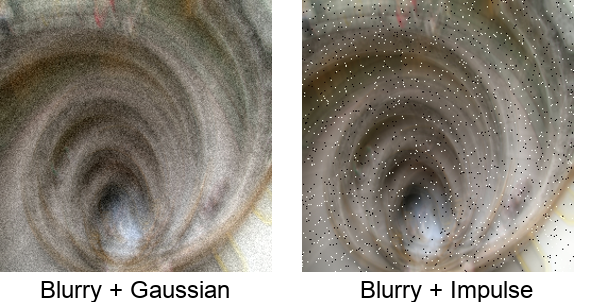}
    \includegraphics[width=0.95\linewidth]{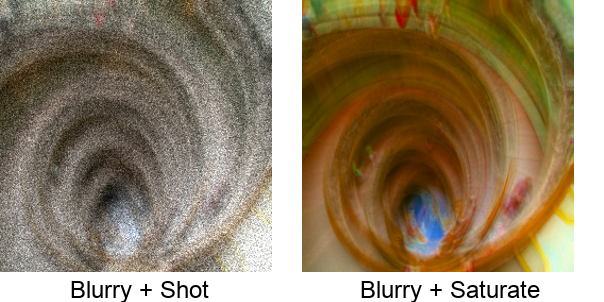}
    \caption{Examples of blurry images with realistic noise. The clean image is taken from~\cite{LaiEtAl2016Comparative}; the simulation of noise follows the procedure in~\cite{HendrycksDietterich2019Benchmarking}. }
    \label{fig:img_noise_examples}
\end{figure}
While early works test their methods on synthetic datasets with Gaussian noise (often with $\sigma = 0.01$ following \cite{KrishnanFergus2009Fast}), only few papers have explicitly handled large, realistic noise, such as impulse/shot noise, or pixel saturation~\cite{TaiLin2012Motion,ZhongEtAl2013Handling,PanEtAl2016Robust,DongEtAl2017Blind,GongEtAl2017Self,ChenEtAl2020OID}; see examples in \cref{fig:img_noise_examples}. In handling practical noise, a common thread is to learn or design a robust loss term $\ell$ that is less sensitive to large/outlying pixel errors, e.g., by learning a pixel mask together with $\mb k$ and $\mb x$~\cite{ZhongEtAl2013Handling,PanEtAl2016Robust,GongEtAl2017Self,ChenEtAl2020OID,ChenEtAl2021Blind}, or by using carefully-defined robust statistical losses~\cite{DongEtAl2017Blind}. 

After 2015, data-driven DL-based methods for BID have emerged, targeting both the uniform and non-uniform settings. There are primarily two families of methods, parallel to those for solving linear inverse problems~\cite{OngieEtAl2020Deep}: 1) \emph{end-to-end approach}. Deep neural networks (DNNs) are directly trained to predict the kernel, the sharp image or both. We refer the reader to the excellent surveys~\cite{KohEtAl2021Single,ZhangEtAl2022Deep}, and the Github repository~\cite{Vasu2021Image} with an updated list of relevant papers; 2) \emph{hybrid approach}. This includes many possibilities: DNNs are pretrained to model priors on $\mb k$ and $\mb x$~\cite{PanEtAl2021Physics,AsimEtAl2020Blind,LiEtAl2018Learning} or to replace algorithmic components to solve \cref{eq:deblur_regfit_generic} (i.e., plug-and-play methods, e.g.~\cite{ZhangEtAl2019Deep}); DNNs are directly trained as components of unrolled numerical methods for solving \cref{eq:deblur_regfit_generic}~\cite{SchulerEtAl2016Learning,AljadaanyEtAl2019Douglas,LiEtAl2019Deep}. Again, we recommend the two surveys and the Github repository for comprehensive coverage. These data-driven methods are apparently powered and meanwhile limited by the capacities of the training datasets used; the difficulty in constructing expressive and realistic training sets and, hence, poor generalization remain the key challenges~\cite{KohEtAl2021Single}.

\subsection{Deep image prior (DIP) for BID}  \label{sec:dip_bid}
Deep image prior (DIP), as its name suggests, hypothesizes that natural images, or, in general, natural visual objects, can be parameterized as the output of trainable DNNs~\cite{UlyanovEtAl2020Deep}. Specifically, any visual object of interest, $\mc O$, is written as $\mc O = G_{\mb \theta} (\mb z)$:  $G_{\mb \theta}$ is a structured DNN (often convolutional DNN to have a bias toward natural visual structures) that can be thought of as a generator, and $\mb z$ is the seed (i.e., input) to $G_{\mb \theta}$. Often, $G_{\mb \theta}$ is trainable and $\mb z$ is randomly initialized and then fixed. 

Visual inverse problems (VIPs) involve estimating a visual object $\mc O$ from an observation $\mb y \approx f\paren{\mc O}$, where $f$ models the observation (i.e., forward) process and the approximation sign $\approx$ indicates the potential existence of observational and modeling noise. Traditionally, VIPs are often posed as regularized data-fitting: 
\vspace{-1em}
\begin{equation} 
        \min_{\mc O} \; \underbrace{\ell\paren{\mb y, f\paren{\mc O}}}_{\text{data fitting}} +  \underbrace{\lambda R\paren{\mc O}}_{\text{regularizer}}, 
\end{equation} 
of which problem~\eqref{eq:deblur_regfit_generic} is a specialization for SSBD. Imposing DIP onto $\mc O$ naturally leads to 
\begin{equation} 
    \min_{\mb \theta} \; \ell\paren{\mb y, f \circ G_{\mb \theta} (\mb z)} + \lambda R \circ G_{\mb \theta} (\mb z), 
\end{equation} 
where $\circ$ denotes function composition, and the regularizer $R$ that encodes other priors is sometimes omitted. This simple idea has fueled surprisingly competitive methods for solving numerous computational vision and imaging tasks, ranging from basic image processing~\cite{UlyanovEtAl2020Deep,HeckelHand2019Deep,HeckelSoltanolkotabi2019Denoising,WangEtAl2019Image,TranEtAl2021Explore}, to advanced computational photography~\cite{GandelsmanEtAl2019Double,SitzmannEtAl2020Implicit,TancikEtAl2020Fourier,MaEtAl2021Unsupervised,williams2019deep}, and to sophisticated medical and scientific imaging applications~\cite{DarestaniHeckel2021Accelerated,LawrenceEtAl2020Phase,BostanEtAl2020Deep,TayalEtAl2021Phase,ZhouHorstmeyer2020Diffraction,ZhuangEtAl2022Practical}; see the recent survey~\cite{QayyumEtAl2021Untrained}. 

\begin{figure*}[!htbp] 
    \centering 
    \includegraphics[width=0.8\linewidth]{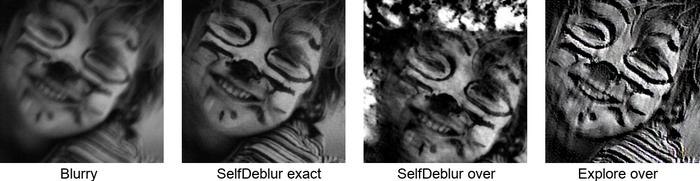}
    \includegraphics[width=0.8\linewidth]{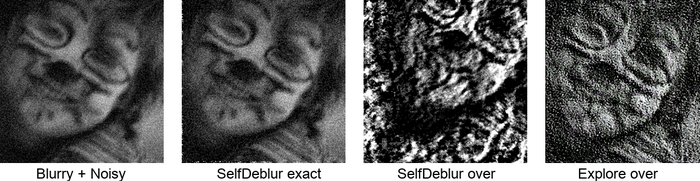}
    \includegraphics[width=0.8\linewidth]{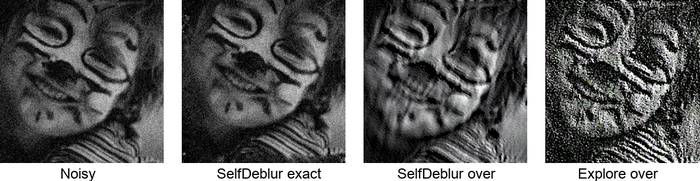}
    \caption{Deblurring performance of \selfdeblur\,\cite{RenEtAl2020Neural} and \explore\,\cite{TranEtAl2021Explore} on blurry only (1st row), blurry and noisy (2nd row), and noisy only (3rd row) images. The noise, if present, is Gaussian noise with $\sigma = 0.08$. The columns are the observed image (1st column), recovery result of \selfdeblur\,with \emph{exact} specification of the kernel size, recovery result of \selfdeblur\, with \emph{over} specification of the kernel size, and recovery result of \explore\,with \emph{over} specification of the kernel size, respectively. Note that the pretrained models from \explore\,allow only a fixed kernel size $64 \times 64$. } 
    \label{fig:prob_selfdeblur}
\end{figure*} 
When applying the DIP idea to BID, due to the asymmetric roles played by the kernel $\mb k$ and the image $\mb x$, it is natural to parameterize them separately following the Double-DIP idea~\cite{GandelsmanEtAl2019Double} to obtain: 
\begin{multline} \label{eq:double_dip_basic}
\min_{\mb \theta_{\mb k}, \mb \theta_{\mb x}} \; \ell\paren{\mb y, G_{\mb \theta_{\mb k}}(\mb z_{\mb k}) \ast G_{\mb \theta_{\mb x}}(\mb z_{\mb x})} + \\
\lambda_{\mb k} R_{\mb k} \circ G_{\mb \theta_{\mb k}}(\mb z_{\mb k}) + \lambda_{\mb x} R_{\mb x} \circ G_{\mb \theta_{\mb x}}(\mb z_{\mb x}), 
\end{multline} 
i.e., DIP reformulation of problem~\eqref{eq:deblur_regfit_generic}. This is the exact recipe followed by two previous works~\cite{WangEtAl2019Image,RenEtAl2020Neural}; they differ by their choices of $G_{\mb \theta_k}$ and $G_{\mb \theta_x}$, as well as the regularizers $R_{\mb k}$ and $R_{\mb x}$. We focus on reviewing \selfdeblur\;~\cite{RenEtAl2020Neural} here, as our method mostly builds on top of it and the evaluation in \cite{WangEtAl2019Image} is very limited. 
\begin{itemize} 

    \item  \cite{RenEtAl2020Neural} (\selfdeblur): $\ell$ is the MSE. For the generators, $G_{\mb \theta_{\mb x}}$ is convolutional U-Net similar to above, while $G_{\mb \theta_{\mb k}}$ is a $2$-layer fully connected network. The disparate generators are to encode the asymmetry between the kernel and the image, and reflect the fact that the kernel tends to be much simpler than the image itself. Softmax and sigmoid final activations are then applied to $G_{\mb \theta_{\mb k}}$ and $G_{\mb \theta_{\mb x}}$, respectively. In addition, $R_{\mb x}$ is the classical TV regularizer that helps the method to work in the presence of low-level noise also. In summary, 
    \begin{empheq}[box=\fbox]{align}  \label{eq:selfdeblur_main_setting}
        & \min_{\mb \theta_{\mb k}, \mb \theta_{\mb x}} \; \norm{\mb y- G_{\mb \theta_{\mb k}}(\mb z_{\mb k}) \ast G_{\mb \theta_{\mb x}}(\mb z_{\mb x})}_2^2 \nonumber \\
        & \qquad \qquad \qquad \qquad + \lambda_{\mb x} \norm{\nabla_{\mb x} G_{\mb \theta_{\mb x}}(\mb z_{\mb x})}_1, \\
        & G_{\mb \theta_{\mb k}} \text{\small : $2$-layer MLP, softmax final activation} \nonumber \\
        & G_{\mb \theta_{\mb x}} \text{\small : conv. U-Net, sigmoid final activation} \nonumber   
    \end{empheq}
    From \cref{fig:prob_selfdeblur}, it is evident that \selfdeblur\,works well only when $\mb y$ is blurry only and the kernel size is exactly specified. When there is considerable noise or the kernel-size is overspecified, \selfdeblur\,breaks down abruptly. 
\end{itemize} 

To move beyond the uniform blur model in~\cref{eq:bd_model} and construct a model that hopefully generalizes across different datasets, \explore\,\cite{TranEtAl2021Explore} proposes learning an abstract blur operator $\mc F$ from a rich set of sharp-blurry image pairs. 
Once $\mc F$ is learned, for any given blurry image $\mb y$, the clean image $\mb x$ and the abstract kernel $\mb k$ are estimated via a generalized version of problem~\eqref{eq:double_dip_basic}:
\begin{multline}  \label{eq:explore_main_setting}
    \min_{\mb \theta_{\mb k}, \mb \theta_{\mb x}} \; \ell\paren{\mb y, \mc F\paren{ G_{\mb \theta_{\mb x}}(\mb z_{\mb x}), G_{\mb \theta_{\mb k}}(\mb z_{\mb k})}} +  \\
     \qquad \lambda_{\mb k} \norm{G_{\mb \theta_{\mb k}}(\mb z_{\mb k})}_2 + \lambda_{\mb x} \norm{\nabla G_{\mb \theta_{\mb x}}(\mb z_{\mb x})}_{2/3},
\end{multline}
Although \explore\,is a powerful and bold idea, but it is unclear if they really learn generalizable blur models, as well as if \cref{eq:explore_main_setting} is a good implementation of the double DIP idea. Our quick test shows that it does not work on a simple uniform blur case; see the $4$-th column of \cref{fig:prob_selfdeblur}, especially when there is noise.  

\cite{AsimEtAl2020Blind} proposes three formulations for BID based on deep generative models in the same line of \cref{eq:double_dip_basic}, but with pretrained generator(s).
Since this method requires the pretrained kernel generator $G_{\mb \theta_{\mb k}}$ from certain motion blur datasets, we will not compare with this method later. 

None of the three DIP-for-BID works~\cite{WangEtAl2019Image,RenEtAl2020Neural,TranEtAl2021Explore} discussed above addresses the practicality issues around unknown kernel size, substantial noise, and model stability. Next, we propose several crucial modifications to \selfdeblur\,that tackle these issues altogether.

\section{Our Method}\label{sec:method}
Our method follows the double-DIP idea as formulated in \cref{eq:double_dip_basic}, and builds on the two prior works~\cite{WangEtAl2019Image} and \cite{RenEtAl2020Neural} (\selfdeblur), especially the latter. In \cref{sec:method_ingredients}, we describe six crucial ingredients of our method, and argue why they are necessary for the success. We then present our whole algorithm pipeline in \cref{sec:method_pipeline}. 

\subsection{Crucial ingredients}
\label{sec:method_ingredients}

\subsubsection{Overspecifying the size of \texorpdfstring{$\mb k$}{k}}
\label{sec:method_over_k}
As we discussed in \cref{sec:bd_bid}, most SOTA single-instance methods are evaluated on synthetic datasets, such as \cite{LevinEtAl2011Understanding} and \cite{LaiEtAl2016Comparative}, where reasonably tight upper bounds of kernel sizes are available. However, on more realistic datasets such as \cite{NahEtAl2020NTIRE,NahEtAl2021NTIRE,RimEtAl2020Real} and particularly in real-world applications, no such tight bounds are available. 

\begin{figure}[!htbp]
    \centering
     \includegraphics[width=1.\linewidth]{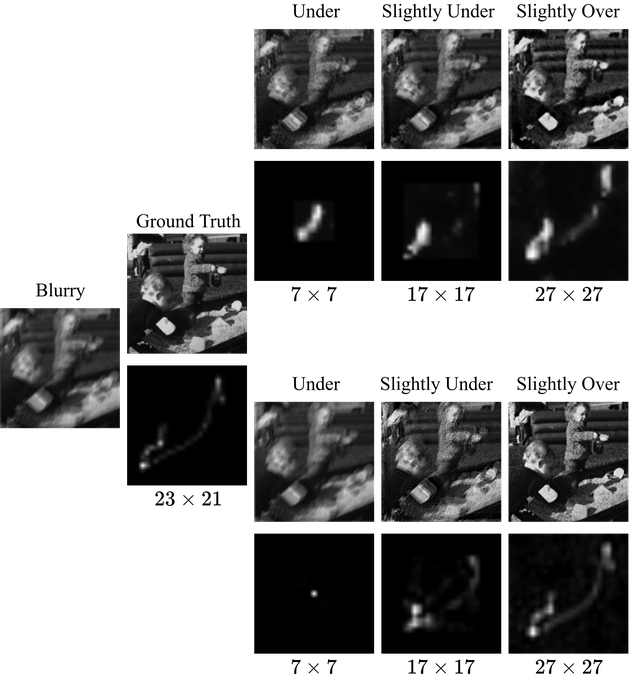}  
    \caption{Illustration of the problem with under-specification of the kernel size. We take \selfdeblur\,(top group) and our method (bottom group; details in \cref{sec:method_pipeline}) with different kernel-size specifications ($7 \times 7$, $17 \times 17$, $27 \times 27$, respectively), in contrast to the ``true''---we estimate by locating the nonzero support of the kernel---kernel-size $23 \times 21$.  }
    \label{fig:demo_kernel_underspec}
\end{figure} 
In general, recovering $\mb k$ is not possible when the kernel size is underspecified. In fact, recovery of $\mb x$ is also not possible in this situation; consider the following argument for 1D cases. 
\begin{example} 
Assume that $\mb k \in \bb R^3$, $\mb x \in \bb R^5$, and $\mb y \in \bb R^5$ due to truncation. So 
\begin{equation} 
\mb y = \mc T\paren{\mb k \ast \mb x }
 = 
 \underbrace{
\begin{bmatrix} 
x_2 & x_1  &  \\
x_3 & x_2  & x_1 \\
x_4 & x_3  & x_2 \\
x_5 & x_4  & x_3 \\
    & x_5  & x_4 
\end{bmatrix}
 }_{\mb M_{\mb x}}
\begin{bmatrix} 
k_1 \\ k_2  \\ k_3 
\end{bmatrix}.  
\end{equation}
Now, suppose that the kernel size is specified as $2$ and also $\mb x$ is correctly recovered with a kernel estimate $\mb k' \in \bb R^2$. Then, depending on the convention of the truncation, one of following products 
\begin{equation} \label{eq:example_under_spec}
    \begin{bmatrix} 
        x_1 &  \\
        x_2 & x_1 \\
        x_3 & x_2 \\
        x_4 & x_3 \\
        x_5 & x_4 
    \end{bmatrix} 
    \begin{bmatrix} 
        k_1' \\ k_2'
    \end{bmatrix} 
    \quad \text{or} \quad 
    \begin{bmatrix} 
        x_2  & x_1 \\
        x_3  & x_2 \\
        x_4  & x_3 \\
        x_5  & x_4 \\
           & x_5 
    \end{bmatrix} 
    \begin{bmatrix} 
        k_1' \\ k_2'
    \end{bmatrix}.  
\end{equation} 
should reproduce $\mb y$. But for generic $\mb x$, the matrix $\mb M_{\mb x}$ is column full-rank and hence $\mb y$ lies in the $3$-dimensional column space of $\mb M_{\mb x}$, i.e., $\mathrm{col}(\mb M_{\mb x})$. Both products in \cref{eq:example_under_spec} can fail to reproduce $\mb y$, as they produce points in $2$-dimensional subspaces of $\mathrm{col}(\mb M_{\mb x})$ only. Due to the contradiction, recovery of $\mb x$ is generally not possible with the length-$2$ kernel specification. 
\end{example}
Indeed, as shown in \cref{fig:demo_kernel_underspec}, when the kernel is significantly under-specified, the estimated kernel is disparate from the true kernel. When the under-specification is slight, we can at best recover part of the true kernel. In both cases, the estimated images are still blurry to different degrees. 

On the other hand, \cref{fig:demo_kernel_underspec} also shows that with slight kernel-size overspecification, we manage to estimate the kernel and image with reasonably good quality. In theory, overspecification at least allows the possibility of the recovering the kernel padded with zeros. However, shortness of the kernel is also crucial in SSBD. Intuitively, when overspecification is substantial, there may be a fundamental identifiability issue, i.e., it is likely that $\mb y = \mc T\paren{\mb k \ast \mb x} = \mc T \paren{\mb k' \ast \mb x'}$ for a $\mb k'$ that is substantially larger in size than $\mb k$, where $\mc T$ is the truncation operator defined in \cref{eq:bd_true_model}. So the question is what level of overspecification is safe: small enough to avoid the potential identifiability issue, while large enough to allow typical blur kernels. 

\begin{figure}[!htbp] 
    \centering 
    \includegraphics[width=0.95\linewidth]{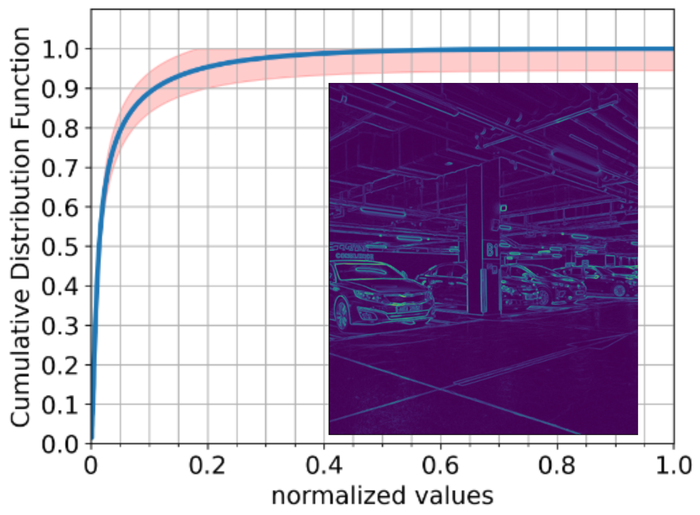}
    \caption{Cumulative distribution function (CDF) of pixel-wise gradient norms over typical natural images. This is estimated from $234$ images randomly sampled from the \texttt{RealBlur} dataset~\cite{RimEtAl2020Real}. For each image, we obtain the gradient map by convolving the image with the standard Sobel filter. After calculating the pixel-wise $\ell_2$ norms, we normalize these values into $[0, 1]$ by dividing using the largest value and then estimate the CDF. The blue curve is the mean CDF and the shallow region indicates the standard deviation over the $234$ images.}
    \label{fig:natural_img_grad_sparse}
\end{figure}
Regarding the identifiability of SSBD with the model $\mb y = \mb k \ast \mb z$ where $\mb z$ is sparse with respect to the canonical basis, \cite{ChoudharyMitra2014Sparse} presents a strong negative result: for all $n_{\mb k}, n_{\mb z} \ge 5$, there always exist non-identifiable pairs for any sparsity pattern assumed on $\mb z$ (distilled from their Section III.B and Theorem 2); \cite{ChoudharyMitra2018Properties} provides a more quantitative version of the result (Theorem 3). Unfortunately, it remains open up to date if these non-identifiable cases are rare events\footnote{In particular, if they form a measure-zero set. }. Nonetheless, all existing identifiability results based on other assumptions on $\mb k$ and $\mb z$ (particularly subspace-constrained and subspace-sparse assumptions as in~\cite{LiEtAl2017Identifiability,KechKrahmer2017Optimal}) roughly state that 
\begin{equation} \label{eq:identifiable_dog}
   \mathrm{DoF} \paren{\mb y} \ge \mathrm{DoF} \paren{\mb k} + \mathrm{DoF} \paren{\mb z} 
\end{equation}
is the identifiability limit, where $\mathrm{DoF}$ stands for degrees of freedom. For SSBD, this can be mapped to\footnote{The result in \cref{eq:identifiable_dog} assumes a circular convolution model: $\mb y = \mb a \circledast  \mb z$, but it is well known that the linear convolution can be written as circular convolution by appropriate zero-padding to the two convolving components. } 
\begin{equation}
    \mathrm{SIZE} \paren{\mb y} \ge \mathrm{SIZE} \paren{\mb k} + \mathrm{NNZ} \paren{\mb z},  
\end{equation} 
where $\mathrm{NNZ}$ denotes the number of non-zeros. For BID, $\nabla \mb x$ is assumed to be sparse, we thus have 
\begin{equation}  \label{eq:k_overspec_key}
    \mathrm{SIZE} \paren{\mb y} \ge \mathrm{SIZE} \paren{\mb k} + \mathrm{NNZ} \paren{\abs{\nabla \mb x}}, 
\end{equation} 
where $\abs{\nabla \mb x}$ denotes the element-wise gradient magnitude for image $\mb x$. So \cref{eq:k_overspec_key} tells us that a reasonable upper bound for kernel size depends on the typical sparsity level of gradient norms of natural images that we deal with in BID. 

\cref{fig:natural_img_grad_sparse} provides the mean cumulative distribution function estimated over a subset of natural images from the \texttt{RealBlur} dataset~\cite{RimEtAl2020Real}. On average, $80\%$ of the gradient norms are below $5\%$ of the largest gradient norm, and $50\%$ below $1\%$ of the largest gradient norm. So if we set $1\%$ as the cutoff threshold, the numerical sparsity level of $\abs{\nabla \mb x}$ is below $0.5$, i.e., no more than half of the pixel values are nonzero after the cutoff. Thus, \emph{we over-specify the size of $\mb k$ as half of the size of $\mb y$ in both directions}. This is a safe choice: if we allow extremely ``thin'' images and kernels consisting of single columns only, this still allows recovery. For general rectangular images and kernels, we could be slightly more aggressive in the over-specification. \emph{As far as we are aware, our setting represents the first time that the kernel size has been set in this ``aggressive'' regime. }

\subsubsection{Overspecifying the size of \texorpdfstring{$\mb x$}{x}} 
\label{sec:method_over_x}
\begin{figure}[!htbp]
    \centering
    \includegraphics[width=0.95\linewidth]{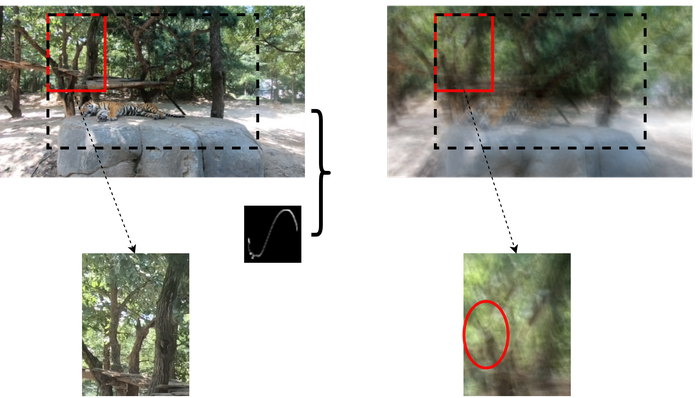}
    \caption{Illustration of the truncated linear convolution, and the necessity of appropriately specifying the size of $\mb x$. The black dashed box delineates the actual field of view (FOV) of the camera; the left column is the clean image, and the right the blurry image due to the horizontal $S$-shaped blur kernel. Note that inside the enlarged window of the blurry image, there are ``ghost'' branches from outside the FOV. Hence, if we specify the size of $\mb x$ exactly as the FOV, we are not able to recovery the clean scene inside the FOV due to the ``ghost'' visual components near the four boundaries. }
    \label{fig:demo_kernel_overspec}
\end{figure}

\begin{figure}[!htbp] 
    \centering 
    \includegraphics[width=0.95\linewidth]{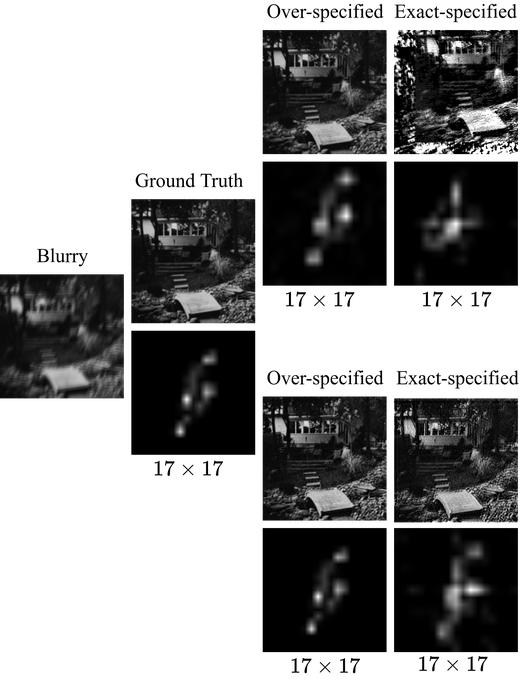}
    \caption{Illustration of the necessity of overspecifying the size of $\mb x$. We take \selfdeblur\,(top group) and our method (bottom group; details in \cref{sec:method_pipeline}). The kernel size is specified as $17 \times 17$, slightly larger than the actual size $15 \times 11$ for both methods. In the over-specified cases, the size of $\mb x$ is specified as $(n_{\mb k} + n_{\mb y} - 1) \times (m_{\mb k} + m_{\mb y} - 1)$. In the exactly-specified cases, the size of $\mb x$ is specified as $n_{\mb y} \times m_{\mb y}$. Both methods return reasonable kernel and image estimates when the size of $\mb x$ is overspecified, and both produce estimates with visible artifacts when the size of $\mb x$ is exactly-specified---the artifacts by \selfdeblur\,are significant.}
    \label{fig:overspecified}
\end{figure}

Suppose that $\mb k \in \bb R^{n_{\mb k} \times m_{\mb k}}$ and $\mb y \in \bb R^{n_{\mb y} \times m_{\mb y}}$. By the truncated linear convolution model of~\cref{eq:bd_true_model} (illustrated in \cref{fig:demo_kernel_overspec}), the part of $\mb x$ that can contribute to the values of $\mb y$ has a size of
\begin{equation} \label{eq:x_overspec_formula}
     \paren{n_{\mb k} + n_{\mb y} - 1} \times \paren{m_{\mb k} + m_{\mb y} - 1},
\end{equation} 
which is the appropriate size that we should specify for $\mb x$. Physically, underspecification, e.g., specifying the size of $\mb x$ identical to that of $\mb y$ is likely to lead to recovery failures, as illustrated in \cref{fig:demo_kernel_overspec,fig:overspecified}. 
While the majority of previous works follow \cref{eq:x_overspec_formula} in specifying the size of $\mb x$, e.g., \cite{SunEtAl2013Edge}, \cite{PanEtAl2016Blind}, \cite{DongEtAl2017Blind}, and \selfdeblur\, \cite{RenEtAl2020Neural}, a small number of them set the size of $\mb x$ same as that of $\mb y$, e.g., \cite{XuEtAl2013Unnatural}) and \explore\, \cite{TranEtAl2021Explore}. We follow \cref{eq:x_overspec_formula} in our setting. 

\begin{figure}[!htbp]
    \centering
    \includegraphics[width=0.95\linewidth]{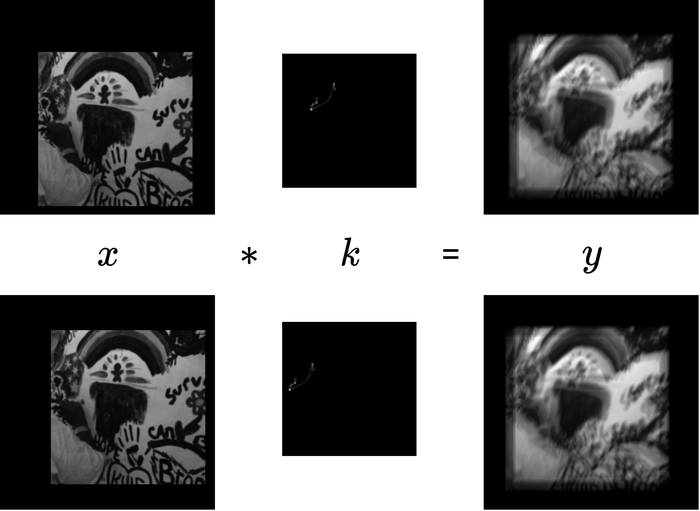} \\
    \vspace{1.5em}
    \includegraphics[width=0.95\linewidth]{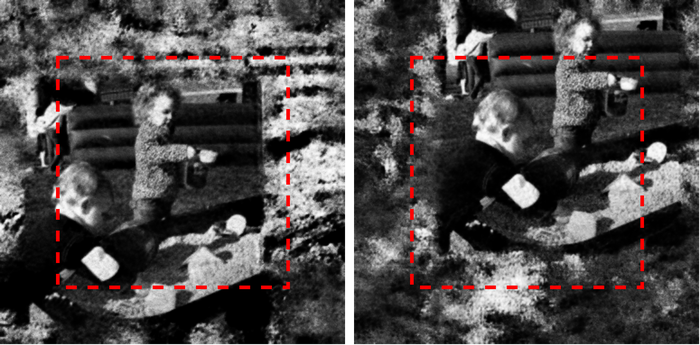}
    \caption{Illustration of the bounded shift effect (top group), and the issue caused by central cropping as implemented in \selfdeblur\,(bottom group). Due to the simultaneous over-specification of the kernel and image sizes, the kernel and the image contents (i.e., the nonzero parts) can shift in opposite directions in $\bb R^2$---so long as they do not shift outside the boundaries, that leads to equivalent $\paren{\mb k, \mb x}$ pairs to produce the same blurry image $\mb y$. Due to the uncertainty of the locations of kernel and image contents, central cropping (which is used in \selfdeblur) may include estimation noise from the background, as indicated by the red cropping boxes. }
    \label{fig:bounded_shift}
\end{figure} 
However, we do not know $n_{\mb k}$ and $m_{\mb k}$ exactly. By our overspecification strategy for $\mb k$ described in \cref{sec:method_over_k}, the actual size we use, i.e., $\lceil \frac{1}{2} n_{\mb y} \rceil \times \lceil \frac{1}{2} m_{\mb y} \rceil$, can be substantially larger than $n_{\mb k} \times m_{\mb k}$. So the size we specify for $\mb x$ now becomes 
\begin{equation} 
    (\lceil \frac{1}{2} n_{\mb y} \rceil + n_{\mb y} -1) \times  (\lceil \frac{1}{2} m_{\mb y} \rceil + m_{\mb y} - 1). 
\end{equation} 
The simultaneous overspecification of $\mb k$ and $\mb x$ causes another problem: the bounded shift effect. 

Recall that if $\mb k$ and $\mb x$ are 1-D infinite sequences, $\mb k \ast \mb x = \paren{\frac{1}{\alpha} \mb k_{-\tau}} \ast \paren{\alpha \mb x_{\tau}}$ for all $\alpha \ne 0$ and $\tau \in \bb Z$. In other words, there are both scale and shift ambiguities if we want to recover $\mb k$ and $\mb x$ from $\mb y = \mb k \ast \mb x$. There are similar ambiguities for 2-D $\mb k$ and $\mb x$ for BID. With the truncated convolution model of \cref{eq:bd_true_model} on finite sequences, we do not have the shift ambiguity if the size of either $\mb k$ or $\mb x$ is exactly-specified. But, when both sizes are over-specified as we propose here, we expect the bounded shift ambiguity, as shown in \cref{fig:bounded_shift}: even if we successfully recover $\mb k$ and $\mb x$, their contents are embedded, not necessarily centered, in the larger background regions that we overspecify. 

So we need a post-processing step to locate the contents of $\mb k$ and $\mb x$ after we obtain the overspecified versions of both; we propose an effective post-processing step in \cref{sec:post_processing_x}. We note that \selfdeblur\,uses the same $\paren{n_{\mb y} + n_{\mb k} - 1} \times \paren{m_{\mb y} + m_{\mb k} - 1}$ rule as ours to overspecify the size of $\mb x$, but their $\mb n_{\mb k} \times \mb m_{\mb k}$ is close to the true kernel size as they mostly evaluate only on synthetic data. Thus, the bounded shift ambiguity is not quite visible, and they simply centrally crop $\mb x$ to obtain the final estimated image. Once we move to real-world images where substantial overspecification of the kernel size is unavoidable, the central cropping strategy may cut out part of the image content augmented with non-physical estimation noise, as we show in \cref{fig:bounded_shift}. 

\subsubsection{The loss and regularizers}
\label{sec:loss_reg}
As summarized in \cref{eq:selfdeblur_main_setting}, \selfdeblur\,uses the standard MSE loss $\ell$ and TV regularization, i.e., $R\paren{\mb x}=\norm{\nabla_{\mb x} G_{\mb \theta_{\mb x}}(\mb z_{\mb x})}_1$. Here, we propose changing both the loss and the regularizer to make the method effective and robust even in the presence of substantial noise that may be beyond Gaussian. 

For the loss, we switch to the famous Huber loss~\cite{Huber1964Robust} 
\begin{equation} 
    \ell_{\mathrm{Huber}, \delta}(u) = 
    \begin{cases} 
          \frac{1}{2} u^2   &  \abs{u} \le \delta, \\
          \delta\paren{\abs{u} - \frac{1}{2}\delta}  & \text{otherwise}. 
    \end{cases} 
\end{equation} 
The Huber loss penalizes less of large values compared to the MSE, and hence in regression problems the overall loss becomes less dominated by large errors. This implies that the regression models estimated from Huber loss minimization be less sensitive to outlying data points that tend to cause large regression errors. For BID, outlying pixels could be caused by, e.g., large noise (e.g., shot noise) and pixel saturation. This choice enables our method to work beyond the regime of low-level Gaussian noise that the majority of previous works, including \selfdeblur, have focused on. 

\begin{figure}[!htbp]
    \centering
    \includegraphics[width=\linewidth]{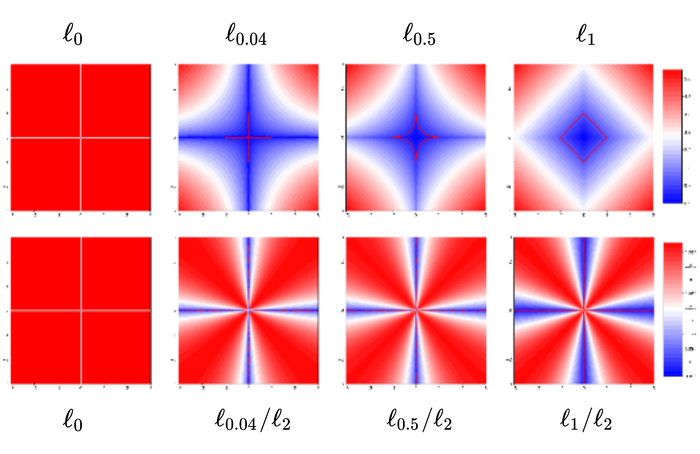}
    \caption{Landscapes of different surrogates for the $\ell_0$ function on $\bb R^2$. The normalized metrics $\ell_p/\ell_2$ are uniformly closer to $\ell_0$ than their unnormalized counterparts---$\ell_p$ norms where $p \in (0, 1]$. The approximation of $\ell_p/\ell_2$ to $\ell_0$ becomes increasingly sharper as $p$ goes down to $0$. }
    \label{fig:comp_sparse_norms}
\end{figure} 
\begin{table}[!htbp]
    \centering 
    \caption{Performance of $\ell_1/\ell_2$ vs $\ell_1$ as regularization with the \emph{optimal} regularization parameter $\lambda_{\mb x}$'s. We take all test cases from the Levin dataset, and for each image, we search for the best $\lambda_{\mb x}$ (in terms of best peak PSNR) over the selections: $1$, $5\mathrm{e}{-1}$, $2\mathrm{e}{-1}$, $1\mathrm{e}{-1}$, $5\mathrm{e}{-2}$, $2\mathrm{e}{-2}$, $1\mathrm{e}{-2}$, $5\mathrm{e}{-3}$, $2\mathrm{e}{-3}$, $1\mathrm{e}{-3}$, $5\mathrm{e}{-4}$, $2\mathrm{e}{-4}$, $1\mathrm{e}{-4}$, $5\mathrm{e}{-5}$, $2\mathrm{e}{-5}$, $1\mathrm{e}{-5}$ for $\ell_1/\ell_2$ and $\ell_1$ regularizers, respectively. We report the mean peak PNSRs and mean $\lambda_{\mb x}$'s (and the standard deviations inside parentheses) over the whole dataset for both low-level ($\sigma = 1\mathrm{e}{-3}$) and high-level ($\sigma = 5\mathrm{e}{-2}$) Gaussian noise.}
    \label{table:reg_norm_perf}
    \setlength{\tabcolsep}{0.5mm}{
    \begin{tabular}{c| c| c| c| c}
    \hline
    &
    \multicolumn{2}{c|}{Low Level} &
    \multicolumn{2}{c}{High Level}
    \\
    \hline
    &
    \multicolumn{1}{c|}{PSNR} &
    \multicolumn{1}{c|}{$\lambda$}
    &
    \multicolumn{1}{c|}{PSNR} &
    \multicolumn{1}{c}{$\lambda$}
    \\
    \hline
    $\ell_1/\ell_2$
    & 32.64 \tiny({0.69})
    & 0.0001 \tiny({0.018})
    & 27.74 \tiny({0.23})
    & 0.0002 \tiny({0.0019})
    \\
    
    \hline
    $\ell_1$
    & 31.12 \tiny({0.52})
    & 0.002 \tiny({0.07})
    & 24.34 \tiny({0.78})
    & 0.02 \tiny({0.10})
    \\
    \hline
    \end{tabular}
    }
    \end{table}
\begin{figure}[!htbp] 
    \centering 
    \includegraphics[width=0.9\linewidth]{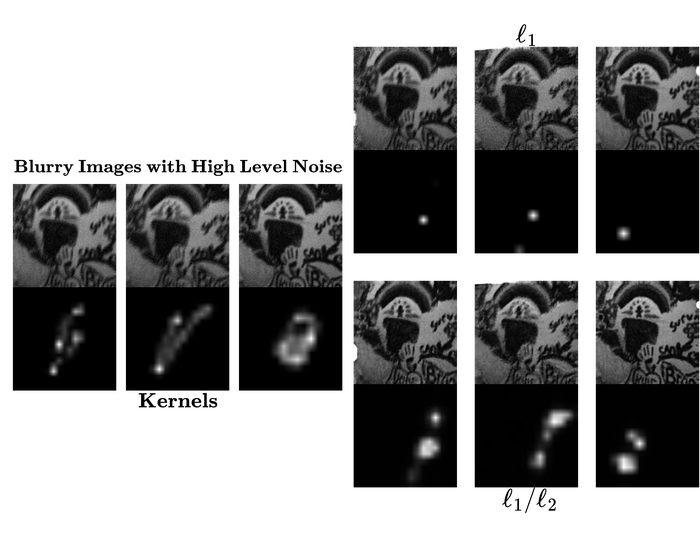}
    \caption{Illustration of the benefit of $\ell_1/\ell_2$ over $\ell_1$ in avoiding trivial solutions in the high-noise regime (Gaussian noise with $\sigma = 0.1$). \textbf{Left}: blurry and noisy images with their corresponding kernels; \textbf{Right-Top}: recovered images and kernels with $\ell_1$ regularization; \textbf{Right-Bottom}: recovered images and kernels with $\ell_1/\ell_2$ regularization. The $\ell_1$ regularization leads to single-blob kernel estimates that resemble the trivial $\mb \delta$ function, and the estimated images are also similar to the original blurry and noisy images. In contrast, the recovered images from the $\ell_1/\ell_2$ regularization are much sharper. }
    \label{fig:l1l2_trivial_sol}
\end{figure} 
For the regularizer, we choose the $\ell_1/\ell_2$ version 
\begin{equation} 
    R\paren{\mb x} = \frac{\norm{\nabla_{\mb x} G_{\mb \theta_{\mb x}}(\mb z_{\mb x})}_1}{\norm{\nabla_{\mb x} G_{\mb \theta_{\mb x}}(\mb z_{\mb x})}_2}
\end{equation} 
for three reasons/benefits: 1) \emph{scaling invariance and perturbation robustness}. To encode the sparsity prior on $\nabla \mb x$, a natural choice is the $\ell_0$ function, which is scale-invariant but sensitive to perturbations. $\ell_1$ is a popular surrogate for $\ell_0$ and robust to perturbations, but is scale equivariant. $\ell_1/\ell_2$ is scale-invariant and robust to small perturbations. \cref{fig:comp_sparse_norms} visualizes the differences between these functions; 2) \emph{insensitivity of the estimation performance to the regularization parameter $\lambda_{\mb x}$}. Empirically, we find that with $\ell_1/\ell_2$ regularizer we can fix the $\lambda_{\mb x}$ level to obtain good performance across low- and high-level Gaussian noise, whereas the $\ell_1$ regularizer requires setting $\lambda_{\mb x}$ to different orders of magnitude across different noise levels for good performance. Moreover, $\ell_1/\ell_2$ regularization leads to consistently superior performance. Details are included in \cref{table:reg_norm_perf}; and 3) \emph{avoiding trivial solutions}. As reviewed in \cref{sec:bd_bid}, the original motivation of replacing the $\ell_1$ with $\ell_1/\ell_2$ is to avoid the trivial solution $\mb k= \mb \delta$ when using the simplex normalization on $\mb k$~\cite{KrishnanEtAl2011Blind}. Although the simplex normalization is still used in \selfdeblur, the ``double-DIP'' parametrization together with gradient descent can potentially impose additional structural biases. So, a priori, it is unclear if we still need to worry about finding the trivial solution. \cref{fig:l1l2_trivial_sol} shows this concern remains: when the blurry images are also substantially noisy, the $\ell_1$ regularizer tends to produce single-blob estimates that resemble finite-supported $\mb \delta$ functions coupled with blurry image estimates. In contrast, the $\ell_1/\ell_2$ regularizer leads to much cleaner images, and also kernels that at least capture certain aspects of the groundtruth kernels. 

\subsubsection{The DIP models} \label{sec:dip_models}
\begin{figure*}[!htbp]
    \centering
    \includegraphics[width=0.9\linewidth]{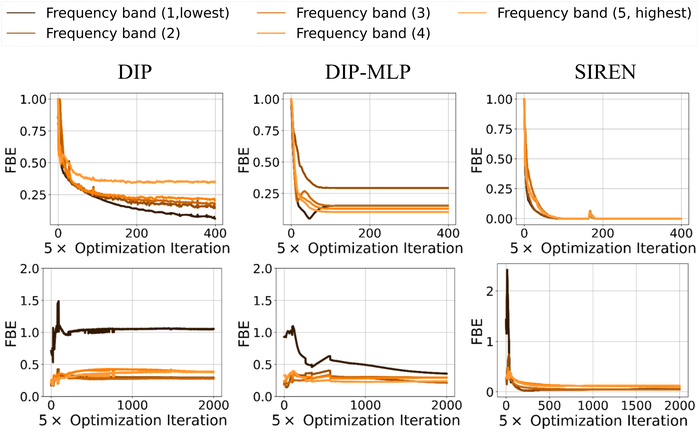}
    \caption{Evolution of DIP, DIP-MLP, and SIREN representation of kernels during kernel estimation. \textbf{Top}: simple regression of a motion blur kernel, i.e., $\min_{\wh{\mb k}}\; \|\mb k - \wh{\mb k}\|^2_2$ where $\wh{\mb k}$ is the estimated kernel represented by each of the three models; \textbf{Bottom}: non-blind kernel estimation, i.e., $\min_{\wh{\mb k}}\; \|\mb y - \wh{\mb k} \ast \mb x \|^2_2$ where again $\wh{\mb k}$ is the estimated kernel represented by each of the three models, and $\mb y$ and $\mb x$ are known. To evaluate the progress of each setting, we calculate the frequency band error (FBE), inspired by the frequency band correspondence (FBC) in~\cite{ShiEtAl2022Measuring}\protect\footnotemark: For each setting, we calculate the point-wise relative estimation error over the Fourier domain $\abs*{ \mc F(\mb k) - \mc F(\wh{\mb k})}/\abs*{\mc F(\mb k)}$, and then divide the Fourier frequencies into five bands radially (the same division used in~\cite{ShiEtAl2022Measuring}) and compute the per-band average. We term this metric frequency band error (FBE), and plot the evolution of the FBEs of all five frequency bands against the optimization iteration. It is evident that in both kernel estimation settings, SIREN recovers all frequency bands much faster and reliably than DIP and DIP-MLP. }
    \label{fig:evolution_kernel}
\end{figure*} 
\footnotetext{We note in passing that the reason we do not use FBC directly is that it may be misleading: the correspondence ratio as they define it can be larger than $1$, so in principle the average approaching $1$ does not imply that recovery is good. When checking their code (\url{https://github.com/shizenglin/Measure-and-Control-Spectral-Bias}), we find that they actually truncate values greater than $1$, which could make the metric more misleading. }
\begin{figure}[!htbp] 
    \centering 
    \includegraphics[width=0.95\linewidth]{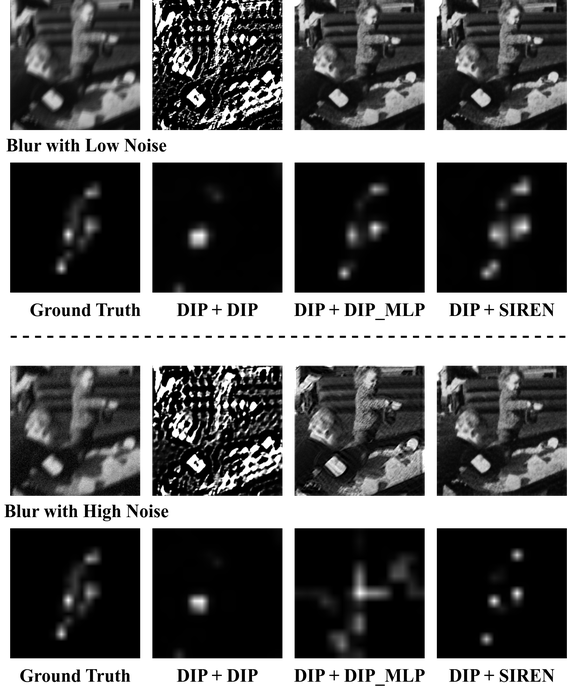}
    \caption{Performance of different model combinations for $(\mb k, \mb x)$. \textbf{Top}: with low Gaussian noise ($\sigma = 0.001$);  \textbf{Bottom}: with high Gaussian noise ($\sigma = 0.05$). Our combination, DIP (for $\mb x$) + SIREN (for $\mb k$), leads to more faithful kernel and image estimation in both low- and high-noise regimes that DIP+DIP (as in~\cite{WangEtAl2019Image}) and DIP+DIP-MLP (as in \selfdeblur). For fair comparison, we only change the model combination and leave all other settings as our default. }
    \label{fig:diffNoiseModels}
\end{figure} 
\begin{figure}[!htbp] 
    \centering 
    \includegraphics[width=\linewidth]{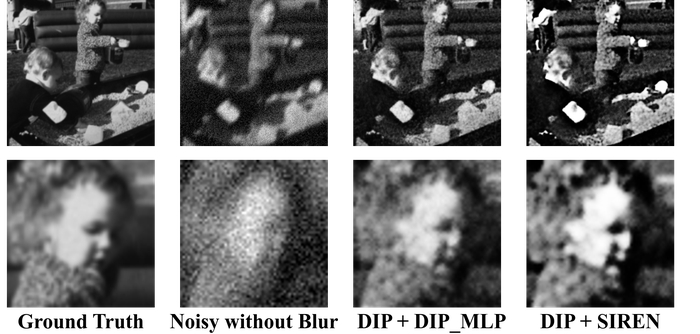}
    
    \caption{Performance of the DIP+DIP-MLP (as in \selfdeblur) and DIP+SIREN (ours) model combinations for a noisy image (Gaussian with $\sigma = 0.1$) without blur. The DIP+SIREN combination leads to sharper estimation of the image compared to DIP+DIP-MLP. For fair comparison, we only change the model combination and leave all other settings as our default.}
    \label{fig:highNoiseDenoise}
\end{figure}
As discussed around \cref{eq:double_dip_basic} and detailed in the DNN choices in \cref{eq:selfdeblur_main_setting,eq:explore_main_setting}, the DIP models to parameterize $\mb k$ and $\mb x$ should encode the right structural priors for them and reflect the asymmetry between $\mb k$ and $\mb x$. Same as \selfdeblur, we choose a convolutional U-Net $G_{\mb \theta}$ for $\mb x$. For $\mb k$, we choose the sinusoidal representation networks (SIREN) \cite{SitzmannEtAl2020Implicit} over the MLP architecture used in \selfdeblur. 

Same as DIP, SIREN also parametrizes visual objects using DNNs. Unlike DIP where the DNN outputs the visual object, in SIREN \emph{the DNN represents the visual object itself}. For example, SIREN models a continuous grayscale image as $\mc I: [0, 1]^2 \mapsto \bb R$, i.e., a real-valued function on the compact domain $[0, 1]^2 \subset \bb R^2$, and then produces a finite-resolution version of $\mc I$ via discretization. The DNN in SIREN is a modified MLP architecture that takes two coordinate inputs and returns a single value (for grayscale image) or three values (for RGB images). 

Practical blur kernels can have substantial high-frequent components in the Fourier domain, e.g., most motion blur kernels that consist of convoluted curves (see \cref{fig:kernel_size_examples}), and narrow Gaussian-shaped defocus kernels. The reason for choosing SIREN over DIP to represent $\mb k$ is that SIREN and similar coordinate encoding networks are empirically observed to learn high-frequency components of visual objects better than DIP \cite{SitzmannEtAl2020Implicit,TancikEtAl2020Fourier}; see also \cref{fig:evolution_kernel}, where we show quantitatively that on two simplified kernel estimation problems, SIREN allows recovering all frequency bands, particularly the high frequency band, of the true kernel much more efficiently and reliably than DIP with the default encoder-decoder (dubbed as DIP) and with the MLP architecture (dubbed as DIP-MLP) for $G_{\mb \theta}$. When we plug SIREN into BID, the DIP (for $\mb x$)+SIREN (for $\mb k$) model combination easily outperforms other combinations, i.e., DIP+DIP (as in~\cite{WangEtAl2019Image}) and DIP+DIP-MLP (as in \selfdeblur), especially when substantial noise is present, as shown in \cref{fig:diffNoiseModels}. Moreover, we also observe the benefit of SIREN in terms of improving the model stability: \cref{fig:highNoiseDenoise} shows that when the image is only contaminated by high noise, the DIP+SIREN combination tends to return a sharper image estimate than that of DIP+DIP-MLP. 

\subsubsection{Early stopping (ES)}
\label{sec:method_ES}
\begin{figure}[!htbp] 
    \centering 
    \includegraphics[width=0.8\linewidth]{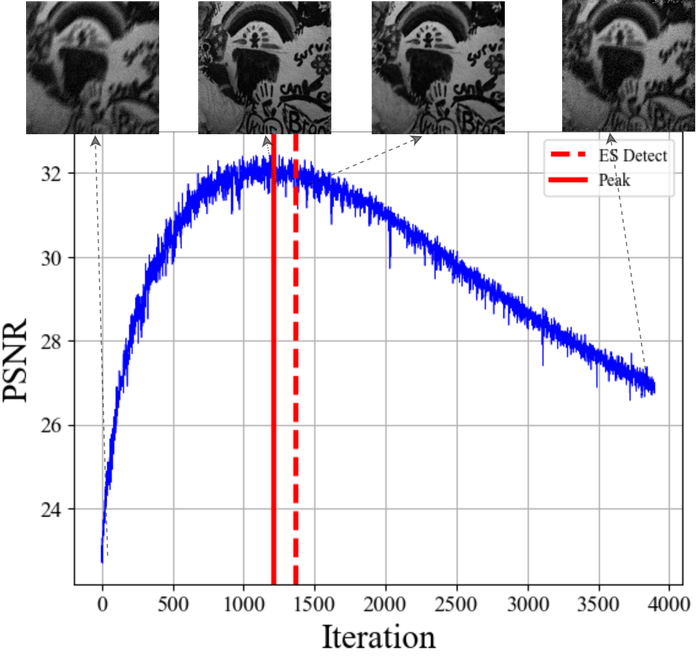}
    \caption{Illustration of the overfitting issue of \selfdeblur\,with the setting in~\eqref{eq:selfdeblur_main_setting}. The estimation quality of $\mb x$ first climbs to a peak and then plunges due to overfitting to the noise. The early stopping (ES) method for DIP developed in our prior paper~\cite{WangEtAl2021Early} can successfully detect stopping points that lead to near-peak performance. }
    \label{fig:prob_early_learning_then_overfitting}
\end{figure}
\begin{figure}[!htbp]
    \centering
    \includegraphics[width=\linewidth]{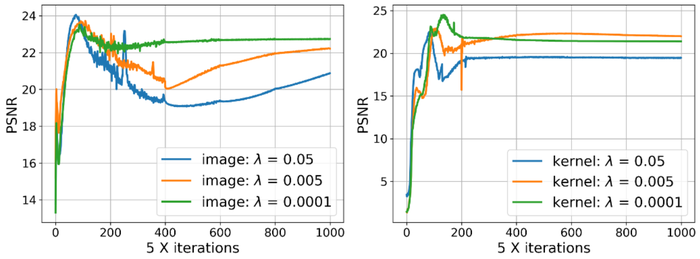}
    \caption{PSNR curves of our method on $(\mb k, \mb x)$ reconstruction with different regularization parameters (Gaussian noise with $\sigma = 0.05$). Overfitting is persistent across the different regularization levels. Moreover, the peak iterations of $\mb k$ and $\mb x$ curves for each $\lambda$ are roughly equationed, and so we only use the $\mb x$ curves for ES detection. }
    \label{fig:overfitting_lambda}
\end{figure} 
Besides the three common practicality issues for BID that we have addressed so far, there is one more specific to the double-DIP approach: overfitting. As shown in \cref{fig:prob_early_learning_then_overfitting}, the estimation quality (measured by PSNR with respect to the groundtruth image $\mb x$) of \selfdeblur\, first climbs to a peak and then degrades as the iteration goes on. 

To understand what happens here, we can think about the double-DIP loss itself:
\begin{equation} 
\ell\paren{\mb y, G_{\mb \theta_{\mb k}}(\mb z_{\mb k}) \ast G_{\mb \theta_{\mb x}}(\mb z_{\mb x})}
\end{equation}
from \cref{eq:double_dip_basic}. In practice, the image $\mb y$ is both blurry and noisy, and the DIP models $G_{\mb \theta_{\mb k}}(\mb z_{\mb k})$ and $G_{\mb \theta_{\mb x}}(\mb z_{\mb x})$ are substantially overparametrized. So if we perform global optimization, $\mb y = G_{\mb \theta_{\mb k}}(\mb z_{\mb k}) \ast G_{\mb \theta_{\mb x}}(\mb z_{\mb x})$ for typical losses, such as MSE. Thus, the final $G_{\mb \theta_{\mb x}}(\mb z_{\mb x})$ likely accounts for noise also besides the desired image content, which leads to the final quality degradation. The bell-shaped quality curve is explained by the implicit bias of first-order optimization methods used to perform the loss minimization: over-parametrized DNN models trained with first-order methods tend to learn structured visual contents much faster than learn unstructured noise; see \cite{HeckelSoltanolkotabi2020Compressive} and \cite{HeckelSoltanolkotabi2019Denoising} for complete theories on simplified models. 
The previous double-DIP-based works \cite{WangEtAl2019Image}, \selfdeblur\,\cite{RenEtAl2020Neural}, \explore\,\cite{TranEtAl2021Explore} do not address this issue, as they work with negligible noise levels that avoid the overfitting. To deal with practical noise that can be substantial, we need to address it in this paper. 

To get a good reconstruction, we can either control the DNN capacities by proper regularization, or stop the iteration early around the peak performance---early stopping (ES); see our prior works \cite{LiEtAl2021Self} and \cite{WangEtAl2021Early} for summaries of related work. We have shown in the couple of papers that the regularization strategy suffers from serious practicality issues; \cref{fig:overfitting_lambda} shows that overfitting is persistent across different levels of regularization (with our choice of $\ell_1/\ell_2$ regularizer as detailed above). So, we advocate ES-based solution instead, and adopt the windowed-moving-variance-based ES (WMV-ES) method developed in \cite{WangEtAl2021Early} that proves effective and lightweight for DIP and its variants on numerous application scenarios. As the name suggests, WMV-ES calculates the windowed moving variance curve of the intermediate reconstructions, and detects the first major valley of the WMV curve as the recommended ES point. For our purpose, we observe that the $\mb k$ and $\mb x$ PSNR curves are often automatically ``synchronized" and reach the peaks roughly around the same iteration. Thus, we only keep track of reconstructed images, not the kernels.  \cref{fig:prob_early_learning_then_overfitting} shows this simple method can effectively detect a near-peak stopping point with little loss of the reconstruction quality. 

\subsubsection{Post-processing to locate \texorpdfstring{$\wh{\mb x}$}{x}}
\label{sec:post_processing_x}
As discussed in \cref{sec:method_over_x} and illustrated in \cref{fig:bounded_shift}, the simultaneous overspecification of the sizes of $\mb k$ and $\mb x$ leads to the bounded shift effect on $\mb k$ and $\mb x$, and hence the estimated $\mb k$ and $\mb x$ may not be centered. 
So we need an algorithm to automatically locate the estimated image $\wh{\mb x}$, assumed of the same size as $\mb y$. Once we can locate $\wh{\mb x}$ and thereof estimate the shift from the center, we can use shift-symmetry between $\mb k$ and $\mb x$ to locate $\wh{\mb k}$ also if desired. 

To locate $\wh{\mb x}$, we propose a simple sliding-window strategy: we use the noisy and blurry image $\mb y$ as a template, and slide it across the output, overspecified image from $G_{\mb \theta_{\mb x}}$. The similarity of each of windowed patch from $G_{\mb \theta_{\mb x}}$ and $\mb y$ is calculated using structural similarity index measure (SSIM) to emphasize the perceptual nearness, and the patch with the largest SSIM value is eventually extracted as $\wh{\mb x}$.

\subsection{Our algorithm pipeline}
\label{sec:method_pipeline}
\begin{algorithm}[!htbp]
    \caption{BID with unknown kernel size and substantial noise (uniform kernel)}
    \label{alg:framework_ubid} 
    \begin{algorithmic}[1]
    \Require blurry and noisy image $\mb y$, kernel size $n_{\mb k} \times m_{\mb k}$ (default: $\lceil n_{\mb y}/2 \rceil \times \lceil m_{\mb y}/2 \rceil$), random seed $\mb z_{\mb x}$ for $\mb x$, randomly initialized network weights $\mb \theta_{\mb k}^{(0)}$ and $\mb \theta_{\mb x}^{(0)}$, optimal image estimate $\mb x^\ast = G_{\mb \theta_{\mb x}^{(0)}}(\mb z_{\mb x})$, regularization parameter $\lambda_{\mb x}$, iteration index $i = 1$, WMV-ES window size $W = 100$, WMV-ES patience number $P = 200$ (high noise) and $P= 500$ (low noise), WMV-ES empty queue $\mc Q$, WMV-ES $\mathrm{VAR}_{\min} = \infty$ ($\mathrm{VAR}$: variance)
    \Ensure estimated image $\wh{\mb x}$ %
    \While{not stopped}
    \State take an ADAM step to optimize \cref{eq:ours_main_setting} and obtain $\mb \theta_{\mb k}^{(i)}$, $\mb \theta_{\mb x}^{(i)}$, and $\mb x^{(i)} = G_{\mb \theta_{\mb x}^{(i)}}(\mb z_{\mb x})$ 
    \State push $\mb x^{(i)}$ to $\mc Q$, pop $\mb Q$ if $\abs{\mc Q} > W$ 
    \If{$\abs{\mc Q} = W$}
    \State compute $\mathrm{VAR}$ of elements inside $\mc Q$ 
    \If{$\mathrm{VAR} < \mathrm{VAR}_{\min}$} 
    \State $\mathrm{VAR}_{\min} \leftarrow \mathrm{VAR}$, $\mb x^\ast \leftarrow \mb x^{(i)}$ 
    \EndIf
    \EndIf 
    \If{$\mathrm{VAR}_{\min}$ does not decrease over $P$ iterations}
    \State exit and return $\mb x^\ast$
    \EndIf
    \State $i = i + 1$
    \EndWhile
    \State extract $\wh{\mb x}$ of size $n_{\mb y} \times m_{\mb y}$ from $\mb x^\ast$ using the sliding-window method (\cref{sec:post_processing_x}) 
    \end{algorithmic}
\end{algorithm}
In summary, given the blurry and noisy image $\mb y \in \bb R^{n_{\mb y} \times m_{\mb y}}$, we specify the kernel size as $n_{\mb k} \times  m_{\mb k} = \lceil n_{\mb y}/2\rceil \times \lceil m_{\mb y}/2 \rceil$ by default when the kernel size is unknown (\cref{sec:method_over_k})---which concerns most practical scenarios, and as given values when an estimate is available. According to the property of linear convolution, we set the size of the image $\mb x$ as $(n_{\mb y} + n_{\mb k} - 1) \times (m_{\mb y} + m_{\mb k} - 1)$ (\cref{sec:method_over_x}). We choose $\ell$ as the Huber loss (with $\delta = 0.05$), and the $\ell_1/\ell_2$ regularizer to promote sparsity in the gradient domain of the estimated image (\cref{sec:loss_reg}). Moreover, we choose the DIP model for the image, and the SIREN model for the kernel. In contrast to the key optimization objective of \selfdeblur\, as summarized in \cref{eq:selfdeblur_main_setting}, our method aims to solve 
\begin{empheq}[box=\fbox]{align}  \label{eq:ours_main_setting}
    & \min_{\mb \theta_{\mb k}, \mb \theta_{\mb x}} \; \ell_{\mathrm{Huber}}\paren{\mb y,  \paren{\mc D \circ K_{\mb \theta_{\mb k}}} \ast G_{\mb \theta_{\mb x}}(\mb z_{\mb x})} \nonumber \\
    & \qquad \qquad \qquad \qquad + \lambda_{\mb x} \frac{\norm{\nabla_{\mb x} G_{\mb \theta_{\mb x}}(\mb z_{\mb x})}_1}{\norm{\nabla_{\mb x} G_{\mb \theta_{\mb x}}(\mb z_{\mb x})}_2}, \\
    & K_{\mb \theta_{\mb k}} \text{\small : $2$-layer MLP, $2$ coordinate inputs}, \nonumber \\ 
    & \quad \quad \quad \text{\small $1$ output with sigmoid activation} \nonumber \\
    & \mc D \text{\small : discretization operator} \nonumber \\
    & G_{\mb \theta_{\mb x}} \text{\small : conv. U-Net, sigmoid final activation} \nonumber   
\end{empheq}
where for the MLP model $K_{\mb \theta_{\mb k}}: \bb R^2 \mapsto \bb R$ represents the kernel $\mb k$ as a continuous function, and $\mc D$ denotes the discretization process that produces a finite-resolution kernel (\cref{sec:dip_models}). The overfitting issue, especially when there is substantial noise, is handled by the WMV-ES method described in \cref{sec:method_ES}, and bounded shift effect as described in \cref{sec:method_over_x} is handled by the sliding-window-based detection method detailed in \cref{sec:post_processing_x}. The complete BID pipeline is summarized in \cref{alg:framework_ubid}. 
\begin{figure}[!htbp]
    \centering 
    \includegraphics[width=\linewidth]{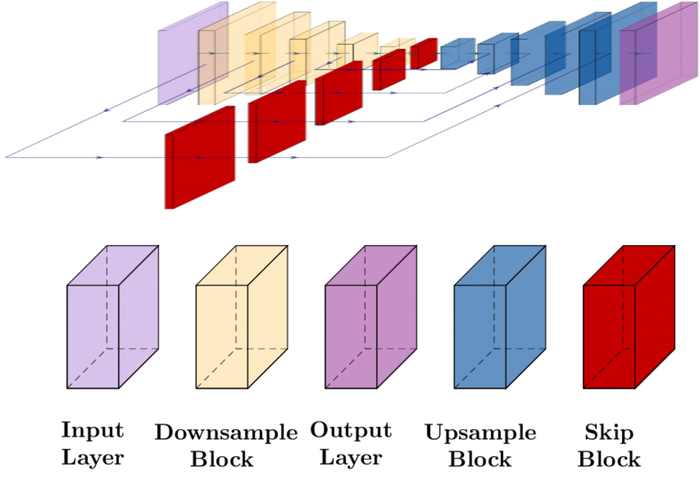}
    \caption{The default network architectures of the DIP model used in our method. Details inside the blocks are as follows. \textbf{Downsample Block}: convolution $\rightarrow$ downsample $\rightarrow$ batchnorm $\rightarrow$ leakyReLU $\rightarrow$ convolution $\rightarrow$ batchnorm $\rightarrow$ leakyReLU; \textbf{Upsample Block}: batchnorm $\rightarrow$ convolution $\rightarrow$ batchnorm $\rightarrow$ leakyReLU $\rightarrow$ convolution $\rightarrow$ batchnorm $\rightarrow$ leakyReLU $\rightarrow$ upsample; \textbf{Skip Block}: convolution $\rightarrow$ batchnorm $\rightarrow$ leakyReLU. }
    \label{fig:dip-model-details}
\end{figure} 
\begin{figure}[!htbp]
    \centering 
    \includegraphics[width=\linewidth]{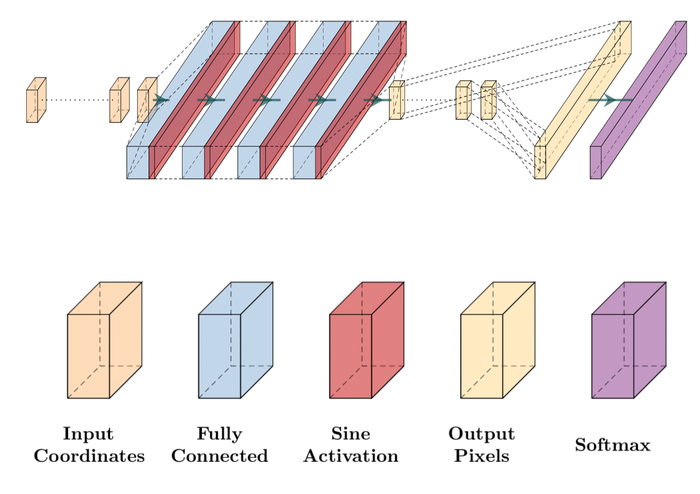}
    \caption{The default network architectures of the SIREN model used in our method. }
    \label{fig:siren-model-details}
\end{figure}
\cref{fig:dip-model-details} and \cref{fig:siren-model-details} visualize the DIP and SIREN models that we use for our method throughout the paper. 
\section{Experiments}\label{sec:expriments}
In this section, we first compare our method with $5$ SOTA single-instance BID methods on synthetic blurry and noisy images (\cref{sec:exp_synthetic_data}). We perform quantitative evaluations of all these methods in terms of their stability to: 1) kernel-size overspecification, 2) substantial noise, and 3) model ``overspecification", i.e., BID methods applied to image with noise only, corresponding to the three major practicality issues that we pinpoint in \cref{sec:introduction}. Once we confirm the superiority of our method on the synthetic data\footnote{The existing synthetic BID datasets are too small to support training data-driven methods.}, we move to real-world datasets, and benchmark our method against \selfdeblur\,and $3$ representative SOTA data-driven BID methods (\cref{sec:exp_realworld}). 

\subsection{Experiment setup}
\paragraph{Training details for our method}  
We use PyTorch to implement our method. We optimize the objective in \cref{eq:ours_main_setting} using the ADAM optimizer, with initial learning rates (LRs) $1\mathrm{e-}2$ for $\mb \theta_{\mb x}$ and $1\mathrm{e-}4$ for $\mb \theta_{\mb k}$ on synthetic data and $1\mathrm{e-}3$ for $\mb \theta_{\mb x}$ and $1\mathrm{e-}5$ for $\mb \theta_{\mb k}$ on real-world data. The disparate LRs allow the image estimate to update relatively more rapidly that the kernel estimate. All other parameters are as defaulted in \texttt{torch.optim.Adam}. We use a predefined LR schedule (using \texttt{MultiStepLR} in pytorch): both LRs decay by a factor of $\gamma=0.5$ once the iteration reaches any of the $[2000, 3000, 5000, 8000]$ milestones. The maximum number of iterations is set as $10,000$. By default, we use our WMV-ES to select the final estimates of $\mb k$ and $\mb x$. For all other settings, we strictly follow what are stated in \cref{alg:framework_ubid} unless otherwise declared. 

\paragraph{Synthetic and real-world datasets}
For synthetic datasets, we choose the popular datasets released by \cite{LevinEtAl2011Understanding} (dubbed as \texttt{LEVIN11}\footnote{Available at \url{https://webee.technion.ac.il/people/anat.levin/papers/LevinEtalCVPR09Data.rar}}) and \cite{LaiEtAl2016Comparative} (dubbed as \texttt{LAI16}\footnote{Available at \url{http://vllab.ucmerced.edu/wlai24/cvpr16_deblur_study/}}), respectively. Blurry images are directly synthesized following \cref{eq:bd_true_model} (without noise). Since groundtruth images and kernels are known in both datasets, we can explicitly control the level of kernel over-specification and the type and level of the noise. Moreover, we can also synthesize noise-only images to test the model stability. So \texttt{LEVIN11} and \texttt{LAI16} are ideal for us to evaluate and compare BID methods on all three kinds of stability that we care about. \texttt{LEVIN11} contains $4$ grayscale images of size $256 \times 256$ and $8$ different kernels with size ranging from $13\times 13$ to $27\times 27$, leading to $32$ blurry images. \texttt{LAI16} has $25$ RGB natural images of size around $1000 \times 700$ and $4$ kernels with larger sizes than \texttt{LEVIN11}: $31\times 31$, $51\times 51$, $53\times 53$, $75\times 75$, respectively, leading to $100$ blurry images.\footnote{\texttt{LAI16} has $4$ trajectories to synthesize non-uniform motion blur also, which we do not consider in this paper. Moreover, it also includes $100$ real-world blurry images without groundtruth kernels.} For both datasets, we use all the images in our subsequent experiments. 

For real-world datasets, we take the \texttt{NTIRE2020}~\cite{NahEtAl2020NTIRE}\footnote{Available at (registration needed to download the dataset): \url{https://competitions.codalab.org/competitions/22233\#learn\_the\_details}. We suspect that this is a superset of the REDS (REalistic and Dynamic Scenes) dataset (available at \url{https://seungjunnah.github.io/Datasets/reds.html}), at least with the same generation procedure as that of REDS. } and the \texttt{RealBlur}~\cite{RimEtAl2020Real}\footnote{Available at: \url{http://cg.postech.ac.kr/research/realblur/}} dataset. The blurry images in \texttt{NTIRE2020} are temporal averaging of consecutive frames from video sequences captured by high-speed cameras, totaling $24000$ and $3000$ blurry images in the training and validation sets, respectively\footnote{\texttt{NTIRE2020} is developed for data-driven approaches that require an extensive training set. }. Both camera shakes and object motions are involved, and temporal averaging emulates the blurring process due to temporal integration during exposure~\cite{NahEtAl2019NTIRE}. Since the exposure time is very short to ensure the high frame rate, \texttt{NTIRE2020} only covers well-lit scenes. In contrast, \texttt{RealBlur} emphasizes low-light environments that often involve a long exposure time and hence substantial blur. It captures sharp-blurry image pairs of static scenes with a customized dual-camera system, and only involves camera shakes as the source of relative motions. In total, \texttt{RealBlur} contains $4556$ pairs of sharp-blurry image pairs, covering $232$ low-light static scenes. For our experiments, we do not use the entire datasets but instead focus on $125$ selected cases that reflect the difficulty and diversity of real-world BID; see~\cref{sec:exp_realworld_selection} for details. 

\paragraph{Evaluation metrics}
Since we have the groundtruth clean images for both the synthetic and real-world data, we quantify and compare the performance of all selected BID methods using reference-based image quality assessment metrics. Besides the standard PSNR (peak signal-to-noise ratio) and SSIM (similarity structural index metric) metrics, we also take the information-theoretic VIF (visual information fidelity~\cite{SheikhBovik2006Image}) and DL-based metric LPIPS (learned perceptual image patch similarity, \cite{zhang2018unreasonable}) that have shown good correlation with human perception of image quality. We report all four metrics in all our quantitative results below.  
\paragraph{Model size and speed}
For our method, the total number of parameters is about $2.3$ million, and on average, it takes about $10$ minutes (on an Nvidia V100 GPU) to reconstruct a sharp image of size $1000 \times 1000$. \selfdeblur\, gets a similar number of parameters and is slightly faster ($\sim 8$ minutes). In this paper, we prioritize quality over speed, and hence we do not perform a systematic benchmark of speed, especially with respect to data-driven methods, for which inference only takes a single forward pass. Our recent work~\cite{LiEtAl2022Random} addresses the speed issue of DIP; we leave the potential integration as future work.  

\begin{figure}[!htbp]
  \centering
   \includegraphics[width = 0.90\linewidth]{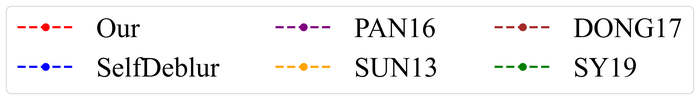}
    \includegraphics[width = 0.49\linewidth]{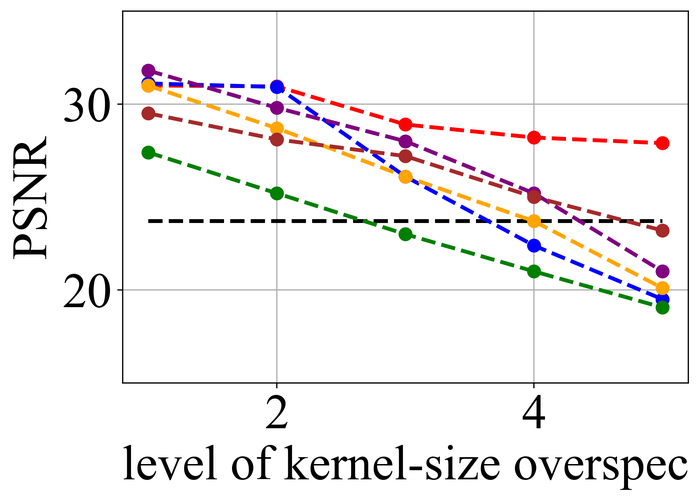}
    \includegraphics[width = 0.49\linewidth]{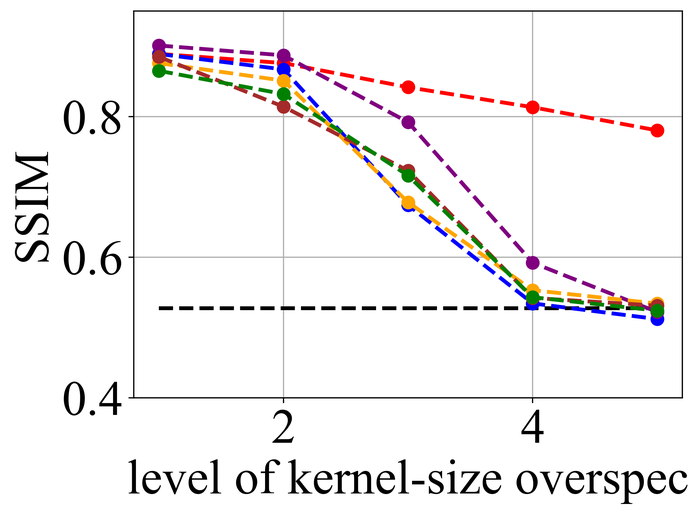}
 \\ 
    \includegraphics[width = 0.49\linewidth]{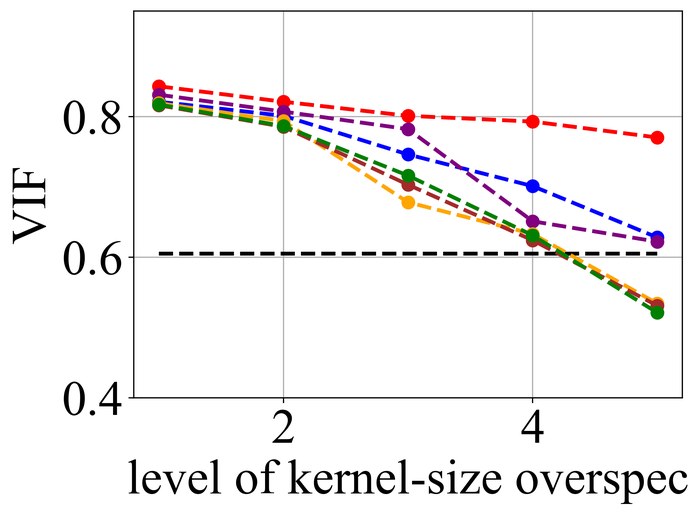}
    \includegraphics[width = 0.49\linewidth]{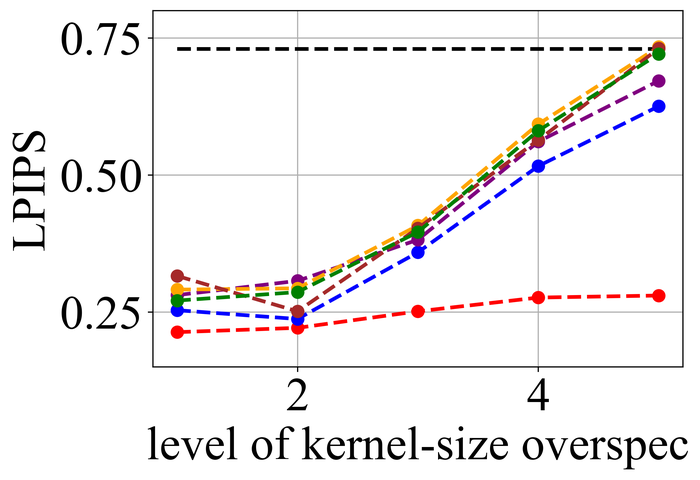}
  \caption{Comparison of the performance of the $6$ selected single-instance BID methods on \texttt{LEVIN11} with various levels of kernel-size overspecification. For PSNR, SSIM, and VIF, higher the better. For LPIPS, lower the better. The dashed lines indicate the performance baselines where the blurry image $\mb y$ and the groundtruth image $\mb x$ are directly compared. }
  \label{fig:bd_results_scale_levin}
\end{figure}
\begin{figure}[!htbp]
  \centering
    \includegraphics[width = 0.90\linewidth]{figs/Figure_scale/legend.png}
    \includegraphics[width = 0.49\linewidth]{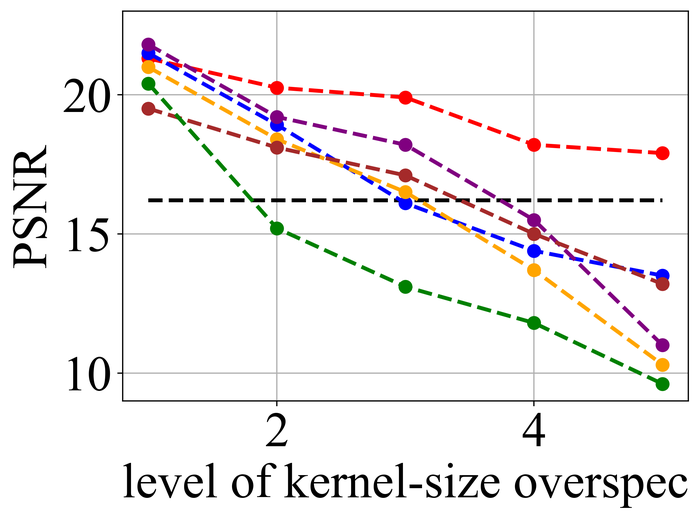}
    \includegraphics[width = 0.49\linewidth]{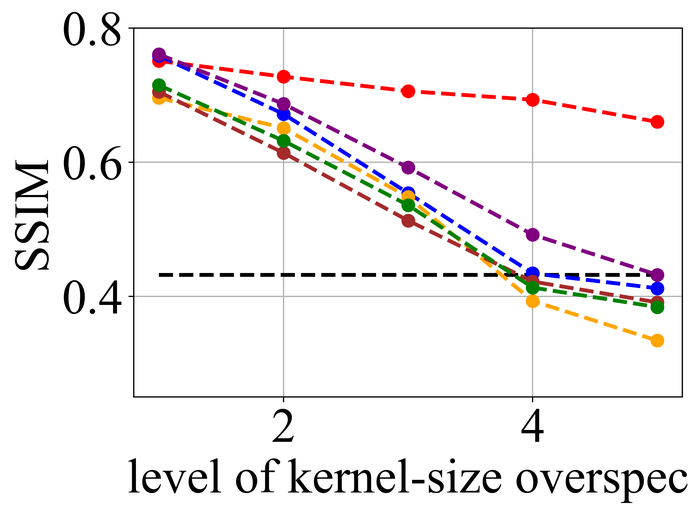}
  \\
  \includegraphics[width = 0.49\linewidth]{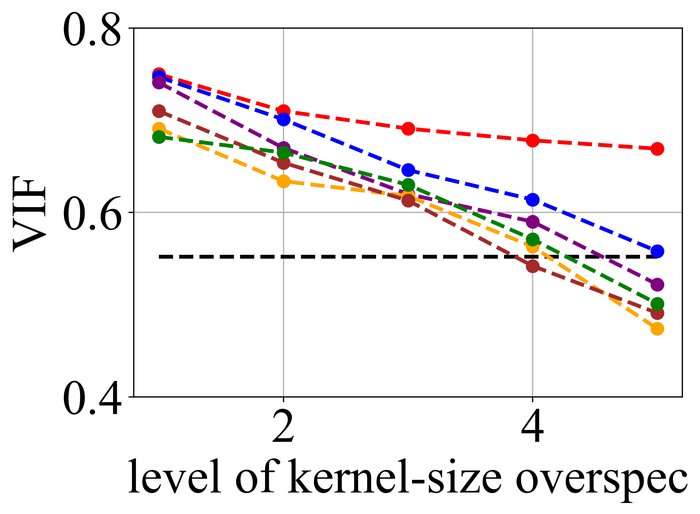}
    \includegraphics[width = 0.49\linewidth]{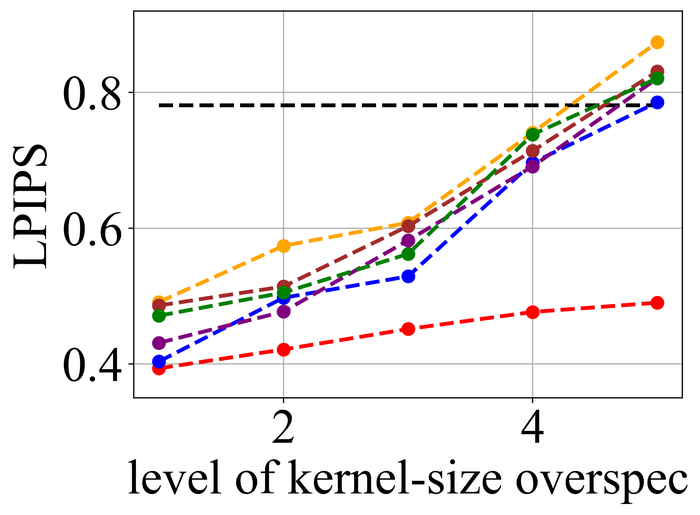}
  \caption{Comparison of the performance of the $6$ selected single-instance BID methods on \texttt{LAI16} with various levels of kernel-size overspecification. For PSNR, SSIM, and VIF, higher the better. For LPIPS, lower the better. The dashed lines indicate the performance baselines where the blurry image $\mb y$ and the groundtruth image $\mb x$ are directly compared. }
  \label{fig:bd_results_scale_lai}
\end{figure}
\subsection{Results on synthetic datasets} \label{sec:exp_synthetic_data}

Among single-instance methods, we pick \cite{SunEtAl2013Edge} (\texttt{SUN13}\footnote{Code available at: \url{http://cs.brown.edu/~lbsun/deblur2013/deblur2013iccp.html}}) that is among the top performing methods according to the 2016 survey paper~\cite{LaiEtAl2016Comparative}, and \cite{PanEtAl2016Blind} (\texttt{PAN16}\footnote{Code available at: \url{https://jspan.github.io/projects/dark-channel-deblur/index.html}}) that introduces the dark channel prior to BID and has been popular since 2016. We also select \cite{DongEtAl2017Blind} (\texttt{DONG17}\footnote{Code available at: \url{https://www.dropbox.com/s/qmxkkwgnmuwrfoj/code\_iccv2017\_outlier.zip?dl=0}}) which is a SOTA method that handles pixel corruptions, and \cite{SiYaoEtAl2019Understanding} (\texttt{SY19}\footnote{Code available at: \url{https://github.com/lisiyaoATbnu/low\_rank\_kernel}}) among the first single-instance BID works addressing unknown kernel sizes. \selfdeblur\,\footnote{Code available at: \url{https://github.com/csdwren/SelfDeblur}} \cite{RenEtAl2020Neural} inspires our method and hence is the main competitor. Together with our methods, all of the $6$ methods target the uniform setting in \cref{eq:bd_model}. 

We strive to make the comparison fair while highlighting methods that require no heavy hyperparameter tuning---in practice, we never know the exact level of overspecification or type/level of noise. \emph{So we always use the same set of hyperparameters for each method}. \texttt{SUN13} and \texttt{PAN16} are not designed to handle kernel-size overspecification and substantial noise; we directly use their default hyperparameters as it is unclear how to finetune them to optimize the performance in these novel scenarios. \texttt{SY19} allows kernel-size overspecification and provides a set of hyperparameters for twice kernel-size overspecification. We follow their recommendation for twice overspecification, and search and select an optimal set of hyperparameters over a grid beyond twice overspecification. For \texttt{DONG17} that handles substantial noise and pixel outliers, we use their default hyperparameter setting that is claimed to be general over different datasets. For \selfdeblur, we use their default setting, except that $\lambda_{\mb x}$ set as $\lambda_{\mb x} = 1\mathrm{e-}5$ instead of their default $\lambda_{\mb x} = 1\mathrm{e-}6$. This is because we observe that larger $\lambda_{\mb x}$ is needed to optimize the performance of \selfdeblur\, as the noise level grows. 
For our method, we set $\lambda_{\mb x} = 1\mathrm{e-}5$. All numbers that we report below are averages over images of the respective datasets. 

\subsubsection{Kernel-size overspecification}
We first evaluate the stability of the selected methods under kernel-size overspecification. Since we know the true kernel size for each instance, we divide the overspecification into $5$ levels: level $1$ corresponds to the true kernel size, level $5$ corresponds to half of the image size in both width and height directions---which is the default over-specification level for our method, and levels $2$--$4$ are evenly distributed in between. 

\cref{fig:bd_results_scale_levin,fig:bd_results_scale_lai} summarize the results on \texttt{LEVIN11} and \texttt{LAI16}, respectively. We observe that:
\begin{itemize} 
  \item When there is no kernel-size overspecification (i.e., level 1), \selfdeblur\, \texttt{PAN16}, and our method are among the top three performing methods (sometimes tied with other methods) by all metrics. This confirms the effectiveness of double-DIP ideas for BID; 
  \item As the overspecification level grows, the performance of all methods degrades, but our method is substantially more stable to such overspecification than other methods. In particular, for level-5 overspecification, while all of the other five methods become close or even worse than the baseline performance---where the blurry image $\mb y$ is directly taken to calculate the metrics, our method still performs strongly and shows considerable positive performance margins over the baseline; 
  \item The performance of all methods becomes uniformly lower moving from \texttt{LEVIN11} to \texttt{LAI16}. This is especially obviously on the pixel-based metrics PSNR and SSIM. We suspect there is mostly due to the larger kernel sizes in \texttt{LAI16} ($27 \times 27$ largest in \texttt{LEVIN11} vs $31 \times 31$ smallest in \texttt{LEVIN11}), which mess up large areas of pixels in each location;  
  \item \texttt{SY19}, the only previous single-instance method that explicitly handles kernel-size overspecification, does not perform well---despite our best effort to search for an optimal set of hyperparameters. In their paper~\cite{SiYaoEtAl2019Understanding}, they have reported promising results with twice overspecification on \texttt{LEVIN11}, much less aggressive than our evaluation: for example, for $13 \times 13$ kernels, they have tried $26 \times 26$ overspecification, but here we experiment with $13\times 13$, $42 \times 42$, $71 \times 71$, $100\times 100$, and $128 \times 128$. We suspect that the disappointing performance is due to the sensitivity of their method to hyperparameters across different overspecification levels. 
\end{itemize}
\begin{figure}[!htbp]
  \centering
   \includegraphics[width = 0.90\linewidth]{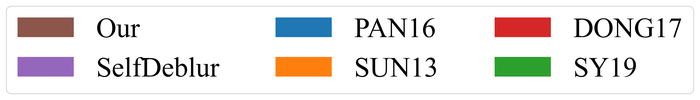}
    \includegraphics[width = 0.49\linewidth]{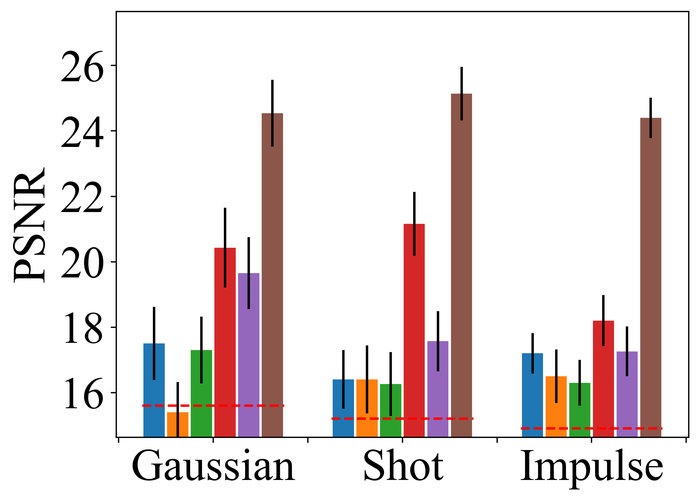}
    \includegraphics[width = 0.49\linewidth]{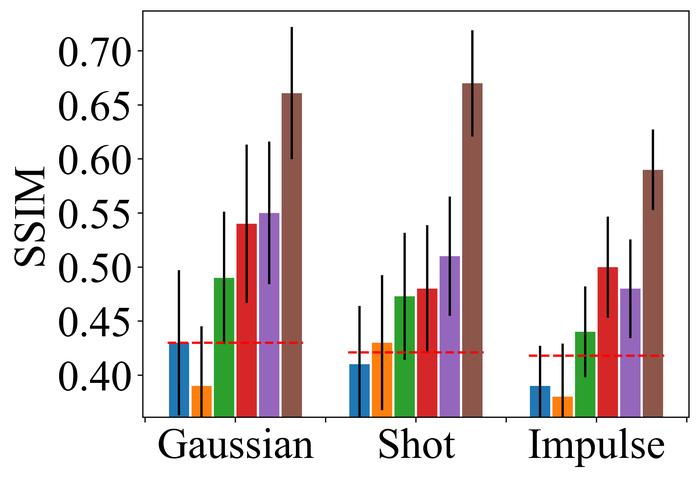}
  \\
    \includegraphics[width = 0.49\linewidth]{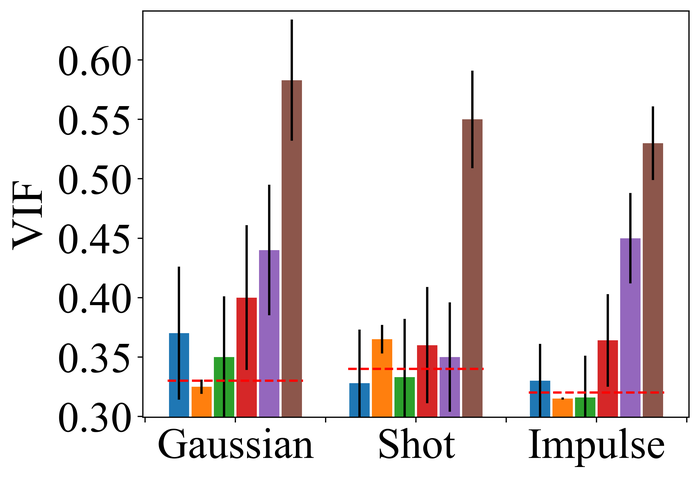}
    \includegraphics[width = 0.49\linewidth]{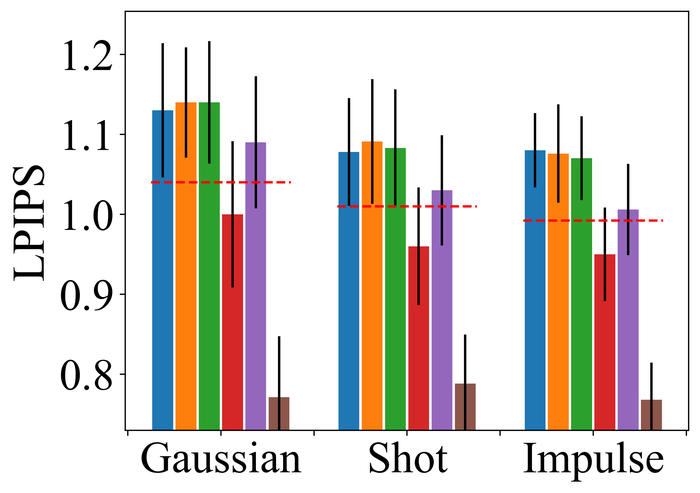} 
  \caption{Comparison of the performance of the $6$ selected single-instance BID methods on \texttt{LAI16} with \textbf{low-level} additive noise: Gaussian ($\sigma = 0.001$), shot ($\eta = 80$), and impulse ($p= 0.01$). For PSNR, SSIM, and VIF, higher the better. For LPIPS, lower the better. The dashes lines indicate the baseline performance where the blurry image $\mb y$ and the groundtruth image $\mb x$ are directly compared.}
  \label{fig:Lai_low_noise}
\end{figure}
\begin{figure}[!htbp]
  \centering
  \includegraphics[width = 0.90\linewidth]{figs/Figure_noise/legend_bar.png}
    \includegraphics[width = 0.49\linewidth]{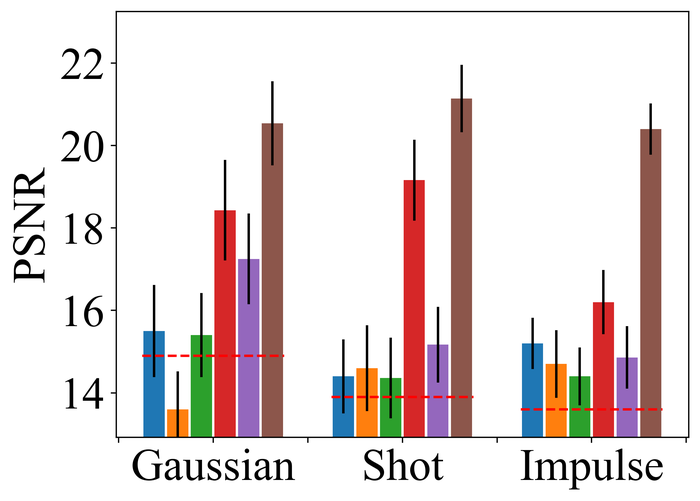}
    \includegraphics[width = 0.49\linewidth]{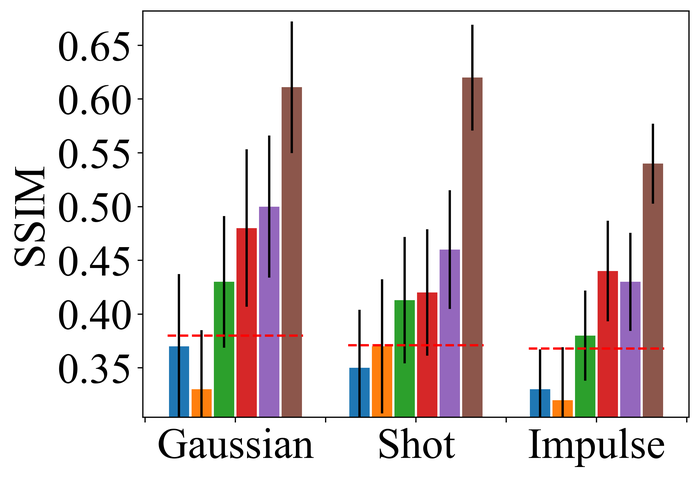}
\\
    \includegraphics[width = 0.49\linewidth]{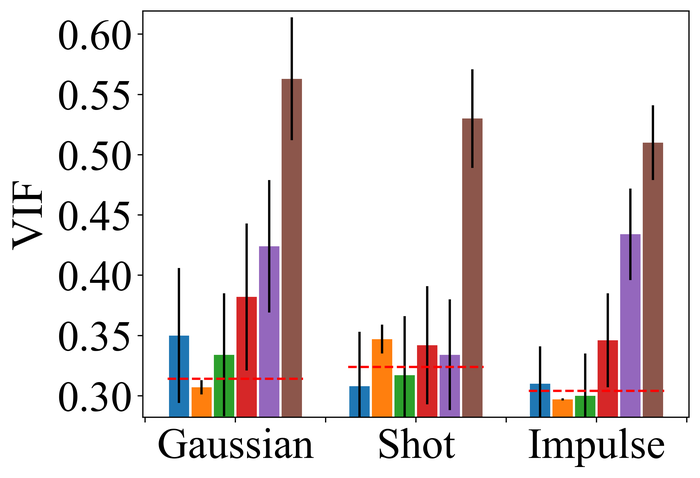}
    \includegraphics[width = 0.49\linewidth]{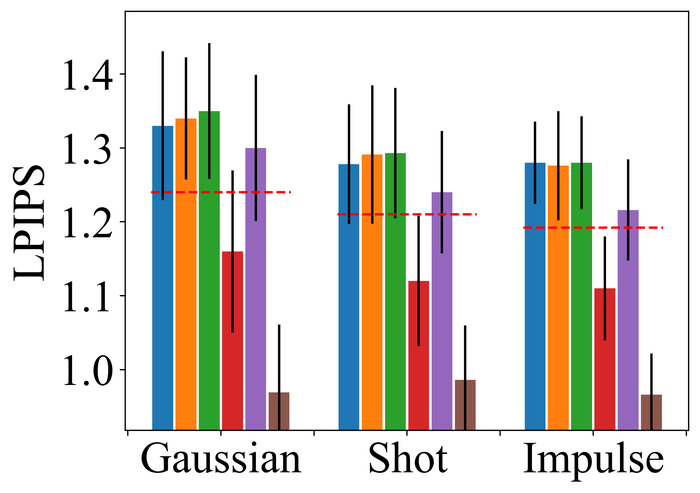} 
  \caption{Comparison of the performance of the $6$ selected single-instance BID methods on \texttt{LAI16} with \textbf{high-level} additive noise: Gaussian ($\sigma = 0.05$), shot ($\eta = 40$), and impulse ($p= 0.05$). For PSNR, SSIM, and VIF, higher the better. For LPIPS, lower the better. The dashes lines indicate the baseline performance where the blurry image $\mb y$ and the groundtruth image $\mb x$ are directly compared.}
  \label{fig:lai_high_noise}
\end{figure}
\begin{figure}[!htbp]
  \centering
   \includegraphics[width = 0.90\linewidth]{figs/Figure_scale/legend.png}
    \includegraphics[width = 0.49\linewidth]{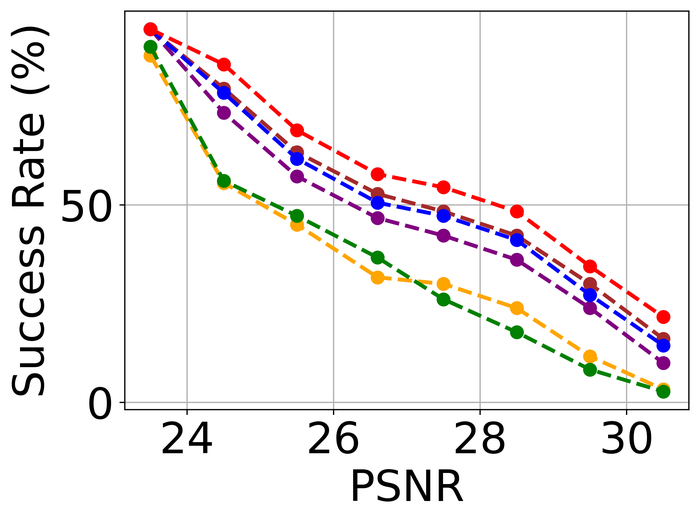}
    \includegraphics[width = 0.49\linewidth]{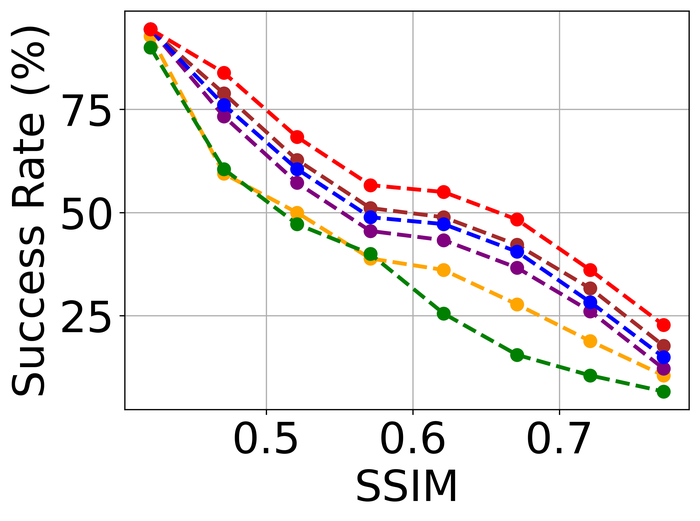}
    \\
    \includegraphics[width = 0.49\linewidth]{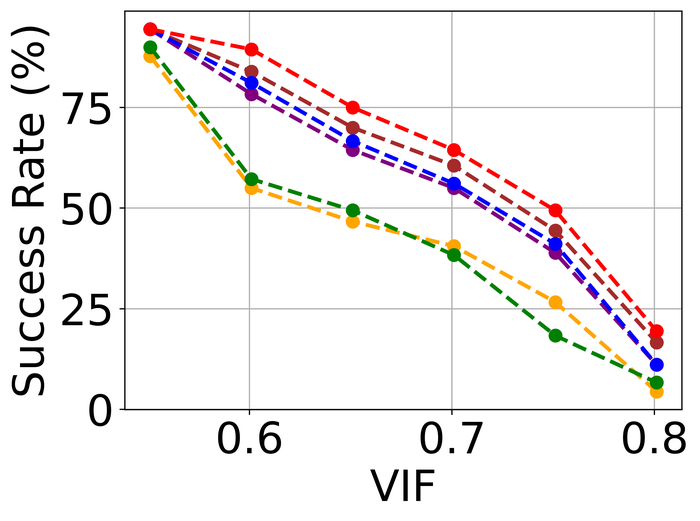}
    \includegraphics[width = 0.49\linewidth]{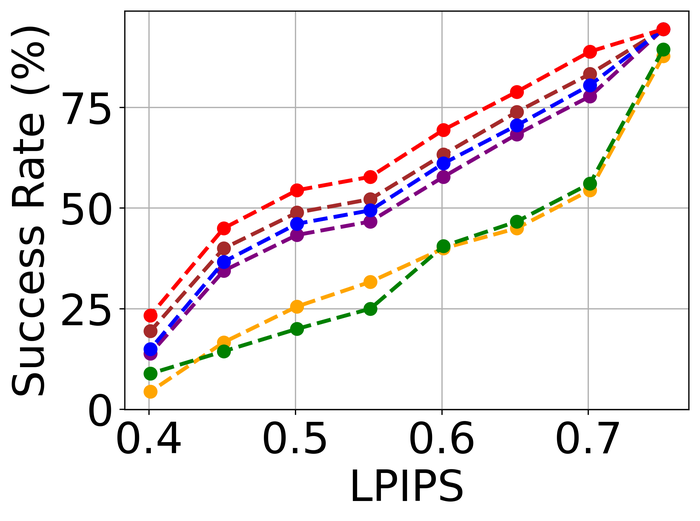}
  \caption{Comparison of the performance of the $6$ selected single-instance BID methods on \texttt{LAI16} with pixel saturation. For any fixed operation point of the evaluation metric (i.e., the horizontal axis), the success rate is defined as the fraction of images recovered at that quality or higher. For PSNR, SSIM, and VIF, higher the better. For LPIPS, lower the better. }
  \label{fig:lai_saturated}
\end{figure}
\subsubsection{Substantial noise}
\label{sec:substantial_noise}
To evaluate the noise stability, we fix the kernel-size overspecification as half of the image size in both directions (i.e., the default for our method) \emph{for all methods}, and focus on \texttt{LAI16}. We consider $4$ types of noise that have been considered in prior works: 
\begin{itemize}
  \item \textbf{Gaussian noise}: zero-mean additive Gaussian noise with standard deviation $\sigma = 0.001$ and $\sigma = 0.05$ for low and high noise levels, respectively; 
  \item \textbf{Impulse noise} (i.e., salt-and-pepper noise): replacing each pixel with probability $p \in [0,1]$ into white ($1$) or black ($0$) pixel with half chance each. Low and high noise levels correspond to $p = 0.005$ and $ p= 0.08$, respectively; 
  \item \textbf{Shot noise} (i.e., pixel-wise independent Poisson noise): for each pixel $x \in [0, 1]$, the noisy pixel is Poisson distributed with rate $\eta x$, where $\eta = 90, 25$ for low and high noise levels, respectively; 
  \item \textbf{Pixel saturation}: each blurry RGB image $\mb y$ in \texttt{LAI16} is first converted into HSV (i.e., hue-saturation-lightness) representation $\mb y_{\mathrm{HSV}}$ with values in $[0, 1]$, and then the saturation channel is rescaled by a factor of $2$, shifted by a factor $0.1$, and then cropped into $[0, 1]$. The resulting HSV representation is then converted back to RBG representation, with all values cropped back into $[0, 1]$.  We further add pixel-wise zero-mean Gaussian noise with standard deviation $\sigma = 0.0001$.  
\end{itemize}
\cref{fig:Lai_low_noise,fig:lai_high_noise} present the results on the first three types of noise, for the low- and high-level, respectively. As expected, all methods perform worse when moving from low- to high-level noise. \texttt{DONG17}, \selfdeblur, and our method are the top three performing methods by all metrics, for both low- and high-level noise. While \selfdeblur \, is even worse than the trivial baseline (i.e., when no BID method is applied) by LPIPS, both \texttt{DONG17} and ours always outperform the baseline---both use robust losses\footnote{In \texttt{DONG17}, the loss consists in applying $h(z) = z^2/2 - \log {(a+e^{bz^2})}/(2b)$ element-wise to $\mb y - \mb k \ast \mb x$, where $a, b > 0$ and so that $h(z) \le 0$. Note that $h(z) \sim  O(z^2)$ as $z \to 0$, and $h(z)$ approaches the constant $0$ when $z$ is large. } that are less sensitive to large errors compared to the standard MSE loss. Our method is the top performer and always win the second best, i.e., \texttt{DONG17}, by large margins by all metrics. 

We observe similar performance trends of these methods in terms of handling pixel saturation, from \cref{fig:lai_saturated}: \selfdeblur, \texttt{DONG17}, and ours are the top three methods, with our method outperforming the other two by considerable margins. Based on these results, we conclude that using robust losses for BID is crucial to achieving robustness to practical noise.

\begin{figure}[!htbp]
  \centering
  \includegraphics[width = 0.90\linewidth]{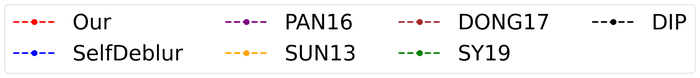}
    \includegraphics[width = 0.49\linewidth]{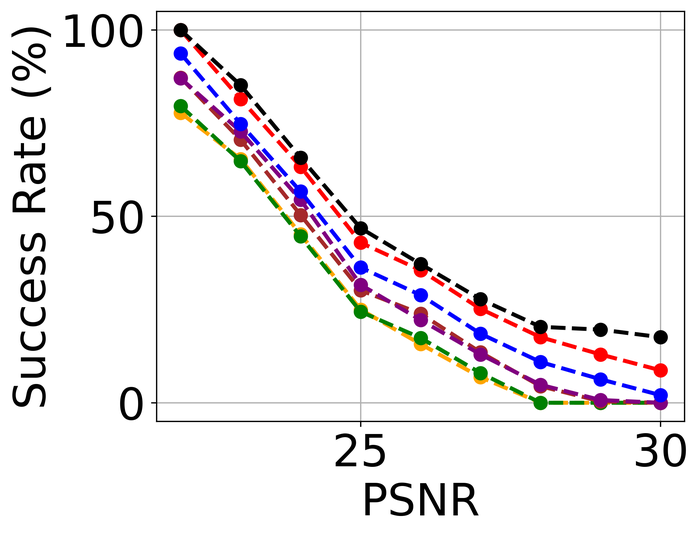}
    \includegraphics[width = 0.49\linewidth]{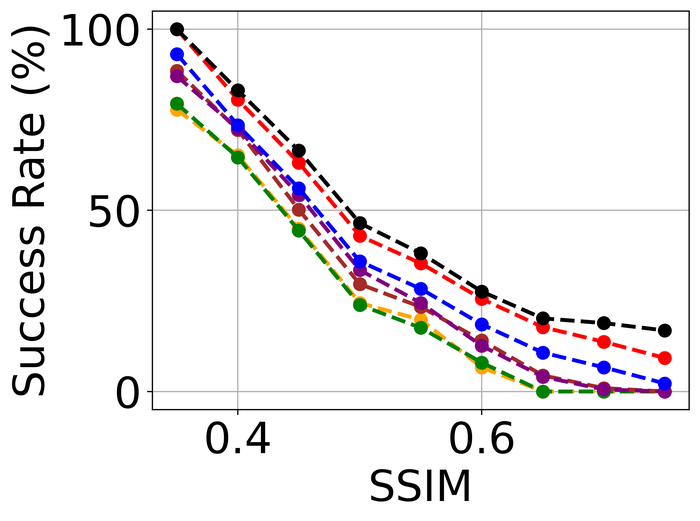}
    \\
    \includegraphics[width = 0.49\linewidth]{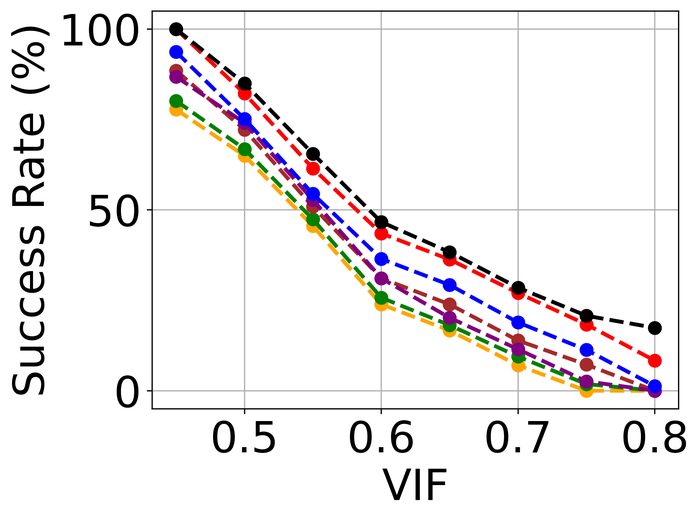}
    \includegraphics[width = 0.49\linewidth]{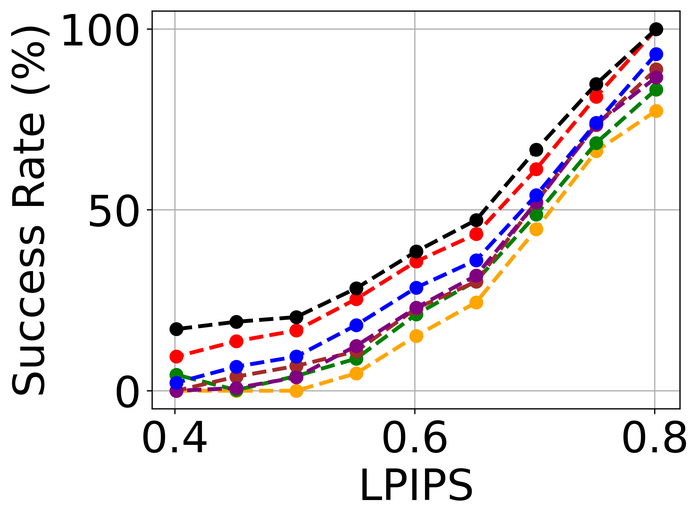} 
  \caption{Comparison of the performance of the $6$ selected single-instance BID methods on \texttt{LAI16} with high-level noise only (no blur). \texttt{DIP} denotes a single-DIP model that does not account for blur at all, i.e., knowing the image is noise-only. 
  The noise is randomly selected from Gaussian ($\sigma=0.1$), shot ($\eta = 40$), and impulse ($p = 0.08$) per image. For any fixed operation point of the evaluation metric (i.e., the horizontal axis), the success rate is defined as the fraction of images recovered at that quality or higher. For PSNR, SSIM, and VIF, higher the better. For LPIPS, lower the better. }
  \label{fig:lai_model_stable_high_noise}
\end{figure}
\begin{figure}[!htbp]
  \centering
    \includegraphics[width = 0.8\linewidth]{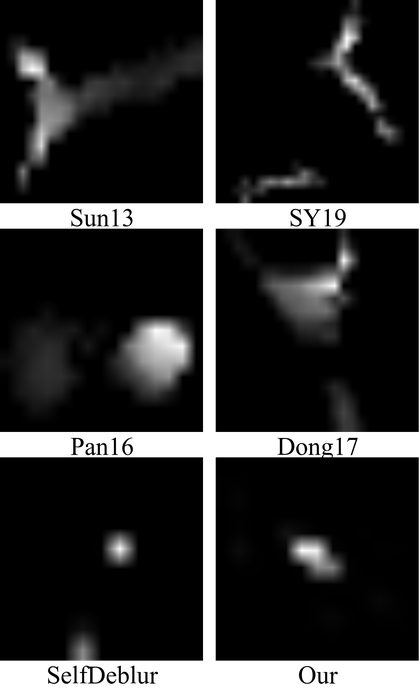}
    \caption {Examples of estimated kernels of the $6$ selected single-instance BID methods on \texttt{LAI16} with high-level noise only (no blur, same setting as in \cref{fig:lai_model_stable_high_noise}). }
  \label{fig:lai_model_stable_high_noise_kernel}
\end{figure}
\subsubsection{Model stability}
To evaluate model stability, we simulate noise-only images without blurs. For each image, we \emph{randomly} pick one of the three types of high-level noise: Gaussian ($\sigma=0.1$), shot ($\eta = 40$), and impulse ($p = 0.08$), and apply it to produce the simulated noisy image. Note that the individual noise levels are considerably higher than those used in \cref{fig:lai_high_noise}. The reason is that we hope to stretch the difficulty level of the test: intuitively, an ideal BID method should tolerate more noise on a noise-only input than on a blurry-and-noisy input.
\emph{As far as we are aware, this is the first evaluation of SOTA BID methods in terms of model stability}. 

The results are presented in \cref{fig:lai_model_stable_high_noise}. There, \texttt{DIP} denotes the single-DIP method that directly models the noise only, i.e., by considering 
\begin{align*}
  \min_{\mb \theta_{\mb x}} \;  \ell_{\mathrm{Huber}} (\mb y, G_{\mb \theta_{\mb x}}(\mb z_{\mb x})) + \lambda_{\mb x} \frac{\norm{\nabla_{\mb x} G_{\mb \theta_{\mb x}}(\mb z_{\mb x})}_1}{\norm{\nabla_{\mb x} G_{\mb \theta_{\mb x}}(\mb z_{\mb x})}_2}. 
\end{align*}
We use exactly the same architecture for $G_{\mb \theta_{\mb x}}$ and the same $\lambda_{\mb x}$ as used in our method. Since this method incorporates the knowledge that the image has no blur, it is not surprising it performs the best. Immediately after, it is evident that \selfdeblur\, and ours are the clear winners by all metrics, and ours leads \selfdeblur\, by visible margins. Moreover, the performance of our method approaches that of \texttt{DIP}, suggesting strong model stability of our method. Unfortunately, although \texttt{DONG17} can tolerate substantial noise together with blur, it does not work well when there is no blur. In fact, the estimated kernels of the four non-Double-DIP methods (i.e., \texttt{SUN13}, \texttt{SY19}, \texttt{PAN16}, \texttt{DONG17}) are far from the delta function---which is the true kernel in this case, as shown in \cref{fig:lai_model_stable_high_noise_kernel}. In contrast, \selfdeblur\, and our method recover kernels that resemble the delta function. Besides the common sparse gradient prior on the image used by all methods, \selfdeblur\, and our method also enforce the DIP on the image. We suspect that their superior model stability can be attributed to the simultaneous use of the two priors instead of only one. We reiterate that we do not finetune the hyperparameters of any method moving from the previous blurry-and-noisy test to the current noise-only test: finetuning may improve certain methods, but is deemed impractical as we often do not have such model knowledge about real data.   

\begin{figure}[!htbp]
  \centering
  \includegraphics[width=\linewidth]{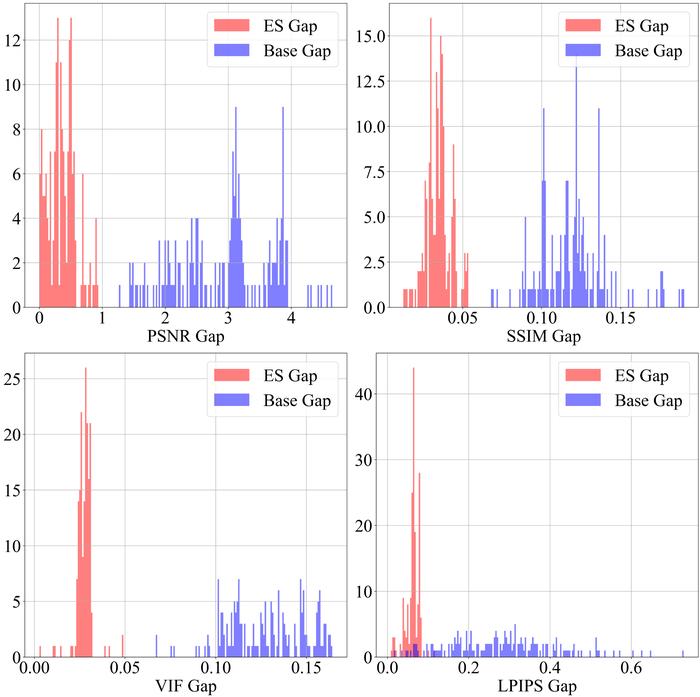}
  \caption{Detection performance of WMV-ES on \texttt{LAI16}. Each plot corresponds to one metric we use, and collects the histogram of the ES-Gap (gap between the true peak performance and the detected performance) and the Base-Gap (gap between the true peak performance and the performance at the last iteration). }
  \label{fig:ES_res}
\end{figure} 
\subsubsection{Early stopping}
As we discussed in \cref{sec:method_ES}, ES is necessary and practical for preventing overfitting when there is substantial noise. Here, we test the WMV-ES method~\cite{WangEtAl2021Early} that we use by default, on \texttt{LAI16} with low- and high-level Gaussian noise (as defined in \cref{sec:substantial_noise}). \cref{fig:ES_res} presents the histograms of ES gap (between the peak performance and the detected performance by the ES method) and the Base gap (between the peak performance and the final performance with overfitting), using all of the four metrics. It is clear that ES is crucial to saving the performance: without ES, the eventual overfitting of double-DIP to noise ruins the recovery, e.g., reducing the PSNR by $3$ points or more for a large portion of the images; with the automatic ES method WMV-ES, we are only slightly off the peak performance---just to be sure, without knowing the groundtruth in practice, we cannot directly stop the algorithm right at the peak performance point. The success of WMV-ES is evident from the clear separation of the histograms between ES Gap and Base Gap, by all of the metrics.

\subsection{Results on real-world datasets}
\label{sec:exp_realworld}

\subsubsection{Competing methods and data preparation} 
\label{sec:exp_realworld_selection}
It is clear by far that the $5$ competing methods that we worked with above are not good choices for real-world BID, due to their sensitivity to kernel-size overspecification and substantial noise. On the other hand, most of the recent SOTA BID methods are data-driven in nature: although they may not be generalizable as limited by the training data, they are attractive as most recent variants directly predict sharp images from blurry images and hence bypass the problems caused by unknown kernel size and even inaccurate blur modeling~\cite{KohEtAl2021Single}. Hence, in this section, we stretch our method, as well as \selfdeblur, by comparing them with $3$ SOTA data-driven methods on the SOTA \texttt{NTIRE2020} and \texttt{RealBlur} BID datasets. 

Scale-recurrent network (\texttt{SRN})~\cite{tao2018scale} and GAN-based \texttt{DeblurGAN-v2}~\cite{kupyn2019deblurgan} are BID models trained on paired blurry-sharp image pairs. The prediction models for both take inspiration from the coarse-to-fine multiscale ideas in traditional BID. In addition, \texttt{DeblurGAN-v2} employs GAN-based discriminators as regularizers to improve the deblurring quality. \texttt{ZHANG20}~\cite{zhang2020deblurring} stresses the practical difficulty in obtaining blurry-sharp training pairs (echoing the discussion of similar difficulty in~\cite{KohEtAl2021Single,ZhangEtAl2022Deep}), and derives a pipeline to learn the blurring and deblurring processes from unpaired blurry and sharp images. For the comparison below, we directly take the pretrained models of the $3$ methods~\footnote{\texttt{SRN} is available at: \url{https://github.com/jiangsutx/SRN-Deblur}; \texttt{DeblurGAN-v2} is available at: \url{https://github.com/VITA-Group/DeblurGANv2}; \texttt{ZHANG20} is available at: \url{https://github.com/HDCVLab/Deblurring-by-Realistic-Blurring}. }. We note that both \texttt{SRN} and \texttt{DeblurGAN-v2} use the \texttt{GoPro} dataset~\cite{NahEtAl2017Deep} as part of their training sets, and \texttt{ZHANG20} builds their own blurry training set \texttt{RWBI}~\cite{zhang2020deblurring}. To the best of our knowledge, \texttt{NTIRE2020} and \texttt{RealBlur} have no overlap with \texttt{GoPro} and \texttt{RWBI}. So we believe our evaluation set makes a good test for real-world generalizability of the $3$ selected methods. 

\begin{figure}[!htbp] 
  \centering 
  \includegraphics[width=0.95\linewidth]{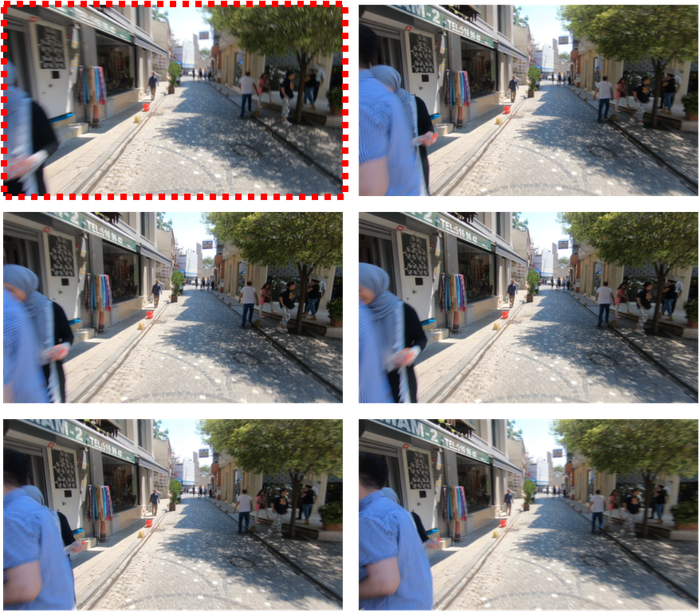}\\
  \vspace{1em}
  \includegraphics[width=0.95\linewidth]{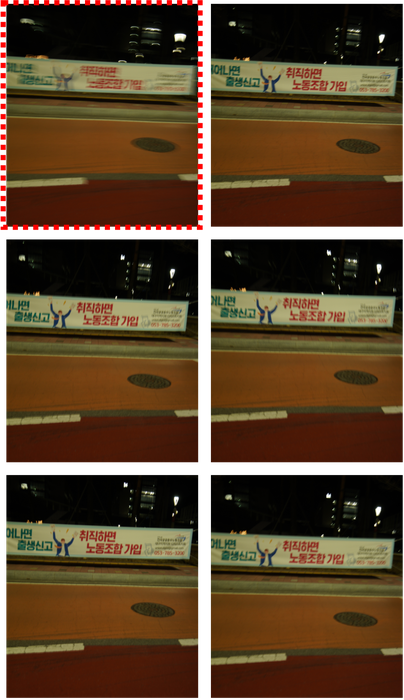}
  \caption{Illustration of image selection from \texttt{NTIRE2020} (top) and \texttt{RealBlur} (bottom), respectively. For images from the same dynamic/static scene, we always select visually the most blurry image (highlighted by red bounding boxes). } 
  \label{fig:real_selection_illus}
\end{figure} 
\begin{figure}[!htbp] 
  \centering 
  \includegraphics[width=0.95\linewidth]{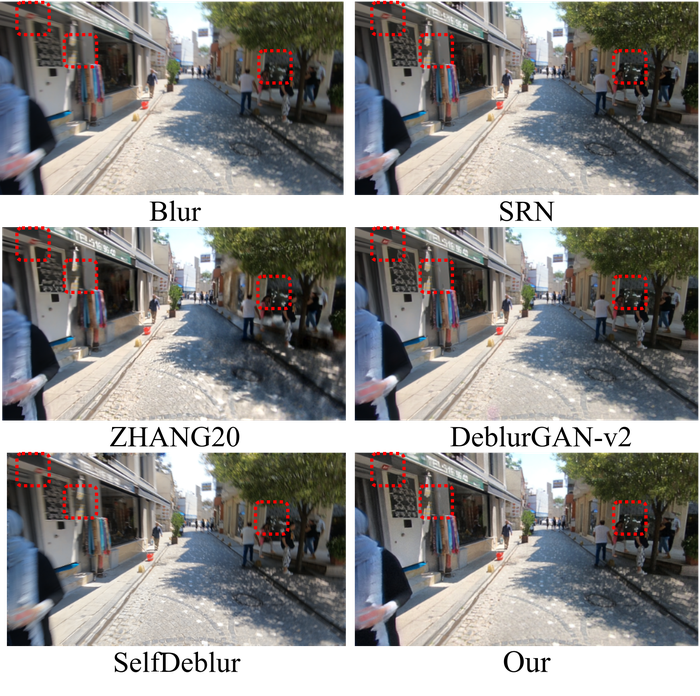}
  \caption{Comparison of deblurring results on a bright scene with high depth contrast} 
  \label{fig:real_table}
\end{figure} 
\begin{figure}[!htbp] 
  \centering 
  \includegraphics[width=0.8\linewidth]{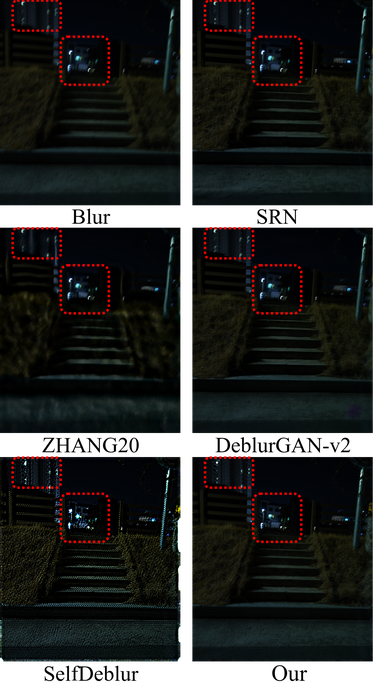}
  \caption{Comparison of deblurring results on a dark scene with high depth contrast} 
  \label{fig:real_table2}
\end{figure} 
\begin{figure}[!htbp] 
  \centering 
  \includegraphics[width=0.8\linewidth]{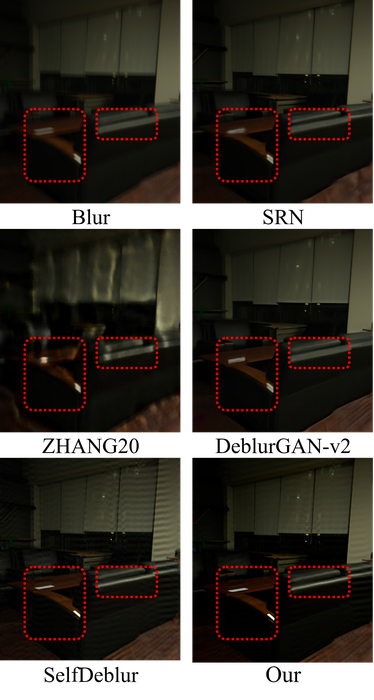}
  \caption{Comparison of deblurring results on a bright scene with low depth contrast} 
  \label{fig:real_table3}
\end{figure} 
\begin{figure}[!htbp] 
  \centering 
  \includegraphics[width=0.8\linewidth]{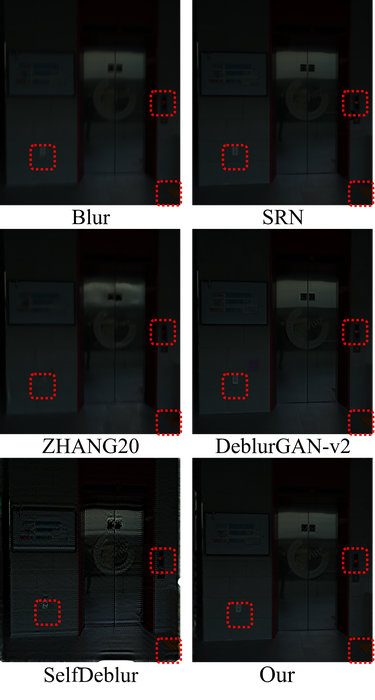}
  \caption{Comparison of deblurring results on a dark scene with low depth contrast} 
  \label{fig:real_table4}
\end{figure} 
\begin{figure}[!htbp] 
  \centering 
  \includegraphics[width=0.8\linewidth]{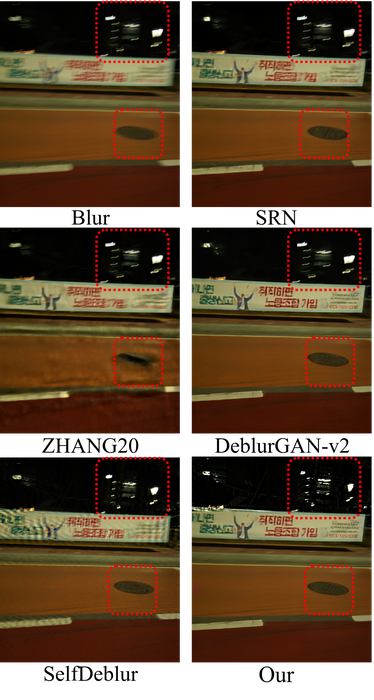}
  \caption{Comparison of deblurring results on a scene with high depth contrast and high brightness contrast} 
  \label{fig:real_table1}
\end{figure} 
\begin{figure}[!htbp] 
  \centering 
  \includegraphics[width=0.8\linewidth]{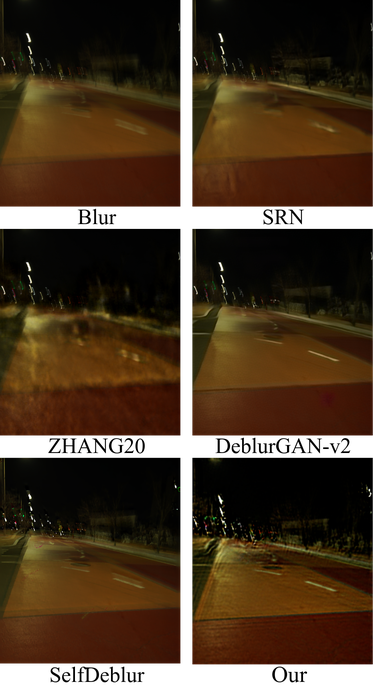}
  \caption{Failure case: Comparison of deblurring results on a scene with high depth contrast and high brightness contrast} 
  \label{fig:real_failure}
\end{figure} 
\begin{table*}[!htbp]
  \caption{Quantitative comparison of deblurring results on the $125$ selected real-world images. For PSNR, SSIM, and VIF, higher the better. For LPIPS, lower the better. S1--S5 represent the $5$ scenarios described in \cref{sec:exp_realworld_selection}. We report in the form of ``mean (standard deviation)" (over the $125$ images) for each method/metric combination. For each line, the first and second best numbers (according to the means) are marked in \textcolor{red}{RED} and \textcolor{Green}{GREEN}, respectively. }
  \centering 
    {%
    \begin{tabular}{c|c|c|c|c|c|c}
      \hline 
      & & SRN & DeblurGAN-v2 & ZHANG20 & SelfDeblur & Ours \\
      \hline
      \multirow{2}{*}{S1}&PSNR& 30.1 (1.159)& \textcolor{red}{31.0 (1.149)} &25.2 (1.188) & 28.2 (1.198)& \textcolor{Green}{30.8 (1.168)}\\
      &SSIM & 0.871 (0.0679) & \textcolor{red}{0.883 (0.0609)} & 0.793 (0.0724) & 0.832 (0.0734)& \textcolor{Green}{0.873(0.0618)}\\
      & VIF & 0.784 (0.0686) & \textcolor{red}{0.801 (0.0647)} & 0.705 (0.0705) & 0.725 (0.0727)& \textcolor{Green}{0.796 (0.0651)}\\
      & LPIPS& 0.972 (0.0966) & \textcolor{Green}{0.827 (0.08869)} & 1.025 (0.104) & 0.987 (0.101)& \textcolor{red}{0.821 (0.0879)} \\
      \hline
      \multirow{2}{*}{S2}&PSNR& 27.1 (1.256)& \textcolor{Green}{27.4 (1.352)} &23.4 (1.449) & 25.9 (1.471)& \textcolor{red}{28.7 (1.236)}\\
      &SSIM & 0.851 (0.0744) & \textcolor{Green}{0.859 (0.0695)} & 0.789 (0.0753) & 0.821 (0.0758)& \textcolor{red}{0.870 (0.0681)}\\
      & VIF & 0.772 (0.0778) & \textcolor{red}{0.783 (0.0758)} & 0.699 (0.0787) & 0.713 (0.0777)& \textcolor{Green}{0.781 (0.0767)}\\
      & LPIPS& 1.021 (0.116) & \textcolor{Green}{0.901 (0.0985)} & 1.076 (0.108) & 1.001 (0.111)& \textcolor{red}{0.811 (0.0947)} \\
      \hline
      \multirow{2}{*}{S3}&PSNR& 28.3 (1.197)& \textcolor{Green}{28.7 (1.139)} &25.2 (1.236) & 26.2 (1.227)& \textcolor{red}{29.4 (1.144)}\\
      &SSIM & 0.866 (0.0647) & \textcolor{Green}{0.867 (0.0608)} & 0.803 (0.0658) & 0.827 (0.0637)& \textcolor{red}{0.872 (0.0589)}\\
      & VIF & 0.761 (0.0772) & \textcolor{red}{0.787 (0.0727)} & 0.701 (0.0766) & 0.731 (0.0776)& \textcolor{Green}{0.780 (0.0679)}\\
      & LPIPS& 1.008 (0.0985) & \textcolor{Green}{0.869 (0.0936)} & 1.076 (0.107) & 0.985 (0.110)& \textcolor{red}{0.839 (0.0911)} \\
      \hline
      \multirow{2}{*}{S4}&PSNR& 26.7 (1.014)& \textcolor{Green}{27.1 (0.985)} &23.3 (1.043) & 25.8 (1.055)& \textcolor{red}{28.5 (0.947)}\\
      &SSIM & 0.849 (0.0542) & \textcolor{Green}{0.851 (0.0498)} & 0.780 (0.0567) & 0.812 (0.0578)& \textcolor{red}{0.861 (0.0481)}\\
      & VIF & 0.756 (0.0621) & \textcolor{Green}{0.767 (0.0592)} & 0.687 (0.0663) & 0.721 (0.0674)& \textcolor{red}{0.776 (0.0574)}\\
      & LPIPS& 1.015 (0.0941) & \textcolor{Green}{0.925 (0.0862)} & 1.050 (0.0927) & 0.996 (0.0674)& \textcolor{red}{0.893 (0.0848)} \\
      \hline
       \multirow{2}{*}{S5}&PSNR& 28.6 (1.352)& \textcolor{Green}{28.7 (1.314)} &24.7 (1.410) & 26.4 (1.400)& \textcolor{red}{29.2 (1.284)}\\
      &SSIM & 0.846 (0.0754) & \textcolor{Green}{0.855 (0.0694)} & 0.781 (0.0762) & 0.818 (0.0771)& \textcolor{red}{0.867 (0.0674)}\\
      & VIF & 0.756 (0.0756) & \textcolor{Green}{0.771 (0.0754)} & 0.692 (0.0784) & 0.710 (0.0793)& \textcolor{red}{0.776 (0.0761)}\\
      & LPIPS& 1.012 (0.1093) & \textcolor{Green}{0.874 (0.1085)} & 1.065 (0.1141) & 0.992 (0.1149)& \textcolor{red}{0.856 (0.0945)} \\
      \hline
  \end{tabular}
    } 
  \label{tab:real_world}
  \end{table*}
As alluded to above, both \texttt{NTIRE2020} and \texttt{RealBlur} have their own strengths and limitations: images in \texttt{NTIRE2020} may contain multiple motions, but are captured in well-lit environments; \texttt{RealBlur} covers many dark scenes, but the scenes are static and relative motions are caused by camera shakes only. In preliminary tests, we find the $3$ selected data-driven methods perform vastly differently across images, even within the same dataset. The dictating factors seem to include contrast of scene depth, contrast of brightness, and the combination thereof: different scene depths likely correspond to different relative motions, especially in the data of \texttt{NTIRE2020}, as well as different levels of defocus blur, while relative to the bright areas, dark areas tend to be less attended to by typical losses. Hence, we choose both \texttt{NTIRE2020} and \texttt{RealBlur}: the former contains a good portion of images with good depth contrast and multiple moving objects, and the latter provides samples with good brightness and depth contrast. 

We select $125$ representative, visually challenging images from the two datasets: for \texttt{NTIRE 2020}, we pick the most blurry frame from each folder that contains a sequence of consecutive frames; similarly, for \texttt{RealBlur}, we pick the most blurry one from images about the same scene. \cref{fig:real_selection_illus} gives a couple of examples to illustrate our selection. The $125 $ images are classified into $5$ scenarios---$25$ images each: (S1) bright scene with high depth contrast (see an example in \cref{fig:real_table}); (S2) dark scene with high depth contrast (see an example in \cref{fig:real_table2}); (S3) bright scene with low depth contrast (see an example in \cref{fig:real_table3}); (S4) dark scene with low depth contrast (see an example in \cref{fig:real_table4}); (S5) scene with high depth contrast and high brightness contrast (see an example in \cref{fig:real_table1}). \texttt{NTIRE2020} only includes bright scenes, and we pick $35$ images from it: $25$ for S1, and $10$ for S3. Then, from \texttt{RealBlur}, we choose $15$ images to complete S3, and $25$ images for each of S2, S4, and S5, respectively. For reproducibility of our results, the IDs of the selected images can be found in our Github repository: \urlstyle{sf} \url{https://github.com/sun-umn/Blind-Image-Deblurring}. 

\subsubsection{Qualitative and quantitative results}
\cref{fig:real_table,fig:real_table1,fig:real_table2,fig:real_table3,fig:real_table4} present $5$ blurry images (\cref{fig:real_table2} and \cref{fig:real_table4} are too dark to reveal enough details; we apply histogram equalization to enhance the contrast and include them in \cref{sec:app_hist_enhance}), each representing one of the $5$ scenarios, and the recovery results from \texttt{SRN}, \texttt{ZHANG20}, \texttt{DeblurGAN-v2},  \selfdeblur, and our method. \cref{tab:real_world} summarizes the quantitative results over the $125$ selected images using the metrics: \texttt{PSNR}, \texttt{SSIM}, \texttt{VIF}, and \texttt{LPIPS}. 

Our method wins in most cases, followed by GAN-based \texttt{DeblurGAN-v2}. In fact, they are the top two in all cases. \texttt{DeblurGAN-v2} leads our method on \texttt{S1} by all metrics except for \texttt{LPIPS}, and on \texttt{S2} and \texttt{S3} only by \texttt{VIF}. This is likely because \texttt{S1} is sampled entirely from \texttt{NTIRE2020} that consists of bright scenes only, similar to the \texttt{GoPro} dataset that \texttt{DeblurGAN-v2} is trained on; only $10$ out of $25$ images from \texttt{S3} are from \texttt{NTIRE2020}. On \texttt{S2}, \texttt{S4}, and \texttt{S5} where each image consists of part of dark scenes, our method is a clear winner. This can be explained by the emphasis of the \texttt{RealBlur} dataset on dark scenes that have different distributions than \texttt{GoPro} that only includes bright scenes. It is remarkable that our method, a non-data-driven method, can performs on par with SOTA data-driven methods on similar data the latter are trained on, and can perform consistently better on novel data. The performance discrepancy of \texttt{DeblurGAN-v2} on different scenarios again underscores how data-driven methods can be limited by the training data, although overall \texttt{DeblurGAN-v2} indeed shows reasonable generalizability to the novel dataset \texttt{RealDeblur}. 

\texttt{ZHANG20}, the worst performer in our evaluation, is trained on the Real-World Blurry Image (\texttt{RWBI}) dataset~\footnote{Available at: \url{https://drive.google.com/file/d/1fHkPiZOvLQSc4HhT8-wA6dh0M4skpTMi/view}} collected by the same group of authors~\cite{zhang2020deblurring}. Visual inspection into \texttt{RWBI} suggests the blurry scenes are mostly similar to those of \texttt{GoPro}: bright scenes, none or few moving objects, substantial camera motions. So it is no surprise that the original paper~\cite{zhang2020deblurring} reports encouraging generalization performance of their pretrained model on \texttt{GoPro}. By contrast, \texttt{NTIRE2020} images are mostly taken about much more complex scenes with multiple moving objects plus synthetic camera motions, and \texttt{RealBlur} emphasizes dark scenes. The significant distribution shift explains the relatively poor performance of their pretrained model in our evaluation, as seen from \cref{tab:real_world} and the visual results in \cref{fig:real_table,fig:real_table1,fig:real_table2,fig:real_table3,fig:real_table4}, and underscores again the generalizability issue around data-driven methods. Note that \texttt{SRN} is originally trained and tested on \texttt{GoPro}, and hence is subject to similar distribution shift and performance drop. But, \texttt{SRN} is trained on sharp-blurry image pairs, whereas \texttt{ZHANG20} on unpaired sharp and blurry images and so the input knowledge is much weaker and the learning task is more challenging, explaining why \texttt{SRN} is stronger in performance and comes close to \texttt{DeblurGAN-v2}. \selfdeblur\,that our method builds on obviously lags behind. From \cref{fig:real_table1,fig:real_table2,fig:real_table3,fig:real_table4}, we can see obvious texture artifacts in the image contents that \selfdeblur\,recovers, as well as boundary noise (especially in \cref{fig:real_table2,fig:real_table4}) due to the improper cropping used by \selfdeblur\,(discussed in \cref{sec:method_over_x}). 

\subsection{Failure cases and limitations}
\label{sec:failure}
We highlight three major factors that can cause failures: 1) substantial depth contrast that makes the uniform model less accurate; 2) kernel size overspecification that makes kernel estimation challenging; 3) inaccurate localization of the estimated $\wh{\mb x}$ that induces boundary noise. Below, we include a couple of failure examples and brief explanations resorting to these factors. 

In \cref{fig:real_table3}, we can see strip artifacts in the window region from both \selfdeblur\, and our method. We suspect that the strips are due to combined effects of 1) and 2) above. This is experimentally confirmed in \cref{fig:enhance_smallK} below: as we reduce the kernel-size overspecification, the strips are gone, but the recovered foreground floor region also becomes over-smooth and misses details.
\begin{figure}[!htbp] 
  \centering \includegraphics[width=0.98\linewidth]{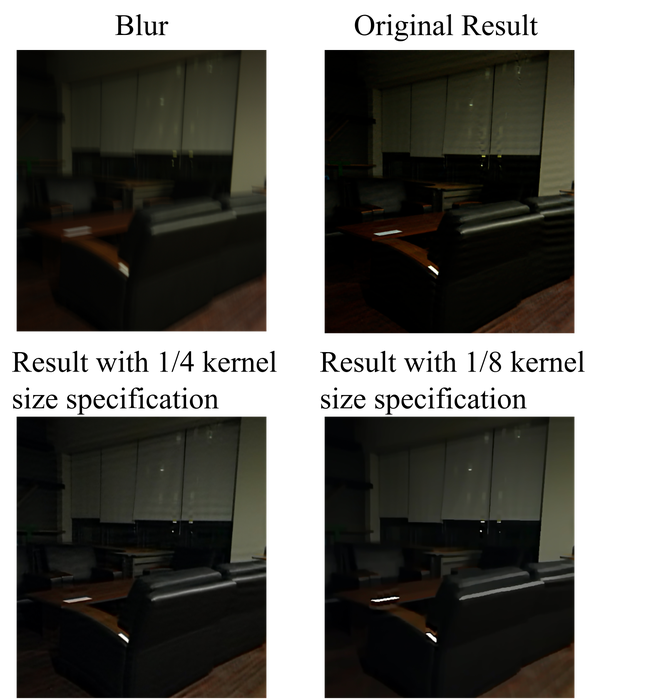}
  \caption{Effect of reducing the kernel-size overspecification in our method for \cref{fig:real_table3}} 
  \label{fig:enhance_smallK}
\end{figure} 

\cref{fig:real_failure} shows a difficult case 
 that fails all methods, including ours. The failure is likely due to: 1) huge depth contrast that violates the uniform model leading to both varying defocus and motion blurs. As is evident, DeblurGAN-v2, \selfdeblur, and ours are among the best performers, but they can only recover reasonable details in the foreground and not the far-away lights; 2) localization of the estimated $\wh{\mb x}$ specific to \selfdeblur and ours. We can see clear spurious light spots near the top-right corners of the reconstructions by both methods.
 
\begin{table}[!htbp]
  \caption{Sensitivity analysis of our method with respect to key hyperparameters. LR: learning rate; $\lambda_{\mb x}$: regularization parameter for the $\ell_1/\ell_2$ regularizer. For PSNR, SSIM, and VIF, higher the better. For LPIPS, lower the better. Default parameters and their results are highlighted in boldface.  }
  \centering 
  \begin{tabular}{c|c|c|c|c|c|c}
     \hline
      LR & $5\mathrm{e-}3$ & $\mb{1\mathrm{e-}3}$ & $5\mathrm{e-}4$ & $1\mathrm{e-}4$ & $5\mathrm{e-}4$ & $1\mathrm{e-}5$\\
      \hline
      PSNR& 26.9 & \textbf{29.3} &28.7 & 27.9& 27.8 &27.8\\
      SSIM & 0.774 & \textbf{0.869} & 0.828 & 0.813& 0.793 & 0.790\\
      VIF & 0.691 & \textbf{0.781} & 0.735 & 0.725& 0.716 & 0.709\\
      LPIPS& 0.972 & \textbf{0.844} & 0.875 & 0.901& 0.921 & 0.927\\
      \hline
  \end{tabular}
  \newline
  \vspace*{5mm}
  \newline
  \begin{tabular}{c|c|c|c|c|c|c}
     \hline
      $\lambda_{\mb x}$  & $5\mathrm{e-}4$ & $1\mathrm{e-}4$ & $\mb{1\mathrm{e-}5}$ & $5\mathrm{e-}5$ & $5\mathrm{e-}6$ & $1\mathrm{e-}6$\\
      \hline
      PSNR& 26.3 &27.7 & \textbf{29.3} & 28.3& 27.7 &27.2\\
      SSIM  & 0.763 & 0.813 & \textbf{0.869} & 0.822& 0.803 & 0.793\\
      VIF  & 0.681 & 0.725 & \textbf{0.781} & 0.745& 0.716 & 0.703\\
      LPIPS & 1.021 & 0.902 & \textbf{0.844} & 0.887& 0.925 & 0.931\\
      \hline
  \end{tabular}
  \label{tab:LR_ablation}
  \end{table}

\subsection{Ablation study}
Learning rates (for $\mb \theta_{\mb k}$ and $\mb \theta_{\mb x}$, respectively) and the regularization parameter $\lambda_{\mb x}$ are the two crucial groups of hyperparameters for our method. Hence, in this ablation study, we focus on these two factors, and perform experiments on the real-world images used in \cref{sec:exp_realworld}. We lock all other hyperparameters to our default setting. 

We lock the LR ratio for $\mb \theta_{\mb x}$ and $\mb \theta_{\mb k}$ to be $100:1$, and hence only specify the LR for $\mb \theta_{\mb x}$ when presenting the results. \cref{tab:LR_ablation} (top) includes the $6$ groups of LRs we have tried, and the resulting performance. When the LR is higher than $1\mathrm{e-}2$, the training fails to converge properly. When we decrease the LR below $1\mathrm{e-}2$, the perform degrades gradually. This is due to that the small LRs entail more iterations to converge, whereas we cap the maximum number of iterations for efficiency. 

The regularization parameter $\lambda_{\mb x}$ controls the trade-off between the data fitting and the enforcement of the sparse gradient prior (see \cref{eq:ours_main_setting}). We also vary $\lambda_{\mb x}$ across $6$ levels, covering the $1\mathrm{e-}4\sim 1\mathrm{e-}6$ range, and summarize the results in \cref{tab:LR_ablation} (bottom). We note that we take the mean of Huber loss over all pixels for the data fitting term, but the $\ell_1/\ell_2$ regularizer scales roughly as $O(\sqrt{\#\text{pixels}})$ which is around $1\mathrm{e}3$ for real-world color images. So the base $\lambda_{\mb x}$ should be $1\mathrm{e-}3$ to cancel out the dimension factor. Our optimal regularization level $1\mathrm{e-}5$ is hence $1\mathrm{e-}2$ in the effective level. Our method is stable when $\lambda_{\mb x}$ is on the $1\mathrm{e-}5$ level, and degrades considerably for levels above or below $1\mathrm{e-}5$.

\section{Discussion}\label{sec:discussion}
In this paper, we have proposed crucial modifications to the recent \selfdeblur\,method~\cite{RenEtAl2020Neural} for BID, and these modifications help successfully tackle the pressing practicality issues around BID: unknown kernel size, substantial noise, and model stability. Systematic evaluation of our method on both synthetic and real-world data confirms the effectiveness of our method. Remarkably, although our method only assumes the simple uniform blur model (i.e., \cref{eq:bd_model}), it performs comparably or superior to SOTA data-driven methods on real-world blurry images---these data-driven methods do not assume explicit forward models and hence are presumably much less constrained, but are limited by the expressiveness of their respective training data that are tricky to collect.

There are multiple directions to extend and generalize the current work. \textbf{First}, the performance of our method on real-world data likely can be further improved if we model non-uniform blur; our forthcoming work~\cite{ZhuangEtAl2023NBID} does exactly this. \textbf{Second}, similar to traditional BID methods that are based on iterative optimization, our method is slow compared to the emerging data-driven methods. One can possibly address this by designing compact DIP models that allow efficient optimization (see, e.g., \cite{LiEtAl2022Random}), and also by initializing the current DIP-based method using SOTA data-driven methods. \textbf{Third}, in principle our method can be readily extended to blind video deblurring, although it seems that one needs to address the increased modeling gap and computational cost. \textbf{Fourth}, the principle of modeling the object of interest by multiple DIP models or variants seems general for solving other inverse problems (see, e.g., our recent application of this to obtain breakthrough results in Fourier phase retrieval~\cite{YangEtAl2022Application,ZhuangEtAl2022Practical}).

\section*{Acknowledgements}
Zhong Zhuang, Hengkang Wang, and Ju Sun are partially supported by NSF CMMI 2038403. We thank the anonymous reviewers and the associate editor for their insightful comments that have substantially helped us improve the presentation of this paper. We thank Le Peng and Wenjie Zhang for allowing us to use the e-scooter image of \cref{fig:question} that they captured. The authors acknowledge the Minnesota Supercomputing Institute (MSI) at the University of Minnesota for providing resources that contributed to the research results reported within this paper.

\section*{Data availability statements}
Part of the code and datasets used during the current study, necessary to interpret, replicate and build upon the findings reported in the article, are available in the Github repository \url{https://github.com/sun-umn/Blind-Image-Deblurring}

{\footnotesize\bibliographystyle{spmpsci}\bibliography{BD}}

\begin{thebibliography}{100}
\providecommand{\url}[1]{{#1}}
\providecommand{\urlprefix}{URL }
\expandafter\ifx\csname urlstyle\endcsname\relax
  \providecommand{\doi}[1]{DOI~\discretionary{}{}{}#1}\else
  \providecommand{\doi}{DOI~\discretionary{}{}{}\begingroup
  \urlstyle{rm}\Url}\fi

\bibitem{AhmedEtAl2014Blind}
Ahmed, A., Recht, B., Romberg, J.: Blind deconvolution using convex
  programming.
\newblock IEEE Transactions on Information Theory \textbf{60}(3), 1711--1732
  (2014).
\newblock \doi{10.1109/tit.2013.2294644}

\bibitem{AljadaanyEtAl2019Douglas}
Aljadaany, R., Pal, D.K., Savvides, M.: Douglas-rachford networks: Learning
  both the image prior and data fidelity terms for blind image deconvolution.
\newblock In: 2019 {IEEE}/{CVF} Conference on Computer Vision and Pattern
  Recognition ({CVPR}). {IEEE} (2019).
\newblock \doi{10.1109/cvpr.2019.01048}

\bibitem{AsimEtAl2020Blind}
Asim, M., Shamshad, F., Ahmed, A.: Blind image deconvolution using deep
  generative priors.
\newblock IEEE Transactions on Computational Imaging \textbf{6}, 1493--1506
  (2020).
\newblock \doi{10.1109/tci.2020.3032671}

\bibitem{BenichouxEtAl2013fundamental}
Benichoux, A., Vincent, E., Gribonval, R.: A fundamental pitfall in blind
  deconvolution with sparse and shift-invariant priors.
\newblock In: IEEE International Conference on Acoustics, Speech and Signal
  Processing. {IEEE} (2013).
\newblock \doi{10.1109/icassp.2013.6638838}

\bibitem{BostanEtAl2020Deep}
Bostan, E., Heckel, R., Chen, M., Kellman, M., Waller, L.: Deep phase decoder:
  self-calibrating phase microscopy with an untrained deep neural network.
\newblock Optica \textbf{7}(6), 559--562 (2020)

\bibitem{Cabrelli1985Minimum}
Cabrelli, C.A.: Minimum entropy deconvolution and simplicity: A noniterative
  algorithm.
\newblock Geophysics \textbf{50}(3), 394--413 (1985).
\newblock \doi{10.1190/1.1441919}

\bibitem{ChanWong1998Total}
Chan, T., Wong, C.K.: Total variation blind deconvolution.
\newblock IEEE Transactions on Image Processing \textbf{7}(3), 370--375 (1998).
\newblock \doi{10.1109/83.661187}

\bibitem{ChenEtAl2019Blind}
Chen, L., Fang, F., Wang, T., Zhang, G.: Blind image deblurring with local
  maximum gradient prior.
\newblock In: {IEEE} Conference on computer vision and pattern recognition
  ({CVPR}). {IEEE} (2019).
\newblock \doi{10.1109/cvpr.2019.00184}

\bibitem{ChenEtAl2020OID}
Chen, L., Fang, F., Zhang, J., Liu, J., Zhang, G.: {OID}: Outlier identifying
  and discarding in blind image deblurring.
\newblock In: European Conference on Computer Vision ({ECCV}), pp. 598--613.
  Springer International Publishing (2020).
\newblock \doi{10.1007/978-3-030-58595-2_36}

\bibitem{ChenEtAl2021Blind}
Chen, L., Zhang, J., Lin, S., Fang, F., Ren, J.S.: Blind deblurring for
  saturated images.
\newblock In: {IEEE} Conference on computer vision and pattern recognition
  ({CVPR}). {IEEE} (2021).
\newblock \doi{10.1109/cvpr46437.2021.00624}

\bibitem{CheungEtAl2020Dictionary}
Cheung, S.C., Shin, J.Y., Lau, Y., Chen, Z., Sun, J., Zhang, Y., Müller, M.A.,
  Eremin, I.M., Wright, J.N., Pasupathy, A.N.: Dictionary learning in
  fourier-transform scanning tunneling spectroscopy.
\newblock Nature Communications \textbf{11}(1) (2020).
\newblock \doi{10.1038/s41467-020-14633-1}

\bibitem{Chi2016Guaranteed}
Chi, Y.: Guaranteed blind sparse spikes deconvolution via lifting and convex
  optimization.
\newblock IEEE Journal of Selected Topics in Signal Processing \textbf{10}(4),
  782--794 (2016).
\newblock \doi{10.1109/jstsp.2016.2543462}

\bibitem{ChoLee2009Fast}
Cho, S., Lee, S.: Fast motion deblurring.
\newblock In: ACM Trans. Graph. {ACM} Press (2009).
\newblock \doi{10.1145/1661412.1618491}

\bibitem{ChoLee2017Convergence}
Cho, S., Lee, S.: Convergence analysis of {MAP} based blur kernel estimation.
\newblock In: {IEEE} International conference on computer vision ({ICCV}).
  {IEEE} (2017).
\newblock \doi{10.1109/iccv.2017.515}

\bibitem{ChoudharyMitra2014Sparse}
Choudhary, S., Mitra, U.: Sparse blind deconvolution: What cannot be done.
\newblock In: IEEE International Symposium on Information Theory. {IEEE}
  (2014).
\newblock \doi{10.1109/isit.2014.6875385}

\bibitem{ChoudharyMitra2018Properties}
Choudhary, S., Mitra, U.: On the properties of the rank-two null space of
  nonsparse and canonical-sparse blind deconvolution.
\newblock {IEEE} Transactions on Signal Processing \textbf{66}(14), 3696--3709
  (2018).
\newblock \doi{10.1109/tsp.2018.2815014}

\bibitem{DarestaniHeckel2021Accelerated}
Darestani, M.Z., Heckel, R.: Accelerated {MRI} with un-trained neural networks.
\newblock IEEE Transactions on Computational Imaging \textbf{7}, 724--733
  (2021).
\newblock \doi{10.1109/tci.2021.3097596}

\bibitem{DingLuo2000fast}
Ding, Z., Luo, Z.Q.: A fast linear programming algorithm for blind
  equalization.
\newblock IEEE Transactions on Communications \textbf{48}(9), 1432--1436
  (2000).
\newblock \doi{10.1109/26.870004}

\bibitem{DongEtAl2017Blind}
Dong, J., Pan, J., Su, Z., Yang, M.H.: Blind image deblurring with outlier
  handling.
\newblock In: {IEEE} International conference on computer vision ({ICCV}).
  {IEEE} (2017).
\newblock \doi{10.1109/iccv.2017.271}

\bibitem{Donoho1981MINIMUM}
Donoho, D.: {ON} minimum entropy deconvolution.
\newblock In: Applied Time Series Analysis II, pp. 565--608. Elsevier (1981).
\newblock \doi{10.1016/b978-0-12-256420-8.50024-1}

\bibitem{EkanadhamEtAl2011blind}
Ekanadham, C., Tranchina, D., Simoncelli, E.: A blind sparse deconvolution
  method for neural spike identification.
\newblock In: Advances in Neural Information Processing Systems (2011)

\bibitem{FangEtAl2014Separable}
Fang, L., Liu, H., Wu, F., Sun, X., Li, H.: Separable kernel for image
  deblurring.
\newblock In: {IEEE} Conference on computer vision and pattern recognition
  ({CVPR}). {IEEE} (2014).
\newblock \doi{10.1109/cvpr.2014.369}

\bibitem{GandelsmanEtAl2019Double}
Gandelsman, Y., Shocher, A., Irani, M.:
  {\textquotedblleft}double-{DIP}{\textquotedblright}: Unsupervised image
  decomposition via coupled deep-image-priors.
\newblock In: {IEEE} Conference on computer vision and pattern recognition
  ({CVPR}). {IEEE} (2019).
\newblock \doi{10.1109/cvpr.2019.01128}

\bibitem{GongEtAl2017Self}
Gong, D., Tan, M., Zhang, Y., van~den Hengel, A., Shi, Q.: Self-paced kernel
  estimation for robust blind image deblurring.
\newblock In: {IEEE} International conference on computer vision ({ICCV}).
  {IEEE} (2017).
\newblock \doi{10.1109/iccv.2017.184}

\bibitem{GongEtAl2016Blind}
Gong, D., Tan, M., Zhang, Y., Hengel, A.V.D., Shi, Q.: Blind image
  deconvolution by automatic gradient activation.
\newblock In: {IEEE} Conference on computer vision and pattern recognition
  ({CVPR}). {IEEE} (2016).
\newblock \doi{10.1109/cvpr.2016.202}

\bibitem{HeckelHand2019Deep}
Heckel, R., Hand, P.: Deep decoder: Concise image representations from
  untrained non-convolutional networks.
\newblock In: International Conference on Learning Representations (2019)

\bibitem{HeckelSoltanolkotabi2019Denoising}
Heckel, R., Soltanolkotabi, M.: Denoising and regularization via exploiting the
  structural bias of convolutional generators.
\newblock arXiv preprint arXiv:1910.14634  (2019)

\bibitem{HeckelSoltanolkotabi2020Compressive}
Heckel, R., Soltanolkotabi, M.: Compressive sensing with un-trained neural
  networks: Gradient descent finds the smoothest approximation.
\newblock arXiv:2005.03991  (2020)

\bibitem{HendrycksDietterich2019Benchmarking}
Hendrycks, D., Dietterich, T.: Benchmarking neural network robustness to common
  corruptions and perturbations.
\newblock In: International Conference on Learning Representations (2019).
\newblock \urlprefix\url{https://openreview.net/forum?id=HJz6tiCqYm}

\bibitem{Huber1964Robust}
Huber, P.J.: Robust estimation of a location parameter.
\newblock The Annals of Mathematical Statistics \textbf{35}(1), 73--101 (1964).
\newblock \doi{10.1214/aoms/1177703732}

\bibitem{HurleyRickard2009Comparing}
Hurley, N., Rickard, S.: Comparing measures of sparsity.
\newblock IEEE Transactions on Information Theory \textbf{55}(10), 4723--4741
  (2009).
\newblock \doi{10.1109/tit.2009.2027527}

\bibitem{JinEtAl2018Normalized}
Jin, M., Roth, S., Favaro, P.: Normalized blind deconvolution.
\newblock In: European Conference on Computer Vision ({ECCV}), pp. 694--711.
  Springer International Publishing (2018).
\newblock \doi{10.1007/978-3-030-01234-2_41}

\bibitem{JoshiEtAl2008PSF}
Joshi, N., Szeliski, R., Kriegman, D.J.: {PSF} estimation using sharp edge
  prediction.
\newblock In: IEEE Conference on computer vision and pattern recognition
  ({CVPR}). {IEEE} (2008).
\newblock \doi{10.1109/cvpr.2008.4587834}

\bibitem{JoshiEtAl2009Image}
Joshi, N., Zitnick, C.L., Szeliski, R., Kriegman, D.J.: Image deblurring and
  denoising using color priors.
\newblock In: IEEE Conference on computer vision and pattern recognition
  ({CVPR}). {IEEE} (2009).
\newblock \doi{10.1109/cvpr.2009.5206802}

\bibitem{KechKrahmer2017Optimal}
Kech, M., Krahmer, F.: Optimal injectivity conditions for bilinear inverse
  problems with applications to identifiability of deconvolution problems.
\newblock {SIAM} Journal on Applied Algebra and Geometry \textbf{1}(1), 20--37
  (2017).
\newblock \doi{10.1137/16m1067469}

\bibitem{KohEtAl2021Single}
Koh, J., Lee, J., Yoon, S.: Single-image deblurring with neural networks: A
  comparative survey.
\newblock Computer Vision and Image Understanding \textbf{203}, 103134 (2021).
\newblock \doi{10.1016/j.cviu.2020.103134}

\bibitem{KomodakisParagios2013MRF}
Komodakis, N., Paragios, N.: {MRF}-based blind image deconvolution.
\newblock In: Asian Conference on Computer Vision ({ACCV}), pp. 361--374.
  Springer Berlin Heidelberg (2013).
\newblock \doi{10.1007/978-3-642-37431-9_28}

\bibitem{KrishnanFergus2009Fast}
Krishnan, D., Fergus, R.: Fast image deconvolution using hyper-laplacian
  priors.
\newblock In: Advances in Neural Information Processing Systems (2009).
\newblock
  \urlprefix\url{https://proceedings.neurips.cc/paper/2009/file/3dd48ab31d016ffcbf3314df2b3cb9ce-Paper.pdf}

\bibitem{KrishnanEtAl2011Blind}
Krishnan, D., Tay, T., Fergus, R.: Blind deconvolution using a normalized
  sparsity measure.
\newblock In: {IEEE} Conference on computer vision and pattern recognition
  ({CVPR}). {IEEE} (2011).
\newblock \doi{10.1109/cvpr.2011.5995521}

\bibitem{KundurHatzinakos1996Blind}
Kundur, D., Hatzinakos, D.: Blind image deconvolution.
\newblock IEEE Signal Processing Magazine \textbf{13}(3), 43--64 (1996).
\newblock \doi{10.1109/79.489268}

\bibitem{KuoEtAl2020Geometry}
Kuo, H.W., Zhang, Y., Lau, Y., Wright, J.: Geometry and symmetry in
  short-and-sparse deconvolution.
\newblock SIAM Journal on Mathematics of Data Science \textbf{2}(1), 216--245
  (2020).
\newblock \doi{10.1137/19m1237569}

\bibitem{kupyn2019deblurgan}
Kupyn, O., Martyniuk, T., Wu, J., Wang, Z.: Deblurgan-v2: Deblurring
  (orders-of-magnitude) faster and better.
\newblock In: Proceedings of the IEEE/CVF International Conference on Computer
  Vision, pp. 8878--8887 (2019)

\bibitem{KoehlerEtAl2012Recording}
Köhler, R., Hirsch, M., Mohler, B., Schölkopf, B., Harmeling, S.: Recording
  and playback of camera shake: Benchmarking blind deconvolution with a
  real-world database.
\newblock In: European Conference on Computer Vision ({ECCV}), pp. 27--40.
  Springer Berlin Heidelberg (2012).
\newblock \doi{10.1007/978-3-642-33786-4_3}

\bibitem{LaiEtAl2016Comparative}
Lai, W.S., Huang, J.B., Hu, Z., Ahuja, N., Yang, M.H.: A comparative study for
  single image blind deblurring.
\newblock In: {IEEE} Conference on computer vision and pattern recognition
  ({CVPR}). {IEEE} (2016).
\newblock \doi{10.1109/cvpr.2016.188}

\bibitem{LawrenceEtAl2020Phase}
Lawrence, H., Bramherzig, D., Li, H., Eickenberg, M., Gabri{\'e}, M.: Phase
  retrieval with holography and untrained priors: Tackling the challenges of
  low-photon nanoscale imaging.
\newblock arXiv preprint arXiv:2012.07386  (2020)

\bibitem{LevinEtAl2011Understanding}
Levin, A., Weiss, Y., Durand, F., Freeman, W.T.: Understanding blind
  deconvolution algorithms.
\newblock IEEE Transactionson Pattern Analysis and Machine Intelligence
  \textbf{33}(12), 2354--2367 (2011).
\newblock \doi{10.1109/tpami.2011.148}

\bibitem{Lewicki1998review}
Lewicki, M.S.: A review of methods for spike sorting: the detection and
  classification of neural action potentials.
\newblock Network: Computation in Neural Systems \textbf{9}(4), R53--R78
  (1998).
\newblock \doi{10.1088/0954-898x_9_4_001}

\bibitem{LiEtAl2018Learning}
Li, L., Pan, J., Lai, W.S., Gao, C., Sang, N., Yang, M.H.: Learning a
  discriminative prior for blind image deblurring.
\newblock In: 2018 {IEEE}/{CVF} Conference on Computer Vision and Pattern
  Recognition. {IEEE} (2018).
\newblock \doi{10.1109/cvpr.2018.00692}

\bibitem{LiEtAl2022Random}
Li, T., Wang, H., Zhuang, Z., Sun, J.: Deep random projector: Accelerated deep
  image prior.
\newblock In: Proceedings of the IEEE/CVF Conference on Computer Vision and
  Pattern Recognition (CVPR), pp. 18176--18185 (2023)

\bibitem{LiEtAl2021Self}
Li, T., Zhuang, Z., Liang, H., Peng, L., Wang, H., Sun, J.: Self-validation:
  Early stopping for single-instance deep generative priors.
\newblock In: British Machine Vision Conference ({BMVC}) (2021)

\bibitem{LiEtAl2019Rapid}
Li, X., Ling, S., Strohmer, T., Wei, K.: Rapid, robust, and reliable blind
  deconvolution via nonconvex optimization.
\newblock Applied and Computational Harmonic Analysis \textbf{47}(3), 893--934
  (2019).
\newblock \doi{10.1016/j.acha.2018.01.001}

\bibitem{LiEtAl2015Unified}
Li, Y., Lee, K., Bresler, Y.: A unified framework for identifiability analysis
  in bilinear inverse problems with applications to subspace and sparsity
  models.
\newblock arXiv:1501.06120  (2015)

\bibitem{LiEtAl2017Identifiability}
Li, Y., Lee, K., Bresler, Y.: Identifiability and stability in blind
  deconvolution under minimal assumptions.
\newblock IEEE Transactions on Information Theory \textbf{63}(7), 4619--4633
  (2017).
\newblock \doi{10.1109/tit.2017.2689779}

\bibitem{LiEtAl2019Deep}
Li, Y., Tofighi, M., Geng, J., Monga, V., Eldar, Y.C.: Deep algorithm unrolling
  for blind image deblurring.
\newblock arXiv:1902.03493  (2019)

\bibitem{LiuEtAl2018Deblurring}
Liu, Y., Dong, W., Gong, D., Zhang, L., Shi, Q.: Deblurring natural image using
  super-gaussian fields.
\newblock In: European Conference on Computer Vision ({ECCV}), pp. 467--484.
  Springer International Publishing (2018).
\newblock \doi{10.1007/978-3-030-01246-5_28}

\bibitem{MaEtAl2021Unsupervised}
Ma, X., Hill, P., Achim, A.: Unsupervised image fusion using deep image priors.
\newblock arXiv:2110.09490  (2021)

\bibitem{MichaeliIrani2014Blind}
Michaeli, T., Irani, M.: Blind deblurring using internal patch recurrence.
\newblock In: European Conference on Computer Vision ({ECCV}), pp. 783--798.
  Springer International Publishing (2014).
\newblock \doi{10.1007/978-3-319-10578-9_51}

\bibitem{MichelashviliWolf2019Speech}
Michelashvili, M., Wolf, L.: Speech denoising by accumulating per-frequency
  modeling fluctuations.
\newblock arXiv:1904.07612  (2019)

\bibitem{NahEtAl2019NTIRE}
Nah, S., Baik, S., Hong, S., Moon, G., Son, S., Timofte, R., Lee, K.M.: {NTIRE}
  2019 challenge on video deblurring and super-resolution: Dataset and study.
\newblock In: 2019 {IEEE}/{CVF} Conference on Computer Vision and Pattern
  Recognition Workshops ({CVPRW}). {IEEE} (2019).
\newblock \doi{10.1109/cvprw.2019.00251}

\bibitem{NahEtAl2017Deep}
Nah, S., Kim, T.H., Lee, K.M.: Deep multi-scale convolutional neural network
  for dynamic scene deblurring.
\newblock In: 2017 {IEEE} Conference on Computer Vision and Pattern Recognition
  ({CVPR}). {IEEE} (2017).
\newblock \doi{10.1109/cvpr.2017.35}

\bibitem{NahEtAl2021NTIRE}
Nah, S., Son, S., Lee, S., Timofte, R., Lee, K.M.: Ntire 2021 challenge on
  image deblurring.
\newblock arXiv:2104.14854  (2021)

\bibitem{NahEtAl2020NTIRE}
Nah, S., Son, S., Timofte, R., Lee, K.M.: {NTIRE} 2020 challenge on image and
  video deblurring.
\newblock arXiv:2005.01244  (2020)

\bibitem{OngieEtAl2020Deep}
Ongie, G., Jalal, A., Metzler, C.A., Baraniuk, R.G., Dimakis, A.G., Willett,
  R.: Deep learning techniques for inverse problems in imaging.
\newblock {IEEE} Journal on Selected Areas in Information Theory \textbf{1}(1),
  39--56 (2020).
\newblock \doi{10.1109/jsait.2020.2991563}

\bibitem{PanEtAl2021Physics}
Pan, J., Dong, J., Liu, Y., Zhang, J., Ren, J., Tang, J., Tai, Y.W., Yang,
  M.H.: Physics-based generative adversarial models for image restoration and
  beyond.
\newblock {IEEE} Transactions on Pattern Analysis and Machine Intelligence
  \textbf{43}(7), 2449--2462 (2021).
\newblock \doi{10.1109/tpami.2020.2969348}

\bibitem{PanEtAl2014Deblurring}
Pan, J., Hu, Z., Su, Z., Yang, M.H.: Deblurring text images via l0-regularized
  intensity and gradient prior.
\newblock In: {IEEE} Conference on computer vision and pattern recognition
  ({CVPR}). {IEEE} (2014).
\newblock \doi{10.1109/cvpr.2014.371}

\bibitem{PanEtAl2016Robust}
Pan, J., Lin, Z., Su, Z., Yang, M.H.: Robust kernel estimation with outliers
  handling for image deblurring.
\newblock In: {IEEE} Conference on computer vision and pattern recognition
  ({CVPR}). {IEEE} (2016).
\newblock \doi{10.1109/cvpr.2016.306}

\bibitem{PanEtAl2016Blind}
Pan, J., Sun, D., Pfister, H., Yang, M.H.: Blind image deblurring using dark
  channel prior.
\newblock In: {IEEE} Conference on computer vision and pattern recognition
  ({CVPR}). {IEEE} (2016).
\newblock \doi{10.1109/cvpr.2016.180}

\bibitem{PerroneFavaro2014Total}
Perrone, D., Favaro, P.: Total variation blind deconvolution: The devil is in
  the details.
\newblock In: {IEEE} Conference on computer vision and pattern recognition
  ({CVPR}). {IEEE} (2014).
\newblock \doi{10.1109/cvpr.2014.372}

\bibitem{QayyumEtAl2021Untrained}
Qayyum, A., Ilahi, I., Shamshad, F., Boussaid, F., Bennamoun, M., Qadir, J.:
  Untrained neural network priors for inverse imaging problems: A survey.
\newblock TechRxiv  (2021).
\newblock \doi{10.36227/techrxiv.14208215}

\bibitem{RavulaDimakis2019One}
Ravula, S., Dimakis, A.G.: One-dimensional deep image prior for time series
  inverse problems.
\newblock arXiv:1904.08594  (2019)

\bibitem{RenEtAl2020Neural}
Ren, D., Zhang, K., Wang, Q., Hu, Q., Zuo, W.: Neural blind deconvolution using
  deep priors.
\newblock In: {IEEE} Conference on computer vision and pattern recognition
  ({CVPR}). {IEEE} (2020).
\newblock \doi{10.1109/cvpr42600.2020.00340}

\bibitem{RimEtAl2020Real}
Rim, J., Lee, H., Won, J., Cho, S.: Real-world blur dataset for learning and
  benchmarking deblurring algorithms.
\newblock In: European Conference on Computer Vision ({ECCV}), pp. 184--201.
  Springer International Publishing (2020).
\newblock \doi{10.1007/978-3-030-58595-2_12}

\bibitem{SchulerEtAl2016Learning}
Schuler, C.J., Hirsch, M., Harmeling, S., Scholkopf, B.: Learning to deblur.
\newblock {IEEE} Transactions on Pattern Analysis and Machine Intelligence
  \textbf{38}(7), 1439--1451 (2016).
\newblock \doi{10.1109/tpami.2015.2481418}

\bibitem{SheikhBovik2006Image}
Sheikh, H., Bovik, A.: Image information and visual quality.
\newblock {IEEE} Transactions on Image Processing \textbf{15}(2), 430--444
  (2006).
\newblock \doi{10.1109/tip.2005.859378}

\bibitem{ShiEtAl2022Measuring}
Shi, Z., Mettes, P., Maji, S., Snoek, C.G.M.: On measuring and controlling the
  spectral bias of the deep image prior.
\newblock International Journal of Computer Vision \textbf{130}(4), 885--908
  (2022).
\newblock \doi{10.1007/s11263-021-01572-7}

\bibitem{SiYaoEtAl2019Understanding}
Si-Yao, L., Ren, D., Yin, Q.: Understanding kernel size in blind deconvolution.
\newblock In: IEEE Winter Conference on Applications of Computer Vision
  ({WACV}). {IEEE} (2019).
\newblock \doi{10.1109/wacv.2019.00224}

\bibitem{SitzmannEtAl2020Implicit}
Sitzmann, V., Martel, J., Bergman, A., Lindell, D., Wetzstein, G.: Implicit
  neural representations with periodic activation functions.
\newblock Advances in Neural Information Processing Systems \textbf{33} (2020)

\bibitem{SunEtAl2013Edge}
Sun, L., Cho, S., Wang, J., Hays, J.: Edge-based blur kernel estimation using
  patch priors.
\newblock In: IEEE International Conference on Computational Photography
  (ICCP). {IEEE} (2013).
\newblock \doi{10.1109/iccphot.2013.6528301}

\bibitem{SunDonoho2021Convex}
Sun, Q., Donoho, D.: Convex sparse blind deconvolution.
\newblock arXiv:2106.07053  (2021)

\bibitem{Szeliski2021Computer}
Szeliski, R.: Computer Vision: Algorithms and Applications, 2nd edn.
\newblock Springer London (2021)

\bibitem{TaiLin2012Motion}
Tai, Y.W., Lin, S.: Motion-aware noise filtering for deblurring of noisy and
  blurry images.
\newblock In: {IEEE} Conference on computer vision and pattern recognition
  ({CVPR}). {IEEE} (2012).
\newblock \doi{10.1109/cvpr.2012.6247653}

\bibitem{TancikEtAl2020Fourier}
Tancik, M., Srinivasan, P., Mildenhall, B., Fridovich-Keil, S., Raghavan, N.,
  Singhal, U., Ramamoorthi, R., Barron, J., Ng, R.: Fourier features let
  networks learn high frequency functions in low dimensional domains.
\newblock In: Advances in Neural Information Processing Systems (2020)

\bibitem{tao2018scale}
Tao, X., Gao, H., Shen, X., Wang, J., Jia, J.: Scale-recurrent network for deep
  image deblurring.
\newblock In: Proceedings of the IEEE conference on computer vision and pattern
  recognition, pp. 8174--8182 (2018)

\bibitem{TayalEtAl2021Phase}
Tayal, K., Manekar, R., Zhuang, Z., Yang, D., Kumar, V., Hofmann, F., Sun, J.:
  Phase retrieval using single-instance deep generative prior.
\newblock In: {OSA} Optical Sensors and Sensing Congress 2021 ({AIS}, {FTS},
  {HISE}, {SENSORS}, {ES}). {OSA} (2021).
\newblock \doi{10.1364/ais.2021.jw2a.37}

\bibitem{TranEtAl2021Explore}
Tran, P., Tran, A., Phung, Q., Hoai, M.: Explore image deblurring via encoded
  blur kernel space.
\newblock In: Proceedings of the {IEEE} Conference on Computer Vision and
  Pattern Recognition ({CVPR}) (2021).
\newblock \doi{10.1109/CVPR46437.2021.01178}

\bibitem{UlyanovEtAl2020Deep}
Ulyanov, D., Vedaldi, A., Lempitsky, V.: Deep image prior.
\newblock International Journal of Computer Vision \textbf{128}(7), 1867--1888
  (2020).
\newblock \doi{10.1007/s11263-020-01303-4}

\bibitem{Vasu2021Image}
Vasu, S.: Image and video deblurring: A curated list of resources for image and
  video deblurring.
\newblock \url{https://github.com/subeeshvasu/Awesome-Deblurring} (2021).
\newblock \urlprefix\url{https://github.com/subeeshvasu/Awesome-Deblurring}.
\newblock Accessed: Dec 12 2021

\bibitem{VembuEtAl1994Convex}
Vembu, S., Verdu, S., Kennedy, R., Sethares, W.: Convex cost functions in blind
  equalization.
\newblock IEEE Transactions on Signal Processing \textbf{42}(8), 1952--1960
  (1994).
\newblock \doi{10.1109/78.301833}

\bibitem{WangEtAl2021Early}
Wang, H., Li, T., Zhuang, Z., Chen, T., Liang, H., Sun, J.: Early stopping for
  deep image prior.
\newblock arXiv:2112.06074  (2021)

\bibitem{WangEtAl2019Image}
Wang, Z., Wang, Z., Li, Q., Bilen, H.: Image deconvolution with deep image and
  kernel priors.
\newblock In: 2019 {IEEE}/{CVF} International Conference on Computer Vision
  Workshop ({ICCVW}). {IEEE} (2019).
\newblock \doi{10.1109/iccvw.2019.00127}

\bibitem{Wiggins1978Minimum}
Wiggins, R.A.: Minimum entropy deconvolution.
\newblock Geoexploration \textbf{16}(1-2), 21--35 (1978).
\newblock \doi{10.1016/0016-7142(78)90005-4}

\bibitem{williams2019deep}
Williams, F., Schneider, T., Silva, C., Zorin, D., Bruna, J., Panozzo, D.: Deep
  geometric prior for surface reconstruction.
\newblock arXiv:1811.10943  (2019)

\bibitem{WipfZhang2014Revisiting}
Wipf, D., Zhang, H.: Revisiting bayesian blind deconvolution.
\newblock Journal of Machine Learning Research \textbf{15}(111), 3775--3814
  (2014)

\bibitem{XuJia2010Two}
Xu, L., Jia, J.: Two-phase kernel estimation for robust motion deblurring.
\newblock In: European Conference on Computer Vision, pp. 157--170. Springer
  Berlin Heidelberg (2010).
\newblock \doi{10.1007/978-3-642-15549-9_12}

\bibitem{XuEtAl2013Unnatural}
Xu, L., Zheng, S., Jia, J.: Unnatural l0 sparse representation for natural
  image deblurring.
\newblock In: {IEEE} Conference on computer vision and pattern recognition
  ({CVPR}). {IEEE} (2013).
\newblock \doi{10.1109/cvpr.2013.147}

\bibitem{YanEtAl2017Image}
Yan, Y., Ren, W., Guo, Y., Wang, R., Cao, X.: Image deblurring via extreme
  channels prior.
\newblock In: {IEEE} Conference on computer vision and pattern recognition
  ({CVPR}). {IEEE} (2017).
\newblock \doi{10.1109/cvpr.2017.738}

\bibitem{YangEtAl2022Application}
Yang, D., Zhuang, Z., Phillips, N.W., KaySong, Zdora, M.C., Harder, R., Cha,
  W., Liu, W., Barmherzig, D.A., Sun, J., Hofmann, F.: Application of
  single-instance deep generative priors for reconstruction of highly strained
  gold microcrystals in bragg coherent x-ray diffraction.
\newblock In preparation  (2022)

\bibitem{YangJi2019Variational}
Yang, L., Ji, H.: A variational {EM} framework with adaptive edge selection for
  blind motion deblurring.
\newblock In: {IEEE} Conference on computer vision and pattern recognition
  ({CVPR}). {IEEE} (2019).
\newblock \doi{10.1109/cvpr.2019.01041}

\bibitem{zhang2020deblurring}
Zhang, K., Luo, W., Zhong, Y., Ma, L., Stenger, B., Liu, W., Li, H.: Deblurring
  by realistic blurring.
\newblock In: Proceedings of the IEEE/CVF Conference on Computer Vision and
  Pattern Recognition, pp. 2737--2746 (2020)

\bibitem{ZhangEtAl2022Deep}
Zhang, K., Ren, W., Luo, W., Lai, W.S., Stenger, B., Yang, M.H., Li, H.: Deep
  image deblurring: A survey.
\newblock International Journal of Computer Vision \textbf{130}(9), 2103--2130
  (2022).
\newblock \doi{10.1007/s11263-022-01633-5}

\bibitem{ZhangEtAl2019Deep}
Zhang, K., Zuo, W., Zhang, L.: Deep plug-and-play super-resolution for
  arbitrary blur kernels.
\newblock In: 2019 {IEEE}/{CVF} Conference on Computer Vision and Pattern
  Recognition ({CVPR}). {IEEE} (2019).
\newblock \doi{10.1109/cvpr.2019.00177}

\bibitem{zhang2018unreasonable}
Zhang, R., Isola, P., Efros, A.A., Shechtman, E., Wang, O.: The unreasonable
  effectiveness of deep features as a perceptual metric (2018)

\bibitem{ZhangEtAl2020Structured}
Zhang, Y., Kuo, H.W., Wright, J.: Structured local optima in sparse blind
  deconvolution.
\newblock IEEE Transactions on Information Theory \textbf{66}(1), 419--452
  (2020).
\newblock \doi{10.1109/tit.2019.2940657}

\bibitem{ZhangEtAl2017Global}
Zhang, Y., Lau, Y., Kuo, H.W., Cheung, S., Pasupathy, A., Wright, J.: On the
  global geometry of sphere-constrained sparse blind deconvolution.
\newblock In: {IEEE} Conference on computer vision and pattern recognition
  ({CVPR}). {IEEE} (2017).
\newblock \doi{10.1109/cvpr.2017.466}

\bibitem{ZhongEtAl2013Handling}
Zhong, L., Cho, S., Metaxas, D., Paris, S., Wang, J.: Handling noise in single
  image deblurring using directional filters.
\newblock In: {IEEE} Conference on computer vision and pattern recognition
  ({CVPR}). {IEEE} (2013).
\newblock \doi{10.1109/cvpr.2013.85}

\bibitem{ZhouHorstmeyer2020Diffraction}
Zhou, K.C., Horstmeyer, R.: Diffraction tomography with a deep image prior.
\newblock Optics Express \textbf{28}(9), 12872 (2020).
\newblock \doi{10.1364/oe.379200}

\bibitem{ZhuangEtAl2023NBID}
Zhuang, Z., Li, T., Wang, H., Zhang, W., Sun, J.: Practical blind image
  deblurring with non-uniform blurs.
\newblock In preparation  (2023)

\bibitem{ZhuangEtAl2022Practical}
Zhuang, Z., Yang, D., Hofmann, F., Barmherzig, D., Sun, J.: Practical phase
  retrieval using double deep image priors.
\newblock arXiv preprint arXiv:2211.00799  (2022)

\end{thebibliography}

\section{Appendix}\label{sec:appendix}
\subsection{List of common acronyms}
\begin{table}[!htbp]
  \centering 
  \caption{List of acronyms (in alphabetic order)}
  \begin{tabular}{l  c}
    \hline
      BID   & blind image deblurring \\
      BD   & blind deconvolution \\
      DIP  & deep image prior \\
      DL & deep learning \\
      DNN  & deep neural network \\
      ES   & early stopping \\
      LPIPS & learned perceptual image patch similarity\\
      LR   & learning rate  \\
      MAP  & maximum a posterior \\ 
      MLP  & multi-layer perceptron \\ 
      MSE   & mean squared error \\
      PSNR   & peak signal-to-noise ratio \\ 
      SIREN  & sinusoidal representation networks \\ 
      SOTA  & state-of-the-art  \\
      SSBD  & short-and-sparse blind deconvolution \\
      SSIM  & structural similarity index measure  \\
      TV   & total-variation \\
      VAR  & variance  \\ 
      VIF  &  visual information fidelity \\
      VIP  & visual inverse problem \\ 
      WMV-ES & windowed-moving-variance-based ES \\
    \hline
  \end{tabular}
\end{table}

\subsection{Contrast-enhanced version of \cref{fig:real_table2,fig:real_table4}} \label{sec:app_hist_enhance}
To reveal more details for images in \cref{fig:real_table2,fig:real_table4} that are about extremely dark scenes, we perform histogram equalization to enhance the contrast and display the results as follows. 
\begin{figure}[!htbp] 
  \centering 
  \includegraphics[width=0.75\linewidth]{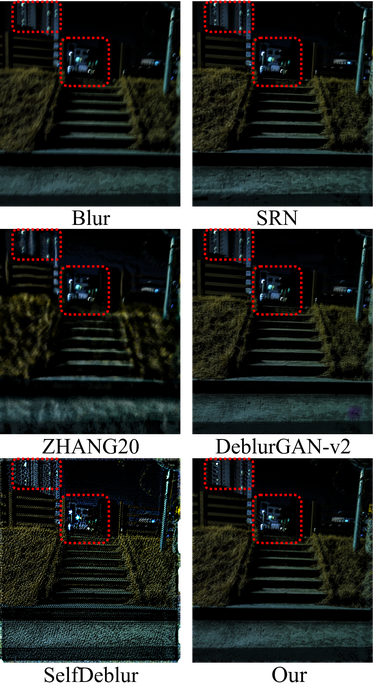}
  \caption{Contrast-enhanced version of \cref{fig:real_table2} after histogram equalization.} 
  \label{fig:real_table2_enhance}
\end{figure} 
\begin{figure}[!htbp] 
  \centering 
  \includegraphics[width=0.75\linewidth]{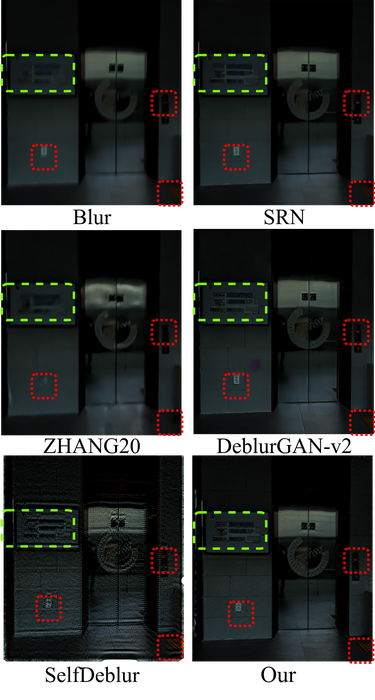}
  \caption{Contrast-enhanced version of \cref{fig:real_table4} after histogram equalization.} 
  \label{fig:real_table4_enhance}
\end{figure} 

\end{document}